\documentclass[a4paper,11pt]{article}
\pdfoutput=1 
\usepackage[symbol]{footmisc} 
\usepackage{jheppub} 

\usepackage{color, colortbl}
\usepackage[T1]{fontenc} 
\usepackage[utf8]{inputenc}
\usepackage{float}
\usepackage[font=footnotesize, justification=centering]{caption}
\usepackage{physics}
\usepackage{epsfig,latexsym}
\usepackage[compat=1.1.0]{ tikz-feynman }
\usepackage{ tikz-feynhand }
\usepackage{amsmath}
\usepackage{amssymb}
\usepackage{cancel}
\usepackage{tensor}
\usepackage{simpler-wick}
\usepackage{slashed}

\DeclareMathAlphabet{\mathpzc}{OT1}{pzc}{m}{it}
\allowdisplaybreaks

\newcommand{\dlo}[1]{\ensuremath{\partial_{#1}}}
\newcommand{\dup}[1]{\ensuremath{\partial^{#1}}}

\renewcommand{\epsilon}{\varepsilon}

\newcommand{\str}[1]{\ensuremath{\text{str}\left(#1\right)}}
\newcommand{\strsans}{\ensuremath{\text{str}}}

\newcommand{\tA}{{\ensuremath{\tilde{A}}}}
\newcommand{\tB}{{\ensuremath{\tilde{B}}}}
\newcommand{\tC}{{\ensuremath{\tilde{C}}}}
\newcommand{\tD}{{\ensuremath{\tilde{D}}}}
\newcommand{\tI}{{\ensuremath{\tilde{I}}}}
\newcommand{\tJ}{{\ensuremath{\tilde{J}}}}
\newcommand{\tK}{{\ensuremath{\tilde{K}}}}
\newcommand{\tL}{{\ensuremath{\tilde{L}}}}

\newcommand{\tM}{{\ensuremath{\tilde{M}}}}

\newcommand{\Str}{\ensuremath{\mathcal{\text{Str}}}}
\newcommand{\degf}{\mathrm{f}}
\newcommand{\degg}{\mathrm{g}}
\newcommand{\gradedcomm}[2]{{\comm{#1}{#2}_{\degf}}}
\newcommand{\quartic}{\kappa}
\newcommand{\quarticSigma}{\kappa}
\newcommand{\quarticSigmaA}{\kappa_1}
\newcommand{\quarticSigmaB}{\kappa_2}

\DeclareMathAlphabet{\mathpzc}{OT1}{pzc}{m}{it}
\allowdisplaybreaks

\title{Loops in supergroups 
}

\author[a,b]{Nathaniel Craig,}
\author[c]{Emanuele Gendy,}
\author[a]{Jessica N. Howard}

\affiliation[a]{Kavli Institute for Theoretical Physics, University of California, Santa Barbara, CA 93106, USA}
\affiliation[b]{Department of Physics, University of California, Santa Barbara, CA 93106, USA}
\affiliation[c]{Technische Universit\"{a}t M\"{u}nchen, Physik-Department, 85748 Garching, Germany}

\emailAdd{ncraig@ucsb.edu}
\emailAdd{emanuele.gendy@tum.de}
\emailAdd{jnhoward@kitp.ucsb.edu}

\abstract{
We study the theory of a scalar in the fundamental representation of the internal supergroup $SU(N|M)$. Remarkably, for $M=N+1$ its tree-level mass does not receive quantum corrections at one loop from either self-coupling or interactions with gauge bosons and fermions. This property comes at the price of introducing both degrees of freedom with wrong statistics and with wrong sign kinetic terms. We detail a method to break $SU(N|M)$ down to its bosonic subgroup through a Higgs-like mechanism, allowing for the partial decoupling of the dangerous modes, and study the associated vacuum structure up to one loop.}

\begin{document}
    \begin{flushright}
    TUM-HEP-1524/24
	\end{flushright}
	\maketitle
	
	\flushbottom
    \section{Introduction}
    \label{sec:introduction}

Symmetries have long guided the pursuit of physics beyond the Standard Model. The space of available symmetries is delineated by the Coleman-Mandula theorem \cite{Coleman:1967ad}, which confines the symmetry group of a massive, interacting theory (subject to various restrictions, including positive-energy representations) to a direct product of the Poincar\'{e} group and an internal symmetry group. 
As with many no-go theorems, its exceptions are as interesting as the rule itself: allowing for spinorial charges leads to spacetime supersymmetry~\cite{Haag:1974qh}, foregoing a mass gap admits conformal symmetry~\cite{Haag:1974qh}, and charging extended objects opens the door to the vast space of generalized symmetries~\cite{Gaiotto:2014kfa}. 
Such symmetries -- whether exact, explicitly broken, or spontaneously broken -- have lent tremendous insight into the structure of the Standard Model and defined the landscape of its possible extensions. But mysteries remain, from the value of the cosmological constant to the mass of the Higgs. 
The fact that these mysteries have thus far resisted explanation in terms of conventional symmetries suggests that it is worth asking whether something might be gained by exploring less conventional candidates.

One such possibility is internal supersymmetry, i.e., an internal symmetry based on a Lie supergroup such as $SU(N|M)$. This is an unconventional symmetry for good reason, as a relativistic theory with a supergroup internal symmetry necessarily features wrong-sign and wrong-statistics ghosts. The unitarity violation implied by these negative-norm states \cite{pauli} is likely fatal to the theory \footnote{See also \cite{Kubo:2023lpz, Kubo:2024ysu} for a modern take on the issue.}, although various attempts have been made at perturbative unitary interpretations in related theories \cite{Lee:1969fy, Cutkosky:1969fq, vanTonder:2006ye, vanTonder:2008ub, Grinstein:2008bg, Anselmi:2017lia, Anselmi:2018kgz, Donoghue:2019fcb}. Non-perturbative evidence for sensible $SU(N|M)$ theories is decidedly mixed: while string theory on stacks of $N$ ordinary D-branes and $M$ negative D-branes gives rise to a $\mathcal{N} = 4$ supersymmetric $U(N|M)$ supergroup gauge theory at low energies and provides a successful prescription for constructing the Seiberg-Witten curve for $\mathcal{N} = 2$ $SU(N|M)$ gauge theories \cite{Dijkgraaf:2016lym}, negative-tension branes come with their own pathologies. Nonetheless, as long as there remains some remote hope for a unitary interpretation, it is worth asking what new insights supergroup internal symmetries might offer in the search for new physics.\footnote{Supergroup symmetries have already appeared in a number of phenomenological settings complementary to the applications in this paper, including Lagrangian formulations of quenched QCD \cite{Bernard:1992mk} and an $SU(2|1)$ completion of the $SU(2) \times U(1)$ electroweak theory \cite{Neeman:1979wp, Fairlie:1979at, Dondi:1979ib, Taylor:1979sm}.} 

Perhaps the most notable virtue is finiteness. While it has long been understood how to handle divergences arising in quantum field theories, finite theories retain the appeal of ultraviolet insensitivity. Optimistically, enlarging the space of (partially or entirely) ultraviolet-insensitive field theories may open new avenues to explaining the smallness of the Higgs mass or cosmological constant. Whereas the finiteness of spacetime supersymmetry arises from cancellations between ordinary bosons and fermions, for internal supersymmetry the cancellation is between ordinary fields and their negative-norm counterparts. This is highly reminiscent of Lee-Wick theories \cite{Lee:1969fy, Lee:1970iw, Grinstein:2007mp}, albeit now controlled by symmetries.

The finiteness of spontaneously broken $SU(N|N)$ gauge theory to all orders in perturbation theory was explored extensively in \cite{Arnone:2000qd, Arnone:2000bv,Arnone:2001iy}, where it was leveraged to provide a gauge-invariant Pauli-Villars-like regulator for pure $SU(N)$ Yang-Mills. Unfortunately, the broader phenomenological applications of this observation are limited by the fact that the smallest representation of $SU(N|N)$ is the adjoint. This raises the natural question of whether $SU(N|M)$ theories with $N \neq M$ enjoy similar finiteness properties. In \cite{Craig:2024rdg}, we demonstrated the one-loop finiteness of corrections to the two-point function of a scalar multiplet in the fundamental of $SU(N|N+1)$ coming from loops of scalar, spinor, and vector multiplets. Here we expand on the results of \cite{Craig:2024rdg} in considerable detail and explore the one-loop vacuum structure of a theory where $SU(N|N+1)$ is spontaneously broken to the bosonic $SU(N) \times SU(N+1) \times U(1)$ subgroup.

We begin by reviewing key features of the $SU(N|M)$ superalgebra and supergroup in Section~\ref{sec:rewievsunm}. In Section~\ref{sec:model} we consider one-loop corrections to the mass of a scalar field transforming in the fundamental of $SU(N|M)$ from a variety of interactions. In particular, in Section~\ref{subsec:scalarcontrib} we first consider corrections from the scalar field's quartic self-coupling before turning to gauge interactions and yukawa couplings in Secs.~\ref{subsec:addinggauge} and \ref{subsec:addingfermions1}, respectively. In each case, we consider the corrections both for exact and softly-broken  $SU(N|M)$ symmetries, finding that one-loop corrections vanish in the former case and are at most logarithmically divergent in the latter case. We then turn to spontaneous symmetry breaking in Section~\ref{sec:breakingSUNM}, exploring the breaking of $SU(N|M)$ down to its bosonic $SU(N)\times SU(M)\times U(1)$ subgroup at both tree level and one-loop. We conclude in Section~\ref{sec:conc}. Various technical results are reserved for a series of appendices. 

\section{Review of $SU(N|M)$}
    \label{sec:rewievsunm}
    
    We start by reviewing the characteristics of the $SU(N|M)$ superalgebra and supergroup \cite{Bars1984}. The defining representation is furnished by matrices of the form
    \begin{align}
		\mathcal{H}=\begin{pmatrix}
			H_N & \theta \\
			\theta^\dagger & H_M
		\end{pmatrix}\ ,
	\end{align}
	where $H_{N}$ ($H_{M}$) is a hermitian $N\times N$ ($M \times M$) matrix with complex bosonic elements (i.e. regular complex numbers), while $\theta$ is a $N\times M$ matrix composed of complex Grassmann numbers. A generic matrix $\mathcal{H}$ of this form can be decomposed as a linear combination of the following generators\footnote{We pick a slightly different convention w.r.t. \cite{Bars1984} for the normalization of the generator relative to the bosonic $U(1)$ for future convenience.}
	\begin{align}
		T^a_{N}&=\begin{pmatrix}
			t^a_N & 0 \\
			0 & 0
		\end{pmatrix}\ ,&T^b_{M}&=\begin{pmatrix}
			0 & 0 \\
			0 & t^b_{M}
		\end{pmatrix}\ ,&
		S^{i}&=\frac{1}{2}\begin{pmatrix}
			0 & s^{i} \nonumber\\
			s^{\dagger i} &  0
		\end{pmatrix}\ ,&
		\tilde{S}^{i}&=\frac{1}{2}\begin{pmatrix}
			0 & \tilde{s}^{i} \\
			\tilde{s}^{\dagger i} &  0
		\end{pmatrix}\ ,\nonumber
	\end{align}
	\begin{align}
		\lambda_U&=\frac{1}{2}\sqrt{\frac{2NM}{M-N}}\left(\begin{array}{c|c}
			\begin{matrix}
				1/N& & 0\\
				&\ddots & \\
				0  & & 1/N
			\end{matrix}
			&0\\ \hline\\
			0 &\begin{matrix}
				1/M& & 0\\
				&\ddots & \\
				0  & & 1/M
			\end{matrix}
		\end{array}\right)\ ,
		\label{eq:sugenerators}
	\end{align}
    where $t^{a}_{N}$ ($t^{b}_{M}$) are the $N^2-1$ ($M^2-1$) generators of $SU(N)$ ($SU(M)$), $s^i$ are the $NM$, $N\times M$ matrices with $-i$ in one entry and 0 everywhere else, and $\tilde{s}^i$ are the $NM$, $N\times M$ matrices with $1$ in one entry and 0 everywhere else.
    It is then clear that $SU(N|M)$ contains a bosonic subgroup $SU(N)\times SU(M)\times U(1)$ generated by $T^a_N$, $T^b_M$ and $\lambda_U$.
	Notice that the total number of generators is $N^2-1+M^2-1+2NM+1=(N+M)^2-1$, the same as $SU(N+M)$. However, in contrast with $SU(N+M)$, to close the superalgebra formed by these generators we need to take anticommutators into account.
    More specifically, we assign a grading $\mathrm{f}(X)$ to each generator $X$ in the following way 
    \begin{align}
		\degf(T^a_N)=\degf(T^b_M)=\degf(\lambda_U)&=0\ ,& 
		\degf(S^i)=\degf(\tilde{S}^i)&=1\ .
	\end{align}
    The definition of a graded commutator then follows straightforwardly 
    \begin{align}
		\comm{X}{Y}_{\degf} \equiv XY-(-1)^{\degf(X)\degf(Y)}YX\ .
	\end{align}
    Such a graded commutator allows us to specify the graded algebra the generators belong to as 
    \begin{align}
		\comm{\lambda_I}{\lambda_J}_{\degf}=i\tensor{f}{_I_J^K}\lambda_K\ ,
	\end{align}
	for generators $\lambda_{I,J,K}$  of $SU(N|M)$ and some structure constants $\tensor{f}{_I_J^K}$.
    The Jacobi identity generalizes to a super-Jacobi one, 
	\begin{align}
		(-1)^{\degf(Z)\degf(X)}[X,[Y,Z]_{\degf}]_{\degf}+(-1)^{\degf(X)\degf( Y)}[Y,[Z,X]_{\degf}]_{\degf}+(-1)^{\degf( Y)\degf(Z)}[Z,[X,Y]_{\degf}]_{\degf}=0\ ,
	\end{align}
	for $X,\, Y,\, Z $ any three generators of $SU(N|M)$.
    A generic matrix $\mathcal{H}$ belonging to the superalgebra $\mathfrak{su}(N|M)$ is then defined as a linear combination of the generators as
    \begin{align}
        \mathcal{H}=\sum_{a=1}^{N^2-1} \omega_a T^a_{N}+\sum_{b=1}^{M^2-1} \omega_b T^b_{M}+\omega_U\lambda_U+\sum_{i=1}^{N M}\theta_i S^i+\sum_{j=1}^{N M}\tilde{\theta}_j \tilde{S}^j\ ,
    \end{align}
    where the $\omega_a$, $\omega_b$ and $\omega_U$ parameters are commuting complex numbers while the $\theta_i$ and $\tilde{\theta}_j$ are Grassmann numbers. 

    Invariants are built using the supertrace\footnote{As explained by \cite{Bars1984}, it is also possible to define a superdeterminant as $\text{sdet}(U)=\exp(\str{\ln U})$, with $U\in SU(N|M)$. However, such invariant cannot be written as a polynomial in the matrix entries, and so we will not consider it in the following when building Lagrangian operators, as it would give rise to non-local terms.}
	\begin{align}
		\str{\mathcal{H}}\equiv\tr(\sigma_3\mathcal{H})=\tr(H_N)-\tr(H_M)\ ,
	\end{align} \label{eq:original_str}
	where 
	\begin{align}
		\sigma_3=\begin{pmatrix}
			\mathbb{I}_{N\times N} & 0 \\
			0 & -\mathbb{I}_{M\times M}
		\end{pmatrix}\ .
	\end{align}
	Indeed, this is the quantity that stays invariant under cyclic permutations of its arguments
	\begin{align}
		\str{X Y}=\str{Y X}\ ,
	\end{align}
	if $X, Y\in \mathfrak{su}(N|M)$ or $X, Y\in SU(N|M)$, as it compensates for the signs picked by anticommuting Grassmann components. Clearly $\str{\mathcal{H}}=0$ for $\mathcal{H}\in \mathfrak{su}(N|M)$, since the supertrace of all the generators in Eq.~\eqref{eq:sugenerators} vanishes.
    Finite elements of the group can be found as usual by exponentiation of the generators, i.e.
	\begin{align}
		U_{ij}=(e^{i\mathcal{H}})_{ij}=\underset{n\to \infty}{\lim}\left[\left(1+\frac{i^n}{n!}\mathcal{H}^n\right)\right]_{ij}\ ,
	\end{align}
	and it can be checked that the group is closed with respect to matrix multiplication, respects associativity, and that for each $U\in SU(N|M)$ there is an inverse 
	\begin{align}
		U^{\dagger}=U^{-1}=e^{-i\mathcal{H}}\ ,
	\end{align}
	also in the group.

    We will refer to the generators in general as $\lambda_I$, and normalize them so that
	\begin{align}
		\str{\lambda_{I}\lambda_J}=\frac{1}{2}g_{IJ}\ ,
		\label{eq:generatornorm}
	\end{align}
	where $g_{IJ}$ is \cite{Bars1984}
	\begin{align}
    g_{IJ}=
		\left(\begin{array}{c|c}
			\begin{array}{ccc}
				\begin{matrix}
					\phantom{-}1&  & &\\
					& \phantom{-}1 & &\\
					& & \phantom{-}1 & \\
					& & & \ddots
				\end{matrix}
				& & \\
				& \pm 1 &  \\
				& & \begin{matrix}
					-1&  & &\\
					& -1 & &\\
					& & -1 & \\
					& & & \ddots
				\end{matrix}\\
			\end{array}&0\\\hline 
			0&
			\begin{array}{cc}
				\begin{matrix}
					0&i  &\\
					-i& 0 &\\
				\end{matrix}& 0\\
				0 &  \begin{matrix}
					0&i  & &\\
					-i& 0 & &\\
					& & \ddots
				\end{matrix}
			\end{array}
		\end{array}
		\right)\ ,
		\label{eq:metric}
	\end{align}
	with the block with $1$ on the diagonal corresponding to the bosonic $SU(N)$ generators $T^a_N$, the $\pm$1 to $\lambda_U$, i.e. the $U(1)$ bosonic generator, the diagonal $-1$ block to the bosonic $SU(M)$ generators $T^b_{M}$ and the bottom right block to the fermionic $S^i$ and $\tilde{S}^i$ generators. The $U(1)$ entry is $-1$ for $N-M>0$ and $+1$ for $N-M<0$\footnote{This is different w.r.t. \cite{Bars1984} because of the different normalization we assigned to the $U(1)$ generator.}, while it vanishes for $N=M$. 
    Given $g_{IJ}$, we can define its inverse $g^{IJ}$ as
	\begin{align}
		g_{IJ}g^{JK}=\delta_I^K\ .
	\end{align}
	An important property is the completeness relation, which generalizes that of $SU(N)$:
	\begin{align}
		(\lambda_I)_{ij}g^{IJ}(\lambda_J)_{kl}=\frac{1}{2}\left(\delta_{il}\delta_{jk}(-1)^{\degf(j)\degf{(k)}}-\frac{1}{N-M}\delta_{ij}\delta_{kl}\right)\ ,
		\label{eq:completenessrel}
	\end{align}
	where the grading of an index is defined as
	\begin{align}
		\degf{(i)}=\begin{cases}
			0 \qquad\text{       if $1\leq i\leq N$}\\
			1\qquad \text{       if $N+1\leq i\leq M$}
		\end{cases}\ .
	\end{align}
    For future convenience, we also introduce a grading for indices in the adjoint representation
    \begin{align}
        \degf(I)=\begin{cases}
			0 \qquad\text{       if $\lambda^I$ is a bosonic generator}\\
			1 \qquad\text{       if $\lambda^I$ is a fermionic generator}
		\end{cases}\ .
    \end{align}
    Then notice that for a generator $(\lambda^I)^i_j$, $\degf(I)=\degf(i)+\degf(j)\mod{2}$, and 
    \begin{align}
        \str{\lambda^I M}=(-1)^{\degf(I)}\str{M\lambda^I}
        \label{eq:gradedciclicity}
    \end{align}
    for any matrix $M\in SU(N|M)$.
    
    Lastly, we define the following notation: given two tensors $A^{IJ}$ and $A^{IJK}$ with indices in the adjoint representation of $SU(N|M)$
    \begin{align}
    	A^{\{IJ\}_\degf}=&\frac{1}{2}\left(A^{IJ}+(-1)^{\degf(I)\degf(J)}A^{JI}\right)\\
    	A^{\{IJK\}_\degf}=&\frac{1}{6}\left(A^{IJK}+(-1)^{\degf(J)\degf(K)}A^{IKJ}+(-1)^{\degf(I)\degf(J)}A^{JIK}+(-1)^{\degf(I)\degf(J)+\degf(I)\degf(K)}A^{JKI}+\right.\nonumber\\
    	&+\left. (-1)^{\degf(K)\degf(I)+\degf(K)\degf(J)}A^{KIJ}+(-1)^{\degf(K)\degf(I)+\degf(K)\degf(J)+\degf(I)\degf(J)}A^{KJI}
    	\right)\ .
    \end{align}
    In general
    \begin{align}
    	A^{\{I_1\dots I_n\}_\degf}=\frac{1}{n!}\sum_{\text{permutations }\sigma}\text{sgn}_\degf(\sigma)A^{\sigma(I_1\dots I_n)}
    \end{align}
    where the graded sign $\text{sgn}_\degf(\sigma)$ of each permutation $\sigma$ is computed by keeping track of all the minus signs one picks to bring the permuted indices $\sigma(I_1\dots I_n)$ back in the order $I_1\dots I_n$.
    
    \section{The model and 1-loop finiteness}
    \label{sec:model}
    Now that the necessary aspects of $SU(N|M)$ have been covered, we can turn to the study of field theories with $SU(N|M)$ global or local symmetries. We begin with the theory of a single scalar field belonging to the fundamental representation of $SU(N|M)$, with particular attention to the structure of loop corrections to the scalar mass. We first show that the scalar self-couplings do not correct the mass at one-loop provided $M = N+1$. We then introduce couplings to spinor and vector multiplets transforming as various representations of $SU(N|M)$ and show that these couplings likewise do not produce corrections to the scalar mass at one-loop when $M=N+1$.
    
	\subsection{Scalar in the fundamental of $SU(N|M)$ }
	\label{subsec:scalarcontrib}
    The main character is a Lorentz scalar belonging to the fundamental representation of $SU(N|M)$. 
    In a natural basis, we can write it as
	\begin{align}
		\Phi_i=\begin{pmatrix}
			\phi_a\\
			\psi_\alpha
		\end{pmatrix}\ ,
		\label{eq:scalarfieldpar}
	\end{align}
	where $\phi_a$ is a regular (bosonic) $N$-component complex scalar and $\psi_\alpha$ is an $M$-component field which is a Lorentz scalar but with fermionic statistics, $\degf(\psi_\alpha)=1$. $\Phi_i$ transforms under $SU(N|M)$ as $\Phi_i \to \tensor{U}{_i^j}\Phi_j$.
	Its renormalizable Lagrangian takes the form
    \begin{align}
		\mathcal{L}_{\Phi}=\dlo{\mu}\Phi^{\dagger i}\dup{\mu}\Phi_i-m^2\Phi^{\dagger i} \Phi_i-\quartic (\Phi^{\dagger i}\Phi_i)^2\ .
        \label{eq:lagrangianphicompact}
	\end{align}
    Notice that we can write all these supergroup invariants in terms of supertraces, since $\Phi^\dagger \cdot \Phi=\str{\Phi\otimes \Phi^\dagger}$, where $\otimes$ indicates here the exterior products of two vectors. 
    It is instructive to decompose the Lagrangian in Eq.~\eqref{eq:lagrangianphicompact} in terms of the parametrization in Eq.~\eqref{eq:scalarfieldpar}:
    \begin{align}
		\mathcal{L}_{\Phi}=&+\dlo{\mu}\phi^{\dagger a}\dup{\mu}\phi_a+\dlo{\mu}\psi^{\dagger \alpha} \dup{\mu}\psi_\alpha-m^2 \phi^{\dagger a}\phi_a-m^2\psi^{\dagger \alpha}\psi_\alpha+\nonumber\\
		&-\quartic\left[(\phi^{\dagger a}\phi_a)^2+\left(\psi^{\dagger \alpha}\psi_\alpha\right)^2+2\phi^{\dagger a}\phi_a\psi^{\dagger \alpha}\psi_\alpha\right]\ .
		\label{eq:expandedscalarlagr}
	\end{align}
    We will initially work with this expandend version of the Lagrangian to gain familiarity with its pieces, and later repeat the computations with its compact version in Eq.~\eqref{eq:lagrangianphicompact}. 
    From Eq.~\eqref{eq:expandedscalarlagr} we can deduce the Feynman rules 
	\begin{table}[H]
		\centering
		\renewcommand{\arraystretch}{1.3}
		\begin{tabular}{cc}
			\begin{minipage}[c]{0.4\textwidth}
				\begin{tikzpicture}[scale=1]
					\begin{feynhand}
						\vertex (a) at (0,0); \vertex (b) at (2,0); 
						\propag[chasca,mom'={$p$}] (a) to [edge label =$\phi$] (b);
						\node at (3.5,0) {\Large{$\frac{i\delta^a_b}{p^2-m^2}$}};
						\node at (0,0.2) {$a$}; \node at (2,0.2) {$b$};
					\end{feynhand} 
				\end{tikzpicture}
				\label{fig:propagscalbos}
			\end{minipage} &
			\begin{minipage}[c]{0.4\textwidth}
				\begin{tikzpicture}[scale=1]
					\begin{feynhand}
						\vertex (a) at (0,0); \vertex (b) at (2,0); 
						\propag[chasca,mom'={\textcolor{black}{$p$}}, red] (a) to [edge label =\textcolor{black}{$\psi$}] (b);
						\node at (3.5,0) {\Large{$\frac{i\delta^\alpha_\beta}{p^2-m^2}$}};
						\node at (0,0.2) {$\alpha$}; \node at (2,0.2) {$\beta$};
					\end{feynhand} 
				\end{tikzpicture}
				\label{fig:propagscalfer}
			\end{minipage} \\[2cm]
			\begin{minipage}[c]{0.5\textwidth}
				\begin{tikzpicture}[scale=1]
					\begin{feynhand}
						\vertex (a) at (1,1); \vertex (a1) at (0,0); 
						\vertex (a2) at (2,0); \vertex (a3) at (0,2); 
						\vertex (a4) at (2,2); 
						\propag[chasca] (a1) to (a); \propag[antsca] (a2) to (a);
						\propag[chasca] (a3) to (a);\propag[antsca] (a4) to (a);
						\node at (4,1) {$-2i\quartic(\delta^a_b\delta^d_c+\delta^a_c\delta^d_b)$};
						\node at (-0.1,2.1) {$a$};\node at (2.1,-0.1) {$c$};
						\node at (2.1,2.1) {$b$};\node at (-0.1,-0.1) {$d$};
					\end{feynhand} 
				\end{tikzpicture}
			\end{minipage} &
			\begin{minipage}[c]{0.4\textwidth}
				\begin{tikzpicture}[scale=1]
					\begin{feynhand}
						\vertex (a) at (1,1); \vertex (a1) at (0,0); 
						\vertex (a2) at (2,0); \vertex (a3) at (0,2); 
						\vertex (a4) at (2,2); 
						\propag[chasca] (a1) to (a); \propag[antsca, red] (a2) to (a);
						\propag[chasca, red] (a3) to (a);\propag[antsca] (a4) to (a);
						\node at (3.5,1) {$-2i\quartic\delta^\alpha_\beta\delta^a_b$};
						\node at (-0.1,2.1) {$\alpha$};\node at (2.1,-0.1) {$\beta$};
						\node at (2.1,2.1) {$b$};\node at (-0.1,-0.1) {$a$};
					\end{feynhand} 
				\end{tikzpicture}
			\end{minipage}
		\end{tabular}
	\end{table}
    \noindent where we reserved the color red for the components $\psi$ with fermionic statistics. 

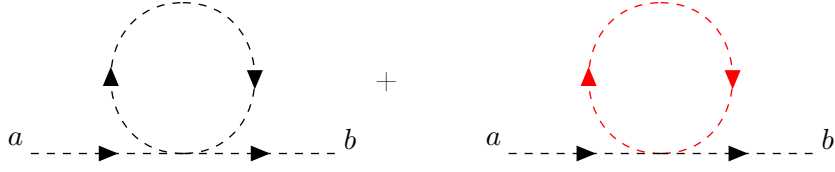
\begin{figure}[H]
		\centering
		\renewcommand{\arraystretch}{1}
		\begin{tabular}{ccc}
			\begin{minipage}[c]{0.26\textwidth}
				\begin{tikzpicture}[scale=2]
					\begin{feynhand}
						\vertex (a) at (0,0); \vertex (b) at (1,0); 
						\vertex (c) at (2,0); \vertex (d) at (1,1); 
						\propag[chasca] (b) to [in=180, out=180, looseness=1.62] (d);
						\propag[chasca] (d) to [in=0, out=0, looseness=1.62] (b);
						\propag[chasca] (a) to [edge label] (b);
						\propag[chasca] (b) to [edge label] (c);
						\node at (-0.1,0.1) {$a$};\node at (2.1,0.1) {$b$};
					\end{feynhand} 
				\end{tikzpicture}
			\end{minipage} &	\begin{minipage}[c]{0.1\textwidth}\centering   +\end{minipage}&
			\begin{minipage}[c]{0.2\textwidth}
				\begin{tikzpicture}[scale=2]
					\begin{feynhand}
						\vertex (a) at (0,0); \vertex (b) at (1,0); 
						\vertex (c) at (2,0); \vertex (d) at (1,1); 
						\propag[chasca, red] (b) to [in=180, out=180, looseness=1.62] (d);
						\propag[chasca, red] (d) to [in=0, out=0, looseness=1.62] (b);
						\propag[chasca] (a) to [edge label] (b);
						\propag[chasca] (b) to [edge label] (c);
						\node at (-0.1,0.1) {$a$};\node at (2.1,0.1) {$b$};
					\end{feynhand} 
				\end{tikzpicture}
			\end{minipage} \\[2cm]
		\end{tabular}
		\caption{One-loop contributions to the $\phi$ mass coming from the Lagrangian in Eq.~\eqref{eq:expandedscalarlagr}.}
        \label{fig:oneloopfromscalars}
	\end{figure}

    Now we can compute the 1-loop correction to the mass of the ordinary scalar $\phi$ due to the coupling $\quartic$. There are two contributions, coming from the diagrams in Fig.~\ref{fig:oneloopfromscalars}.
    They add up to 
	\begin{align}
		\Sigma_{\Phi}=\Sigma_1+\Sigma_2\ .
	\end{align}
	with 
	\begin{align}
		\Sigma_1=2(N+1)\times \quartic \mu^{4-d}\int\frac{d^dp}{(2\pi)^d}\frac{1}{p^2-m^2}=2(N+1)\quartic I(m^2)\ ,
		\label{eq:sigma1}
	\end{align}
	referring to the diagram on the left in Fig.~\ref{fig:oneloopfromscalars}, and 
	\begin{align}
		\Sigma_2=-M\times (2\quartic\mu^{4-d}) \int\frac{d^dp}{(2\pi)^d}\frac{1}{p^2-m^2}=-2 M \quartic I(m^2)\ ,
		\label{eq:sigma2}
	\end{align}
	to the one on the right. Here we used the definition $I(m^2)\equiv\int\frac{d^dp}{(2\pi)^d}\frac{1}{p^2-m^2}$.
	The factor of $2(N+1)$ in Eq.~\eqref{eq:sigma1} comes from the different ways of contracting the $\delta_{ab}$ of the vertex with the external lines, while the diagram on the right can only be contracted in one way and only has a factor of two stemming from the Lagrangian and a factor of $M$ from the tracing inside the loop. The relative minus sign comes from the loop on the right being a fermionic loop.
    Then, we can see that choosing $M=N+1$ the two contributions cancel out completely, $\Sigma_{\Phi}=0$.
    
    Of course, the full symmetry being $SU(N|N+1)$, it is the mass of the whole $\Phi_i$ that must not renormalize at one-loop. This can be seen for example by repeating the same computation for the mass of $\psi$. The only difference here is that now there is a relative minus sign between the two products of $\delta$'s in the four-$\psi$ vertex, i.e. $(\delta^a_b\delta^d_c+\delta^a_c\delta^d_b)\to(\delta^\alpha_\beta\delta^\gamma_\delta-\delta^\alpha_\delta\delta^\gamma_\beta)$ as to get from one to the other we have to exchange two fermionic lines. Then, the two one-loop diagrams will have factors of $-(M-1)$ and $N$, instead of $-(N+1)$ and $M$, respectively, and will cancel again when $M=N+1$. 
    
    Having obtained our result in a basis where we considered the two components of $\Phi$ separately, it is now useful to repeat the computation directly with the Lagrangian in Eq.~\eqref{eq:sugenerators}. It is also a good occasion to familiarize ourselves with (functional) differentiation w.r.t to graded fields, which will turn out to be useful later. For starters, we can obtain the 4-point vertex from Eq.~\eqref{eq:sugenerators} via
    \begin{align}
        &-i \frac{\delta^4}{\delta \Phi_{k}\delta \Phi^{\dagger l}\delta \Phi_{m}\delta \Phi^{\dagger n}} \quartic (\Phi^{\dagger i}\Phi_i\Phi^{\dagger j}\Phi_j)=-2i\quartic \frac{\delta^3}{\delta \Phi_{k}\delta \Phi^{\dagger l}\delta \Phi_{m}}(\Phi_n \Phi^{\dagger j}\Phi_j)=\nonumber\\
        =&-2i\quartic \frac{\delta^2}{\delta \Phi_{k}\delta \Phi^{\dagger l}}(\delta^m_n\Phi^{\dagger j}\Phi_j +(-1)^{\degf(m)}\Phi^{\dagger m}\Phi_n)=-2i\quartic(\delta^m_n \delta^k_l+(-1)^{\degf(m)}\delta^m_l \delta^k_n) \ .
    \end{align}
    When computing the one-loop mass correction we would contract this vertex with a Kronecker delta in flavor space, so that 
    \begin{align}
        i\Sigma(p)\propto -2i\quartic(\delta^m_n \delta^k_l+(-1)^{\degf(m)}\delta^m_l \delta^k_n) \delta_m^l=-2i\quartic \delta^k_n (1+(N-M))
    \end{align}
    where we used the fact that $\delta^j_j(-1)^{\degf(j)}=N-M$. This is the same result as before, and it vanishes for $M=N+1$.
    Alternatively, going to position space and being painfully pedantic with the notation, we can write
    \begin{align}
        G(x_1,x_2)=&-i\quartic \int \dd x \expval{T\{\Phi_l(x_1)\Phi^{\dagger k}(x_2)\Phi^{\dagger i}(x)\Phi_i(x)\Phi^{\dagger j}(x)\Phi_j(x)\}}{0}=\nonumber\\
        =&-i\quartic \int \dd x \expval{T\{\Phi_l(x_1)\Phi^{\dagger i}(x)\Phi_i(x)\Phi^{\dagger j}(x)\Phi_j(x)\Phi^{\dagger k}(x_2)\}}{0}=\nonumber\\
        =&-2i\quartic \int \dd x \left(\expval{\wick{\c2 \Phi_l(x_1)\c2 \Phi^{\dagger i}(x) \c2\Phi_i(x)\c2\Phi^{\dagger j}(x)\c2\Phi_j(x)\c2\Phi^{\dagger k}(x_2)}}{0}+\right.\nonumber \\
        &\left.+\expval{\wick{\c2 \Phi_l(x_1)\c2 \Phi^{\dagger i}(x) \c2\Phi_j(x)\c2\Phi^{\dagger j}(x)(-1)^{\degf(j)}\c2\Phi_i(x)\c2\Phi^{\dagger k}(x_2)}}{0}\right)=\nonumber\\
        =&-2i\quartic \int \dd x \left(D_F(x_1,x)\delta^i_l D_F(x,x)\delta^j_i D_F(x,x_2)\delta_j^k +\right. \nonumber\\
        & \left.+D_F(x_1,x)\delta^i_l D_F(x,x)\delta^j_j(-1)^{\degf(j)}D_F(x,x_2)\delta_i^k\right)=\nonumber\\
        =&-2i \delta^k_l (1+\delta^j_j(-1)^{\degf(j)})\quartic \int \dd xD_F(x_1,x)D_F(x,x)D_F(x,x_2)=\nonumber\\
        =&-2i \delta^k_l (1+(N-M))\quartic \int \dd xD_F(x_1,x)D_F(x,x)D_F(x,x_2)\ ,
    \end{align} 
    again in agreement with our previous result.

We have then found that the one-loop correction to the mass of a scalar vanishes, at the price of introducing fields with the wrong statistics. It is natural to ask what happens to this cancellation when the $SU(N|M)$ symmetry is softly broken by a dimensionful parameter that splits the masses of the $\phi$ and $\psi$ components. Introducing a soft mass term for the wrong-statistics field, 
    \begin{align}
    	\mathcal{L}_{\Phi}\to\mathcal{L}_{\Phi}-m_{\text{soft}}^2\psi^{\dagger \alpha}\psi_\alpha\ ,
        \label{eq:softmassscalar}
    \end{align}
the one-loop contributions to $\delta m^2_\phi$ are the same as in Fig.~\ref{fig:oneloopfromscalars}, modulo the modification of the $\psi_\alpha$ propagator. The effect is unsurprisingly reminiscent of soft breaking terms in theories with spacetime supersymmetry: the corrections to the scalar mass-squared are quadratic in the soft term and only logarithmic in the cutoff. After renormalization using $\overline{\text{MS}}$, the correction to the physical mass of $\phi_a$ takes the form
    \begin{align}
    	\delta m^2_\phi=-2(N+1)\frac{\quartic}{16\pi^2}\left[m_{\text{soft}}^2\left(1+\log(\frac{\mu^2}{m^2+m_{\text{soft}}^2})\right)-m^2\log(1+\frac{m_{\text{soft}}^2}{m^2})\right]\ .
    \end{align}

    \subsection{Adding gauge interactions}
    \label{subsec:addinggauge}
    Now we turn to gauge interactions. Although the one-loop finiteness of scalar self-interactions favors $M = N+1$, for the time being let us still consider generic values of $M$. We introduce a gauge field, $\mathcal{A}_\mu$, belonging to the adjoint representation of $SU(N|M)$.\footnote{To the best of our knowledge, the field theory for $SU(N|M)$ gauge fields was first introduced in \cite{Arnone:2001iy}, in the context of providing a fully gauge-invariant higher-derivative regularization of Yang-Mills.}
    Expanded in terms of generators, $\mathcal{A}_\mu$ takes the form
    \begin{align}
		\mathcal{A}_\mu=\begin{pmatrix}
			A_{\mu}^{1a}t^a_N&B_\mu^i(s_1+\tilde{s}_i)\\
			(B_\mu^\dagger)^i(s_1^\dagger+\tilde{s}^\dagger_i)&A_{\mu}^{2b}t^b_M
		\end{pmatrix}
		+A_\mu^U\lambda_U\ ,
		\label{eq:gaugefields}
	\end{align}
    where the $A^{1,2,U}_\mu$ are bosonic and the $B^i_\mu$ are fermionic. Here an additional difficulty arises. In the scalar case we only met wrong-statistics fields, which we labeled $\psi_\alpha$. In the expansion of $\mathcal{A}_\mu$, this role is taken by the $B^i_\mu$ fields, having fermionic statistics while being integer-spin vector fields. However, in addition to them, we also have $A^{2}_\mu$, which will turn out to have a kinetic term with the wrong sign owing to the negative directions in the metric, see e.g.~Eq.~\eqref{eq:metric}. We will refer to these as wrong-sign ghosts.
    
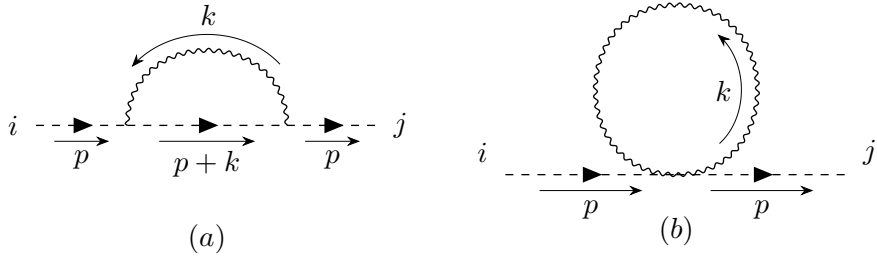
\begin{figure}[H]
		\centering
		\begin{tabular}{cc}
			\begin{minipage}[c][4cm]{0.4\textwidth}
				\begin{tikzpicture}[scale=1.5]
					\begin{feynhand}
						\vertex (a) at (0.8,0); \vertex (a1) at (0,0); 
						\vertex (b) at (2.2,0); 
						\vertex (b1) at (3,0); 
						\propag[chasca, mom'={$p$}] (a1) to  (a); 
						\propag[chasca, mom'={$p+k$}] (a) to (b);
						\propag[bos, mom'={$k$}] (b) to [in=90, out=90, looseness= 1.62] (a);
						\propag[chasca, mom'={$p$}] (b) to (b1); 
						\node at (-0.2,0) {$i$}; \node at (3.2,0) {$j$};
                        \node at (1.5,-1) {$(a)$};
					\end{feynhand} 
				\end{tikzpicture}
			\end{minipage} 
			\begin{minipage}[c][4cm]{0.4\textwidth}
				\begin{tikzpicture}[scale=1.5]
					\begin{feynhand}
						\vertex (a) at (0,0); \vertex (b) at (1.5,0); 
						\vertex (c) at (3,0); \vertex (d) at (1.5,1.5); 
						\propag[bos] (b) to [in=180, out=180, looseness=1.62] (d);
						\propag[bos, revmom'={$k$}] (d) to [in=0, out=0, looseness=1.62] (b);
						\propag[chasca, mom'={$p$}] (a) to [edge label] (b);
						\propag[chasca, mom'={$p$}] (b) to [edge label] (c);
						\node at (-0.2,0.2) {$i$};\node at (3.2,0.2) {$j$};
                       \node at (1.5,-0.5) {$(b)$};
					\end{feynhand} 
				\end{tikzpicture}
			\end{minipage}
		\end{tabular}
        \caption{One-loop contributions to the $\Phi$ mass from gauge interactions.}
        \label{fig:oneloopfromgauge}
	\end{figure}

    Setting these issues aside for the moment, we follow the prescription of \cite{Arnone:2001iy} and introduce Faddeev--Popov-ghost fields $\eta$ and $\bar{\eta}$, belonging to the adjoint of $SU(N|M)$, that we will use to fix the gauge. They can be expressed as
	\begin{align}
		\eta=\begin{pmatrix}
			\eta^1& \rho\\
			\sigma &\eta^2
		\end{pmatrix}\ ,
		\label{eq:ghostfields}
	\end{align}
	and similarly for\footnote{Notice that, as explained in \cite{Arnone:2001iy}, their grading is not trivially the opposite of that of $\mathcal{A}_\mu$. Indeed, in order to obtain the expected behavior for supertraces involving ghosts, we need to introduce an additional grading $\degg(X)$ such that $\degg(\mathcal{A})=\degg(\Phi)=0$ but $\degg(\eta)=\degg(\bar{\eta})=1$, and redefining the commutation of any two fields as
		\begin{align}
  \comm{X}{Y}_{\degf,\degg}=
			XY-(-1)^{\degf(X)\degf(Y)+\degg(X)\degg(Y)}YX\ .
		\end{align}
		However, this will not play any role in the rest of our discussion.} $\bar{\eta}$.
    Although the expansions in Eqs.~\eqref{eq:gaugefields}-\eqref{eq:ghostfields} are useful for visualization, we are now warmed up enough and can deal with the whole fields at once, without having to split between their bosonic and fermionic pieces. 
	The Lagrangian for a theory with an $SU(N|M)$ gauge symmetry and a scalar $\Phi$ belonging to the fundamental of the group can be written as
	\begin{align}
		\mathcal{L}=\mathcal{L}_{G}+\mathcal{L}_{S}+\mathcal{L}_{Gf}+\mathcal{L}_{Gh}\ ,
        \label{eq:fullscalargaugelagrangian}
	\end{align}
	where $\mathcal{L}_G$ is the gauge kinetic term
    \begin{align}
		\mathcal{L}_{G}=-\frac{1}{2}\str{\mathcal{F}_{\mu\nu}\mathcal{F}^{\mu\nu}}=-\frac{1}{4}\mathcal{F}_{\mu\nu}^I\mathcal{F}^{\mu\nu J}g_{IJ}\ ,
	\end{align}
	with
    \begin{align}
		(\mathcal{F}_{\mu\nu})^i_j&=\frac{1}{ig}\comm{\nabla_\mu}{\nabla_\nu}^i_j\ ,&(\mathcal{F}_{\mu\nu})^i_j&=(\mathcal{F}_{\mu\nu})^I(\lambda_I)^i_j\ , & \nabla_\mu&=\dlo{\mu}\delta^i_j+ig(\mathcal{A}_{\mu})^i_j\ ,
	\end{align}
    meaning 
    \begin{align}
        \mathcal{F}_{\mu\nu}^I=\dlo{\mu}\mathcal{A}_\nu^I-\dlo{\nu}\mathcal{A}_\mu^I-g\tensor{f}{_{AB}^I}\mathcal{A}_\mu^A\mathcal{A}_{\nu}^B\ ,
    \end{align}
    and
    \begin{align}
        \mathcal{L}_{G}=&-\frac{1}{4}\mathcal{F}_{\mu\nu}^I\mathcal{F}_{\mu\nu}^Jg_{IJ}=\frac{1}{2}\left(\dlo{\mu}\mathcal{A}_\nu^I\dup{\nu}\mathcal{A}^{\mu J}-\dlo{\mu}\mathcal{A}_\nu^I\dup{\mu}\mathcal{A}^{\nu J}\right)g_{IJ}+\nonumber\\
        &+g\dlo{\mu}\mathcal{A}_{\nu}^I\mathcal{A}^{\mu C}\mathcal{A}^{\nu D}g_{IJ}\tensor{f}{_{CD}^J}-\frac{1}{4}g^2\mathcal{A}^{A}_{\mu}\mathcal{A}^{B}_{\nu}\mathcal{A}^{\mu C}\mathcal{A}^{\nu D}\tensor{f}{_{AB}^I}\tensor{f}{_{CD}^J}g_{IJ}\ .
        \label{eq:gaugeselfint}
    \end{align}
	$\mathcal{L}_S$ is the scalar kinetic term
	\begin{align}
		\mathcal{L}_{S}=\nabla_\mu\Phi^{\dagger i}\nabla^\mu\Phi_i=\dlo{\mu}\Phi^{\dagger i}\dup{\mu}\Phi_i+ig\dlo{\mu}\Phi^{\dagger i} (\mathcal{A}^\mu)_i^j\Phi_j-ig\Phi^{\dagger i} (\mathcal{A}^\mu)_i^j\dlo{\mu}\Phi_j+g^2\Phi^{\dagger i}(\mathcal{A}_\mu\mathcal{A}^\mu)_i^j\Phi_j\ ,
	\end{align}
	while $\mathcal{L}_{Gf}$ is the gauge fixing term
	\begin{align}
		\mathcal{L}_{Gf}=-\frac{1}{\alpha}\str{\left(\dlo{\mu}\mathcal{A}^\mu\right)^2}=-\frac{1}{2\alpha}\dlo{\mu}\mathcal{A}^{I\mu}\dlo{\nu}\mathcal{A}^{J\nu}g_{IJ}\ ,
	\end{align}
	and $\mathcal{L}_{Gh}$ is the ghost term
	\begin{align}
		\mathcal{L}_{Gh}=2\str{\partial^\mu\bar{\eta}\nabla_\mu\eta}\ .
	\end{align}
    Now we see where wrong-sign fields come from in this Lagrangian. The gauge kinetic term can be expanded as
    \begin{align}
        \mathcal{L}_G\supset -\frac{1}{4}(F_{\mu\nu}^1)^{I_1}(F_{\mu\nu}^1)^{I_1}+\frac{1}{4}(F_{\mu\nu}^2)^{I_2}(F_{\mu\nu}^2)^{I_2}
    \end{align}
    where the superscript $1$ or $2$ is used to distinguish the field strength relative to $A^1_\mu$ and $A^2_\mu$, respectively. Clearly, $A^2_\mu$ has a kinetic term with the wrong sign. We will come back to this issue later.
    For the moment, we extract from the Lagrangian in Eq.~\eqref{eq:fullscalargaugelagrangian} the relevant Feynman rules
	\begin{table}[H]
		\centering
		\renewcommand{\arraystretch}{1}
		\begin{tabular}{c}
			\begin{minipage}[c][1cm]{0.8\textwidth}
				\begin{tikzpicture}[scale=1.3]
					\begin{feynhand}
						\vertex (a) at (0,0); \vertex (b) at (2,0); 
						\propag[bos,mom'={$p$}] (a) to [edge label =$\mathcal{A}$] (b);
						\node at (6.5,0) {\Large{$-ig^{IJ}\left[\frac{\eta_{\mu\nu}-(1-\alpha)\frac{p_\mu p_\nu}{p^2+i\epsilon}}{p^2+i\epsilon}\right]$}};
						\node at (0,0.2) {$J,\nu$}; \node at (2,0.2) {$I,\mu$};
					\end{feynhand} 
				\end{tikzpicture}
			\end{minipage} 
		\end{tabular}
	\end{table}
	\begin{table}[H]
		\centering
		\renewcommand{\arraystretch}{1}
		\begin{tabular}{c}
			\begin{minipage}[c][2cm]{0.8\textwidth}
				\begin{tikzpicture}[scale=1]
					\begin{feynhand}
						\vertex (a) at (1,1); \vertex (a1) at (-0.5,1); 
						\vertex (a2) at (2,0); 
						\vertex (a4) at (2,2); 
						\propag[chasca, mom'={$p_1$}] (a1) to [edge label = $\Phi$] (a); \propag[bos] (a2) to [edge label = $\mathcal{A}$] (a);
						\propag[charged scalar, mom'={$p_2$}] (a) to [edge label = $\Phi$] (a4);
						\node at (8.5,1) {\Large{$-i g(\lambda_I)^j_i(p_1+p_2)_{\mu}$}};
						\node at (-0.5,1.2) {$i$}; \node at (2,2.2) {$j$};
						\node at (2,-0.2) {$I,\mu$};
					\end{feynhand} 
				\end{tikzpicture}
			\end{minipage} 
		\end{tabular}
	\end{table}
	\begin{table}[H]
		\centering
		\renewcommand{\arraystretch}{1}
		\begin{tabular}{c}
			\begin{minipage}[c][3cm]{0.8\textwidth}
				\begin{tikzpicture}[scale=1]
					\begin{feynhand}
						\vertex (a) at (1,1); \vertex (a1) at (0,0); 
						\vertex (a2) at (2,0); \vertex (a3) at (0,2); 
						\vertex (a4) at (2,2); 
						\propag[chasca] (a1) to [edge label = $\Phi$] (a); \propag[bos] (a2) to [edge label = $\mathcal{A}$] (a);
						\propag[chasca] (a) to [edge label' = $\Phi$] (a3);\propag[bos] (a4) to [edge label = $\mathcal{A}$] (a);
						\node at (8.5,1) {\Large{$ig^2(\lambda_I)_{j}^k(\lambda_J)_{k}^i\eta_{\mu\nu}$}};
						\node at (-0.2,2.2) {$i$};\node at (2.2,-0.2) {$J, \mu$};
						\node at (2.2,2.2) {$I,\nu$};\node at (-0.2,-0.2) {$j$};
					\end{feynhand} 
				\end{tikzpicture}
			\end{minipage} 
		\end{tabular}
	\end{table}
    \noindent and use them to compute the contribution from the gauge coupling to the mass of the scalar field $\Phi$. At one-loop there are two relevant diagrams, displayed in Fig.~\ref{fig:oneloopfromgauge}. 
    \noindent Let us focus on the first one. Its contribution is 
    \begin{align}
		i\Sigma^a(p)=-g^2\mu^{4-d}\times (\lambda_I)_{i}^kg^{IJ}(\lambda_J)_{k}^j\mathcal{I}_3(p^2,m^2,\alpha)\ ,
	\end{align}
	where we defined 
	\begin{align}
		\mathcal{I}_3(p^2,m^2,\alpha)=\int\frac{\dd[d]{k}}{(2\pi)^d} (2p+k)^\mu (2p+k)^\nu \left[\frac{\eta_{\mu\nu}-(1-\alpha)\frac{k_\mu k_\nu}{k^2+i\epsilon}}{\left(k^2+i\epsilon\right)
			\left[\left(p+k\right)^2-m^2+i\epsilon \right]}\right]\ .
	\end{align}
	Now we can massage the prefactor using the completeness relation Eq.~\eqref{eq:completenessrel}
	\begin{align}
		(\lambda_I)_{i}^kg^{IJ}(\lambda_J)_k^{j}=\frac{1}{2}\left(\delta_{i}^j\delta_{k}^k(-1)^{\degf(k)^2}-\frac{1}{N-M}\delta_{i}^k\delta_{k}^j\right)=\frac{1}{2}\delta_{i}^j\left((N-M)-\frac{1}{N-M}\right)\ ,
	\end{align}
	which again vanishes for $M=N+1$ (actually also for $M=N-1$).

    Let us stop for a second to double-check our result. Indeed, since the loop contains both bosonic and fermionic degrees of freedom hidden in the sums, we may have missed some minus sign when computing it. To check whether this is the case, notice that the previous diagram, when expressed in position space, comes from a term in the perturbative expansion of the form
    \begin{align}
        G(x_1,x_2)=&\,(ig)^2\int \dd x \dd y \bra{0}T\left\{\Phi_{m,x_1}\Phi^{\dagger n}_{x_2}\left(\dlo{\mu}\Phi_x^{\dagger j}(\mathcal{A}^\mu)^i_{j,x}\Phi_{i,x}-\Phi_x^{\dagger j} (\mathcal{A}^\mu)^i_{j,x}\dlo{\mu}\Phi_{i,x}\right)\times\right. \nonumber \\
        &\times \left. \left(\dlo{\nu}\Phi_y^{\dagger k} (\mathcal{A}^\nu)^l_{k,y}\Phi_{l,y}-\Phi_y^{\dagger k} (\mathcal{A}^\nu)^l_{k,y}\dlo{\nu}\Phi_{l,y}\right)\right\}\ket{0}\ .
	\end{align}
    Neglecting derivatives, each term has the schematic form
	\begin{align}
		G(x_1,x_2)&\sim (ig)^2\int \dd x \dd y \bra{0}T\left\{\Phi_{m,x_1}\Phi^{\dagger n}_{x_2}\left(\Phi_x^{\dagger j}(\mathcal{A}^\mu)^i_{j,x}\Phi_{i,x}\right)\left(\Phi_y^{\dagger k} (\mathcal{A}^\nu)^l_{k,y}\Phi_{l,y}\right)\right\}\ket{0}=\nonumber\\
        &=(ig)^2\int \dd x \dd y \bra{0}T\left\{\Phi_{m,x_1}\Phi_x^{\dagger j}(\mathcal{A}^\mu)^i_{j,x}\Phi_{i,x}\Phi_y^{\dagger k} (\mathcal{A}^\nu)^l_{k,y}\Phi_{l,y}\Phi^{\dagger n}_{x_2}\right\}\ket{0}=\nonumber\\
        &=(ig)^2\int \dd x \dd y \bra{0}T\left\{\Phi_{m,x_1}\Phi_x^{\dagger j}(\mathcal{A}^\mu)^i_{j,x}(\mathcal{A}^\nu)^l_{k,y}\Phi_{i,x}\Phi_y^{\dagger k} \Phi_{l,y}\Phi^{\dagger n}_{x_2}\right\}\ket{0}=\nonumber\\
        &\wick{=(ig)^2\int \dd x \dd y \bra{0}\c2\Phi_{m,x_1}\c2\Phi_x^{\dagger j}\c2{(\mathcal{A}^\mu)^i_{j,x}}\c2{(\mathcal{A}^\nu)^l_{k,y}}\c2\Phi_{i,x}\c2\Phi_y^{\dagger k} \c2\Phi_{l,y}\c2\Phi^{\dagger n}_{x_2}\ket{0}}=\nonumber\\
        &=(ig)^2\int \dd x \dd yD_{\Phi}(x_1,x)D_{\mathcal{A}}(x,y)D_{\Phi}(x,y)D_{\Phi}(y,x_2)\ .
	\end{align}
    This means that each contribution to the diagram has the same sign independent of the fermionic or bosonic nature of the lines involved, or rather that the additional minus signs are taken care of by $g^{IJ}$.
    The second diagram instead gives
    \begin{align}
        i\Sigma^b(p)=g^2\mu^{4-d}(\lambda_I)_{i}^kg^{IJ}(\lambda_J)_{k}^j\mathcal{I}_4(p^2,m^2,\alpha)\ ,
	\end{align}
	where 
	\begin{align}
		\mathcal{I}_4(p^2,m^2,\alpha)=\int\frac{\dd[d]{k}}{(2\pi)^d} 
		\eta^{\mu\nu} \left[\frac{\eta_{\mu\nu}-(1-\alpha)\frac{k_\mu k_\nu}{k^2+i\epsilon}}{\left(k^2+i\epsilon\right)}\right]\ .
	\end{align}
    Again, the sum in the prefactor evaluates to 
	\begin{align}
		(\lambda_I)_{i}^kg^{IJ}(\lambda_J)_{k}^j=\frac{1}{2}\delta_{i}^j\left((N-M)-\frac{1}{N-M}\right)\ ,
	\end{align}
	which vanishes for $M=N\pm1$.
	A computation similar to the one presented for the previous diagram shows that here, too, there is no dependence on the grading of the index running inside the loop.
 
    \subsubsection{Soft mass for some $\mathcal{A}_\mu$ components}
    \label{sec:softmassesgaugebosons}
    The gauging of $SU(N|M)$ in the previous section brought with it the introduction of both wrong-statistics and wrong-sign fields. As we did in Section~\ref{subsec:scalarcontrib}, we could explicitly break $SU(N|M)$ by introducing mass terms for some or all of these problematic fields to separate them from the correct-sign, correct-statistics ones. In Section~\ref{sec:breakingSUNM}, we will see a UV-complete model where a soft mass is provided for the wrong-statistics fields via a Higgs mechanism. Here, however, we limit ourselves to exploring what happens to the renormalization of the mass $m_{\Phi}$ of the scalar if we give a soft mass to some of the components of $\mathcal{A}_{\mu}$. To this end, we split $g_{IJ}=g^{(1)}_{IJ}+g^{(2)}_{IJ}$, and add a mass term $\propto \mathcal{A}^{I\mu}\mathcal{A}^{J}_\mu g_{IJ}^{(2)}$. The quadratic Lagrangian then becomes
    \begin{align}
        \mathcal{L}_{G, 1}^0&=-\frac{1}{2}\partial_\mu\mathcal{A}^I_\nu\partial^\mu\mathcal{A}^{J\nu}g^{(1)}_{IJ}+\frac{1}{2}\left(1-\frac{1}{\alpha}\right)\partial_\mu\mathcal{A}^I_\nu\partial^\nu\mathcal{A}^{J\mu}g^{(1)}_{IJ}
        \label{eq:quadraticlagrangianamuboson1} 
        \\
        \mathcal{L}_{G, 2}^0&=-\frac{1}{2}\partial_\mu\mathcal{A}^I_\nu\partial^\mu\mathcal{A}^{J\nu}g^{(2)}_{IJ}+\frac{1}{2}\left(1-\frac{1}{\alpha}\right)\partial_\mu\mathcal{A}^I_\nu\partial^\nu\mathcal{A}^{J\mu}g^{(2)}_{IJ}+\frac{m^2_{\mathcal{A}}}{2}\mathcal{A}^{I\mu}\mathcal{A}^{J}_\mu g_{IJ}^{(2)}\ .
        \label{eq:quadraticlagrangianamufermion1}
    \end{align}
    Again, we would like to check how this modification affects the renormalization of the mass of $\Phi$ at one-loop.
    In the Feynman rules of Section~\ref{subsec:addinggauge} we only need to modify the propagators as:
    \begin{table}[H]
    	\centering
    	\renewcommand{\arraystretch}{1.3}
    	\begin{tabular}{c}
    		\begin{minipage}[c][2cm]{0.8\textwidth}
    			\begin{tikzpicture}[scale=1.3]
    				\begin{feynhand}
    					\vertex (a) at (0,0); \vertex (b) at (2,0); 
    					\propag[bos,mom'={$p$}] (a) to [edge label =$\mathcal{A}^{(1)}$] (b);
    					\node at (6.5,0) {\Large{$-i(g^{(1)})^{IJ}\frac{1}{p^2+i\epsilon}\left[\eta_{\mu\nu}-(1-\alpha)\frac{p_\mu p_\nu}{p^2+i\epsilon}\right]$}};
    					\node at (0,0.2) {$I,\mu$}; \node at (2,0.2) {$J,\nu$};
    				\end{feynhand} 
    			\end{tikzpicture}
    		\end{minipage} 
    	\end{tabular}
    \end{table}
    \begin{table}[H]
    	\centering
    	\renewcommand{\arraystretch}{1.3}
    	\begin{tabular}{c}
    		\begin{minipage}[c][2cm]{0.8\textwidth}
    			\begin{tikzpicture}[scale=1.3]
    				\begin{feynhand}
    					\vertex (a) at (0,0); \vertex (b) at (2,0); 
    					\propag[bos,mom'={$p$}] (a) to [edge label =$\mathcal{A}^{(2)}$] (b);
    					\node at (6.5,0) {\Large{$-i(g^{(2)})^{IJ}\frac{1}{p^2-m_\mathcal{A}^2+i\epsilon}\left[\eta_{\mu\nu}-(1-\alpha)\frac{p_\mu p_\nu}{p^2-\alpha m_\mathcal{A}^2+i\epsilon}\right]$}};
    					\node at (0,0.2) {$I,\mu$}; \node at (2,0.2) {$J,\nu$};
    				\end{feynhand} 
    			\end{tikzpicture}
    		\end{minipage} 
    	\end{tabular}
    \end{table}
    We work in Feynman gauge $\alpha=1$ and perform the computations in $D$ dimensions. The diagrams are given in Fig.~\ref{fig:oneloopfrombosonsafterSSB}.
    \begin{figure}[H]
    	\centering
    	\renewcommand{\arraystretch}{1.3}
    	\begin{tabular}{ccc}
    		\begin{minipage}[c]{0.26\textwidth}
    			\begin{tikzpicture}[scale=1]
    				\begin{feynhand}
    					\vertex (a) at (0.8,1); \vertex (a1) at (0,1); 
    					\vertex (b) at (2.2,1); 
    					\vertex (b1) at (3,1); 
    					\propag[chasca, mom'={$p$}] (a1) to  (a); 
    					\propag[chasca, mom'={$p+q$}] (a) to (b);
    					\propag[bos, mom'={$q$}] (b) to [in=90, out=90, looseness= 1.62, edge label = $2$] (a);
    					\propag[chasca, mom'={$p$}] (b) to (b1); 
    					\node at (-0.2,1) {$i$}; \node at (3.2,1) {$j$};
    					\node at (1.5,-0.5) {$(a)$}; 
    				\end{feynhand} 
    			\end{tikzpicture}
    		\end{minipage} &
    		\begin{minipage}[c]{0.26\textwidth}
    			\begin{tikzpicture}[scale=1]
    				\begin{feynhand}
    					\vertex (a) at (0.8,1); \vertex (a1) at (0,1); 
    					\vertex (b) at (2.2,1); 
    					\vertex (b1) at (3,1); 
    					\propag[chasca, mom'={$p$}] (a1) to  (a); 
    					\propag[chasca, mom'={$p+q$}] (a) to (b);
    					\propag[bos, mom'={$q$}] (b) to [in=90, out=90, looseness= 1.62, edge label = $1$] (a);
    					\propag[chasca, mom'={$p$}] (b) to (b1); 
    					\node at (-0.2,1) {$i$}; \node at (3.2,1) {$j$};
    					\node at (1.5,-0.5) {$(b)$}; 
    				\end{feynhand} 
    			\end{tikzpicture}
    		\end{minipage} &
    		\begin{minipage}[c]{0.26\textwidth}
    			\begin{tikzpicture}[scale=1]
    				\begin{feynhand}
    					\vertex (a) at (0,0); \vertex (b) at (1.5,0); 
    					\vertex (c) at (3,0); \vertex (d) at (1.5,1.5); 
    					\propag[bos] (b) to [in=180, out=180, looseness=1.62] (d);
    					\propag[bos, revmom'={$q$}] (d) to [in=0, out=0, looseness=1.62, edge label = $2$] (b);
    					\propag[chasca, mom'={$p$}] (a) to [edge label] (b);
    					\propag[chasca, mom'={$p$}] (b) to [edge label] (c);
    					\node at (-0.2,0.2) {$i$};\node at (3.2,0.2) {$j$};
    					\node at (1.5,-1.5) {$(c)$}; 
    				\end{feynhand} 
    			\end{tikzpicture}
    		\end{minipage} \\
    	\end{tabular}
    	\caption{One-loop contributions to the $\Phi$ mass after the addition of a soft mass to some components of the gauge bosons $\mathcal{A}_{\mu}^I$.}
    \label{fig:oneloopfrombosonsafterSSB}
    \end{figure}
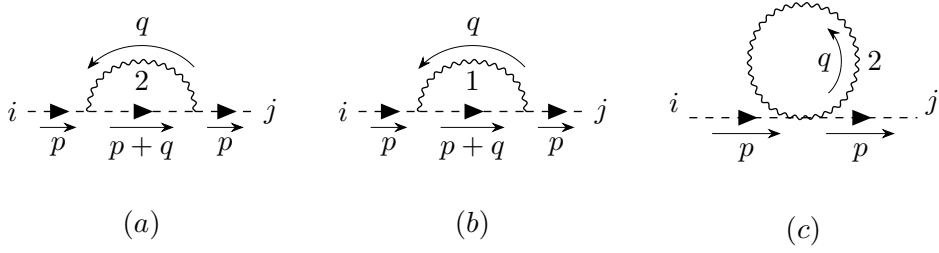
    \noindent We get
    \begin{align}
    	i(\Sigma^{a})^{i}_j(p)=&-\frac{ig^2}{16\pi^2}(\lambda^I)_{j}^k(\lambda^J)_{k}^ig^{(2)}_{IJ} \left[\frac{1}{6}(3m_{\mathcal{A}}^2+3m_{\Phi}^2-p^2)+\frac{1}{\epsilon}(2p^2+m_{\Phi}^2+m_{\mathcal{A}}^2)\right.\nonumber\\
    	&\left.+\int_0^1\dd x \left(p^2(4-6x+3x^2)+2m_{\Phi}^2x+2m_{\mathcal{A}}^2(1-x) \right)\log(\frac{\tilde{\mu}^2}{\Delta})\right]
    \end{align}
    with $\Delta=p^2x(x-1)+m_{\Phi}^2x+m_\mathcal{A}^2(1-x)$. Here we use the convention where $\epsilon \equiv (4 - D)/2$ and $\tilde{\mu}^2 \equiv 4 \pi \mu^2 e^{-\gamma_E}$, with $\gamma_E$ being the usual Euler-Mascheroni constant. Diagram $(b)$ is obtained by sending $g^{(2)}_{IJ}\to g^{(1)}_{IJ}$ and $m_{\mathcal{A}}^2\to 0$ in $i(\Sigma^{a})^i_{j}(p)$:
    \begin{align}
    	i(\Sigma^{b})^i_j(p)=&-\frac{ig^2}{16\pi^2}(\lambda^I)_{j}^k(\lambda^J)_{k}^ig^{(1)}_{IJ} \left[\frac{1}{6}(3m_{\Phi}^2-p^2)+\frac{1}{\epsilon}(2p^2+m_{\Phi}^2)\right.\nonumber\\
    	&\left.+\int_0^1\dd x \left(p^2(4-6x+3x^2)+2m_{\Phi}^2x\right)\log(\frac{\tilde{\mu}^2}{\Delta_0})\right]
    \end{align}
    where $\Delta_0=\Delta\eval_{m^2_{\mathcal{A}}=0}$.
    Finally
    \begin{align}
    	i(\Sigma^{c})_{j}^i(p)=\frac{ig^2}{16\pi^2}(\lambda^I)_{j}^k(\lambda^J)_{k}^ig^{(2)}_{IJ}2m^2_{\mathcal{A}}\left(1+2\left(\frac{1}{\epsilon}+\log(\frac{\tilde{\mu}^2}{m^2_{\mathcal{A}}})\right)\right).
    \end{align}
	The two prefactors $(\lambda^I)_{j}^k(\lambda^J)_{k}^ig^{(1)}_{IJ}$ and $(\lambda^I)_{j}^k(\lambda^J)_{k}^ig^{(2)}_{IJ}$ depend of course on which components of the gauge bosons we decide to give a mass to. For example, let us look at the case where the mass is given to the fermionic components. Define $g_{IJ}^{(F)}$ as the matrix equal to $g_{IJ}$ for the fermionic block and zero elsewhere, and similarly for $g_{IJ}^{(B)}$ and the bosonic blocks.
	Using the completeness relation of $SU(N)$ and $SU(M)$ together with the explicit form of $\lambda_U$ we get
    \begin{align}
    	(\lambda^I)_{j}^k(\lambda^J)_{k}^ig^{(B)}_{IJ}=\begin{pmatrix}
    		\frac{(N^2-1)}{2N} \mathbf{1}_{N\times N}  &0\\
    		0&0
    	\end{pmatrix}+
    	\begin{pmatrix}
    		0&0\\
    		0&-\frac{M^2-1}{2M} \mathbf{1}_{M\times M}
    	\end{pmatrix}+\frac{NM}{2(M-N)}\begin{pmatrix}
    		\frac{1}{N^2}\mathbf{1}_{N\times N}&0\\
    		0&\frac{1}{M^2}\mathbf{1}_{M\times M}
    	\end{pmatrix}\ .
    \end{align}
    The fermionic piece can be computed explicitly to obtain 
    \begin{align}
    	(\lambda^I)_{j}^k(\lambda^J)_{k}^ig^{(F)}_{IJ}=\begin{pmatrix}
    		-\frac{M}{2} \mathbf{1}_{N\times N}  &0\\
    		0&\frac{N}{2} \mathbf{1}_{M\times M}
    	\end{pmatrix}\ .
    \end{align}
    For $M=N+1$ we expect some cancellation to happen. Indeed,
    \begin{align}
    	(\lambda^I)_{j}^k(\lambda^J)_{k}^ig^{(F)}_{IJ}&\to\begin{pmatrix}
    		-\frac{N+1}{2} \mathbf{1}_{N\times N}  &0\\
    		0&\frac{N}{2} \mathbf{1}_{M\times M}
    	\end{pmatrix}\nonumber\\
    	(\lambda^I)_{j}^k(\lambda^J)_{k}^ig^{(B)}_{IJ}&\to\begin{pmatrix}
    		\frac{N+1}{2} \mathbf{1}_{N\times N}  &0\\
    		0&-\frac{N}{2} \mathbf{1}_{M\times M}
    	\end{pmatrix}=-(\lambda^I)_{j}^k(\lambda^J)_{k}^ig^{(F)}_{IJ}\ .
    	\label{eq:fermionicorbosoniccomplrel}
    \end{align}
    Using this result, we find that the sum of the three diagrams is 
    \begin{align}
    	&i\Sigma(p)^{i}_j=-\frac{ig^2}{16\pi^2}(\lambda^I)_{j}^k(\lambda^J)_{k}^ig^{(F)}_{IJ}\left[-\frac{3m_{\mathcal{A}}^2}{2}-\frac{3}{\epsilon}m_{\mathcal{A}}^2-4m_{\mathcal{A}}^2\log(\frac{\tilde{\mu}^2}{m_{\mathcal{A}}^2})\nonumber\right.\\
    	&\left.+\int_0^1 \dd x\left((p^2(4-6x+3x^2)+2m_{\Phi}^2)\log(\frac{\Delta_0}{\Delta})+2m_{\mathcal{A}}^2(1-x)\log(\frac{\tilde{\mu}^2}{\Delta})\right)\right]\ .
    \end{align}
    The result correclty vanishes for $m_{\mathcal{A}}^2\to0$. Renormalization can be performed in the $\overline{\text{MS}}$ scheme, and we see that only a counterterm to $m_{\Phi}^2$ and no field-strength renormalization are needed. More specifically, we need to add two different counterterms for the bosonic and fermionic components of $\Phi$, since the components of $(\lambda^I)_{jk}(\lambda^J)_{ki}g^{(F)}_{IJ}$ are different for the two cases. Of course, this is a consequence of having broken $SU(N|M)$.
    For the bosonic part, then, the physical mass is
    \begin{align}
    	m^2_{\Phi,phys}&=m^2_{\Phi}-\Sigma(m^2_{\Phi})\nonumber\\
    	&=m^2_{\Phi}-\frac{g^2}{16\pi^2}\left(\frac{N+1}{2}\right)\left\{m_{\mathcal{A}}^2\left[-\frac{3}{2}-4\log(\frac{\tilde{\mu}^2}{m_{\mathcal{A}}^2}) \right. \right.\nonumber\\
        &\left.\left. \qquad \qquad \qquad +2 \int_0^1\dd x(1-x)\log(\frac{\tilde{\mu}^2}{m_{\Phi}^2x^2+m_{\mathcal{A}}^2(1-x)})\right]\right.\nonumber\\ 
        &\left. \qquad \qquad \qquad + m_{\Phi}^2\int_0^1 \dd x(4-4x+3x^2)\log(\frac{m_{\Phi}^2 x^2}{m_{\Phi}^2x^2+m_{\mathcal{A}}^2(1-x)})\right\}\ .
    \end{align}
    If we assume $m_{\Phi}^2\ll m_{\mathcal{A}}^2$ we can expand
    \begin{align}
    	\label{eq:masscorrectionfrombosons}
    	m^2_{\Phi,phys}\approx &m^2_{\Phi}+m_{\mathcal{A}}^2\frac{g^2}{16\pi^2}\left(\frac{N+1}{2}\right)\left(1+3\log(\frac{\tilde{\mu}^2}{m^2_\mathcal{A}})\right)\ .	
    \end{align}
    As familiar to theories with more than one scale, this result exhibits possible large logs which would be removed by a careful procedure of matching and running across the $m_\mathcal{A}$ threshold. Nonetheless, the main conclusion would not change, namely that the UV-scale $m_\mathcal{A}$ is fed into the scalar mass, so that we cannot take it to be too large without requiring some fine-tuning. 
    
    Alternatively, we could have given a mass to the wrong sign component $A^2_\mu$ of $\mathcal{A}_\mu$. This means picking $g^{(2)}_{IJ}$ to be $-\delta_{IJ}$ in correspondence of the bosonic $SU(M)$ subgroup, and zero elsewhere, with $g^{(1)}_{IJ}=g_{IJ}-g^{(2)}_{IJ}$. Then the contracted completeness relation gives
    \begin{align}
        (\lambda_I)_{i}^kg^{(1)IJ}(\lambda_J)_{k}^j=-(\lambda_I)_{i}^kg^{(2)IJ}(\lambda_J)_{k}^j=\begin{pmatrix}
            0&0\\
            0&\frac{M^2-1}{2M}\mathbb{I}_{M\times M}
        \end{pmatrix}\ ,
    \end{align} 
    meaning
    \begin{align}
        \delta m^2_{\Phi_i}=m_{\mathcal{A}}^2\frac{g^2}{16\pi^2}\left(1+2\log(\frac{\tilde{\mu}^2}{m^2_\mathcal{A}})\right)\times\begin{cases}
        0 \qquad\text{if $\degf(i)=0$}\\
        \frac{M^2-1}{2M}\qquad\text{if $\degf(i)=1$}\ .
        \end{cases}
    \end{align}

    \subsubsection{Massive Vector Boson Scattering}
    Although turning on soft masses for a vector multiplet in a theory with spacetime supersymmetry does not pose any problems for perturbative unitarity, we are not so fortunate here. Massive non-abelian vector bosons are somewhat notoriously in conflict with perturbative unitarity in the absence of spontaneous symmetry breaking, since the amplitude for scattering their longitudinal modes grows quadratically with energy. This does not depend sensitively on the statistics of the vector fields, leading us to expect that it poses an obstruction to turning on soft breaking terms in the supergroup vector multiplet.

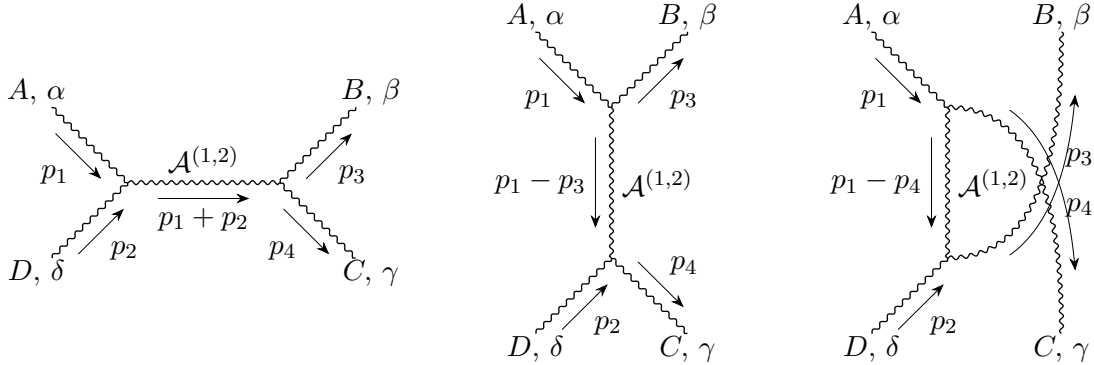
\begin{figure}[H]
    \centering
    \renewcommand{\arraystretch}{1.3}
    \begin{tabular}{p{6cm}p{4cm}p{4cm}}
        \begin{minipage}[c]{0.33\textwidth}
            \begin{tikzpicture}[scale=1]
                \begin{feynhand}
                    \vertex (a) at (-1,0); 
                    \vertex (b) at (1,0); 
                    \vertex (a2) at (-2,-1); 
                    \vertex (a1) at (-2,1); 
                    \vertex (b1) at (2,1); 
                    \vertex (b2) at (2,-1);
                    \propag[bos, mom'={$p_1$}] (a1) to (a);
                    \propag[bos, mom'={$p_2$}] (a2) to (a);
                    \propag[bos, mom'={$p_3$}] (b) to (b1);
                    \propag[bos, mom'={$p_4$}] (b) to (b2);
                    \propag[bos, mom'={$p_1+p_2$}] (a) to [edge label = $\mathcal{A}^{(1,2)}$](b);
                    \node at (-2.2,-1.2) {$D,\, \delta$};
                    \node at (-2.2,1.2) {$A,\,\alpha$};
                    \node at (2.2,1.2) {$B,\, \beta$};
                    \node at (2.2,-1.2) {$C,\, \gamma$};
                \end{feynhand} 
            \end{tikzpicture}
        \end{minipage} &
        \begin{minipage}[c]{0.33\textwidth}
            \begin{tikzpicture}[scale=1]
                \begin{feynhand}
                    \vertex (a) at (0,0); 
                    \vertex (b) at (0,-2); 
                    \vertex (a2) at (-1,1); 
                    \vertex (a1) at (1,1); 
                    \vertex (b1) at (-1,-3); 
                    \vertex (b2) at (1,-3);
                    \propag[bos, mom'={$p_3$}] (a) to (a1);
                    \propag[bos, mom'={$p_1$}] (a2) to (a);
                    \propag[bos, mom'={$p_2$}] (b1) to (b);
                    \propag[bos, mom={$p_4$}] (b) to (b2);
                    \propag[bos, mom'={$p_1-p_3$}] (a) to [edge label = $\mathcal{A}^{(1,2)}$](b);
                    \node at (-1,-3.2) {$D,\, \delta$};
                    \node at (-1,1.2) {$A,\,\alpha$};
                    \node at (1,1.2) {$B,\, \beta$};
                    \node at (1,-3.2) {$C,\, \gamma$};
                \end{feynhand} 
            \end{tikzpicture}
        \end{minipage} &
        \begin{minipage}[c]{0.33\textwidth}
            \begin{tikzpicture}[scale=1]
                \begin{feynhand}
                    \vertex (a) at (0,0); 
                    \vertex (b) at (0,-2); 
                    \vertex (a2) at (-1,1); 
                    \vertex (a1) at (1.5,1); 
                    \vertex (b1) at (-1,-3); 
                    \vertex (b2) at (1.5,-3);
                    \propag[bos, mom'={$p_4$}] (b) to [out=0, in=270] (a1);
                    \propag[bos, mom'={$p_1$}] (a2) to (a);
                    \propag[bos, mom'={$p_2$}] (b1) to (b);
                    \propag[bos, mom={$p_3$}] (a) to [out=0, in=90] (b2);
                    \propag[bos, mom'={$p_1-p_4$}] (a) to [edge label = $\mathcal{A}^{(1,2)}$](b);
                    \node at (-1,-3.2) {$D,\, \delta$};
                    \node at (-1,1.2) {$A,\,\alpha$};
                    \node at (1.5,1.2) {$B,\, \beta$};
                    \node at (1.5,-3.2) {$C,\, \gamma$};
                \end{feynhand} 
            \end{tikzpicture}
        \end{minipage} \\[2cm]
    \end{tabular}
    \caption{Contribution to the scattering of massive vector bosons from factorizable diagrams.}
    \label{fig:mvbscatteringfactorizable}
    \end{figure}

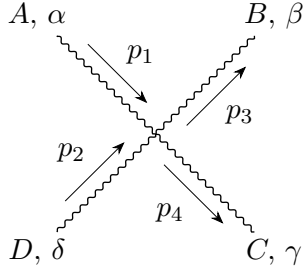
\begin{figure}[H]
    \centering
    \renewcommand{\arraystretch}{1.3}
    \begin{tabular}{p{3cm}}
    \centering
    \begin{minipage}[c][4cm]{1\textwidth}
            \begin{tikzpicture}[scale=1.3]
                \begin{feynhand}
                    \vertex (a) at (1,1); \vertex (a1) at (0,0); 
                    \vertex (a2) at (2,0); \vertex (a3) at (0,2); 
                    \vertex (a4) at (2,2); 
                    \propag[bos, mom=$p_2$] (a1) to (a); 
                    \propag[bos, mom'=$p_4$] (a) to (a2);
                    \propag[bos, mom=$p_1$] (a3) to (a);
                    \propag[bos, mom'=$p_3$] (a) to (a4);
                    \node at (-0.2,2.2) {$A,\,\alpha$};
                    \node at (2.2,2.2) {$B,\,\beta$};
                    \node at (2.2,-0.2) {$C,\,\gamma$};
                    \node at (-0.2,-0.2) {$D,\, \delta$};
                \end{feynhand} 
            \end{tikzpicture}
        \end{minipage}
    \end{tabular}
    \caption{Diagram contributing to the non-factorizable part of massive vector bosons scattering.}
    \label{fig:mvbscatteringnonfactorizable}
    \end{figure}

    We can extract the Feynman rules relevant to longitudinal scattering from the gauge Lagrangian in Eq.~\eqref{eq:gaugeselfint}:
	\begin{figure}[H]
    \centering
    \begin{tabular}{cc}
        \begin{minipage}[c][4cm]{0.5\textwidth}
            \begin{tikzpicture}[scale=1]
                \begin{feynhand}
                    \vertex (a) at (1,1); \vertex (a1) at (-0.5,1); 
                    \vertex (a2) at (2,0); 
                    \vertex (a4) at (2,2); 
                    \propag[bos, mom'={$p_3$}] (a1) to (a); 
                    \propag[bos, mom'={$p_1$}] (a2) to (a);
                    \propag[bos, mom'={$p_2$}] (a4) to (a);
                    \node at (-0.5,1.2) {$C,\,\gamma$}; 
                    \node at (2,2.2) {$B,\,\beta$};
                    \node at (2,-0.2) {$A,\,\alpha$};
                \end{feynhand} 
            \end{tikzpicture}
        \end{minipage} & 
       \begin{minipage}{0.5\textwidth}
       \hspace{-3cm}
       $-g f_{CBA}\left[\eta_{\alpha\gamma}(p_{3\beta}-p_{1\beta})+\eta_{\beta\gamma}(p_{2\alpha}-p_{3\alpha})+\eta_{\alpha\beta}(p_{1\gamma}-p_{2\gamma})\right]$
       \end{minipage}
       \\
    \begin{minipage}[c][4cm]{0.5\textwidth}
            \begin{tikzpicture}[scale=1]
                \begin{feynhand}
                    \vertex (a) at (1,1); \vertex (a1) at (0,0); 
                    \vertex (a2) at (2,0); \vertex (a3) at (0,2); 
                    \vertex (a4) at (2,2); 
                    \propag[bos] (a1) to (a); 
                    \propag[bos] (a2) to (a);
                    \propag[bos] (a) to (a3);
                    \propag[bos] (a4) to (a);
                    \node at (-0.2,2.2) {$D,\,\delta$};
                    \node at (2.2,2.2) {$C,\,\gamma$};
                    \node at (2.2,-0.2) {$B,\,\beta$};
                    \node at (-0.2,-0.2) {$A,\, \alpha$};
                \end{feynhand} 
            \end{tikzpicture}
        \end{minipage} & 
       \begin{minipage}{0.5\linewidth}
       \hspace{-3cm}
        $ \begin{aligned}[b]
              -i g^2 &\left\{f_{DCK}g^{KL}f_{LBA}(\eta_{\delta\beta}\eta_{\alpha\gamma}-\eta_{\delta\alpha}\eta_{\beta\gamma})+\right.\\
             &\left.+
             (-1)^{\degf(B)\degf(C)}f_{DBK}g^{KL}f_{LCA}(\eta_{\delta\gamma}\eta_{\alpha\beta}-\eta_{\delta\alpha}\eta_{\beta\gamma})+\right.\\
             &\left.+(-1)^{\degf(A)(\degf(B)+\degf(C))}f_{DAK}g^{KL}f_{LCB}(\eta_{\delta\gamma}\eta_{\alpha\beta}-\eta_{\delta\beta}\eta_{\alpha\gamma})\right\}
          \end{aligned} $
        \end{minipage}
    \end{tabular}
\end{figure}  
We can use these to compute the scattering of four longitudinally polarized massive vector fields at tree level, considering the case where a mass, $m_\mathcal{A}^2$, is given to the components relative to $g_{IJ}^{(2)}$.
For the polarization vectors, we take\footnote{see e.g. \cite{schwartz2014quantum}, Section~29.2.}
\begin{align}
    \epsilon_1^\mu& =\frac{1}{m_\mathcal{A}}p_1^\mu+\frac{2m_\mathcal{A}}{t-2{m_\mathcal{A}^2}}p_3^\mu& \epsilon_2^\mu& =\frac{1}{m_\mathcal{A}}p_2^\mu+\frac{2m_\mathcal{A}}{t-2{m_\mathcal{A}^2}}p_4^\mu\nonumber\\
    \epsilon_3^\mu& =\frac{1}{m_\mathcal{A}}p_3^\mu+\frac{2m_\mathcal{A}}{t-2{m_\mathcal{A}^2}}p_1^\mu& \epsilon_4^\mu& =\frac{1}{m_\mathcal{A}}p_4^\mu+\frac{2m_\mathcal{A}}{t-2{m_\mathcal{A}^2}}p_2^\mu\ ,
\end{align}
where we use the usual definition for the Mandelstam variables, $s=(p_1+p_2)^2$, $t=(p_1-p_3)^2$, $u=(p_1-p_4)^2$, verifying $s+t+u=4{m_\mathcal{A}^2}$.
The contributing diagrams can be split into the factorizable and contact term contributions.
\paragraph{Factorizable diagrams:}
The relevant diagrams are displayed in Fig.~\ref{fig:mvbscatteringfactorizable}. At high energy, they give the contribution
\begin{align}
    &i\mathcal{M}_{f}(1_{A,\alpha}2_{D,\delta}3_{B,\beta}4_{C,\gamma})\nonumber\\
    &=i g^2 \left\{f_{AID}f_{JBC}\left(g^{IJ}\frac{s(s+2t)}{4{m_\mathcal{A}^2}}+\frac{g^{(2)IJ}t(s+2t)+8 g^{IJ}(s^2-st-t^2)}{4{m_\mathcal{A}^2} t}\right)\right.\nonumber\\
    &+f_{ABI}f_{JCD}\left(g^{IJ}\frac{t(2s+t)}{4{m_\mathcal{A}^2}}+\frac{8 g^{IJ} s+g^{(2)IJ}(2s+t)}{4{m_\mathcal{A}^2}}\right)\nonumber\\
    &\left.+(-1)^{\degf(C)\degf(B)}f_{ACI}f_{JBD}\left(g^{IJ}\frac{t^2-s^2}{4{m_\mathcal{A}^2}}-\frac{8 g^{IJ} (s+t)^2+g^{(2)IJ}t(t-s)}{4{m_\mathcal{A}^2} t}\right)\right\}+\order{1}\ ,
\end{align}
where we only kept the terms that grow with energy and indicated with $\order{1}$ those that do not.
    \paragraph{Contact term:}
    There is only one diagram giving the non-factorizable contribution, displayed in Fig.~\ref{fig:mvbscatteringnonfactorizable}, which yields
    \begin{align}
        &i\mathcal{M}_{nf}(1_{A,\alpha}2_{D,\delta}3_{B,\beta}4_{C,\gamma})\nonumber\\
        &=i g^2 g^{IJ}\left\{f_{AID}f_{JBC}\left(-\frac{s(s+2t)}{4{m_\mathcal{A}^2}}-\frac{s(2s+t)}{{m_\mathcal{A}^2} t}\right)+f_{ABI}f_{JCD}\left(-\frac{t(2s+t)}{4{m_\mathcal{A}^2}}+\frac{t}{{m_\mathcal{A}^2}}\right)\right.\nonumber\\
    &\left. \qquad \qquad \qquad +(-1)^{\degf(C)\degf(B)}f_{ACI}f_{JBD}\left(\frac{s^2-t^2}{4{m_\mathcal{A}^2}}+\frac{(2s^2+st+t^2)}{{m_\mathcal{A}^2} t}\right)\right\}+\order{1}\ ,
    \end{align}
    where again we dropped terms not growing with energy. 
    The total amplitude then reads
    \begin{align}
        &i\mathcal{M}(1_{A,\alpha}2_{D,\delta}3_{B,\beta}4_{C,\gamma})=i\mathcal{M}_{f}(1_{A,\alpha}2_{D,\delta}3_{B,\beta}4_{C,\gamma})+i\mathcal{M}_{nf}(1_{A,\alpha}2_{D,\delta}3_{B,\beta}4_{C,\gamma})\nonumber \\
        &=\frac{ig^2}{4{m_\mathcal{A}^2}}\left\{f_{AID}f_{JBC} \left(g^{(2)IJ}(s+2t)-4g^{IJ}(3s+2t)\right)+f_{ABI}f_{JCD} \left(g^{(2)IJ}+4g^{IJ}\right)(2s+t)\right.\nonumber\\
        &+\left.(-1)^{\degf(C)\degf(B)}f_{ACI}f_{JBD}\left(g^{(2)IJ}(s-t)-4g^{IJ}(3s+t)\right)
        \right\}+\order{1}\ .
        \label{eq:colorordered1}
    \end{align}
    As expected, the strongest high-energy growth of $\order{E^4}$ is canceled between the factorizable and non-factorizable contributions, separately for each color structure. Indeed the highest energy growth behaves as in the massless, gauge-invariant case, where no energy growth is expected. The first correction, then, appears at subleading order. 
    
    Let us now add soft-breaking masses to some of the fermionic degrees of freedom and consider the impact on the scattering of the same degrees of freedom. In this case all the contractions in Eq.~\eqref{eq:colorordered1} containing $g^{(2)IJ}$ vanish, since by assumption $g^{(2)IJ}$ is only non-zero for $I$ and $J$ both fermionic, but the structure constants vanish if all three indices are fermionic. Thus, we are left with 
    \begin{align}
        &i\mathcal{M}^{4f}(1_{A,\alpha}2_{D,\delta}3_{B,\beta}4_{C,\gamma})=\frac{ig^2}{{m_\mathcal{A}^2}}\left\{-f_{AID}f_{JBC} (3s+2t)+f_{ABI}f_{JCD} (2s+t)\right.\nonumber\\
        &\left. \qquad \qquad \qquad \qquad \qquad \qquad -(-1)^{\degf(C)\degf(B)}f_{ACI}f_{JBD}(3s+t)\right\}g^{IJ}+\order{1}\nonumber\\
        =&\frac{2ig^2}{{m_\mathcal{A}^2}}\left[s\left(\str{\lambda_{A}\lambda_{C}\lambda_{D}\lambda_{B}}(-1)^{\degf(B)(\degf(C)+\degf(D))}+\str{\lambda_{A}\lambda_{B}\lambda_{D}\lambda_{C}}(-1)^{\degf(C)\degf(D)}\right)\right.\nonumber\\
        &\left.+t\left(\str{\lambda_{A}\lambda_{D}\lambda_{B}\lambda_{C}}(-1)^{\degf(D)(\degf(B)+\degf(C))}+\str{\lambda_{A}\lambda_{C}\lambda_{B}\lambda_{D}}(-1)^{\degf(B)\degf(C)}\right)\right.\nonumber\\
        &\left.+u\left(\str{\lambda_{A}\lambda_{B}\lambda_{C}\lambda_{D}}+\str{\lambda_{A}\lambda_{D}\lambda_{C}\lambda_{B}}(-1)^{\degf(B)(\degf(C)+\degf(D))+\degf(C)\degf(D)}\right)\right]\nonumber\\
        =&\frac{2ig^2}{{m_\mathcal{A}^2}}\left[s\left(\str{\lambda_{A}\lambda_{C}\lambda_{D}\lambda_{B}}-\str{\lambda_{A}\lambda_{B}\lambda_{D}\lambda_{C}}-\str{\lambda_{A}\lambda_{B}\lambda_{C}\lambda_{D}}+\str{\lambda_{A}\lambda_{D}\lambda_{C}\lambda_{B}}\right)\right.\nonumber\\
        &\left.+t\left(\str{\lambda_{A}\lambda_{D}\lambda_{B}\lambda_{C}}-\str{\lambda_{A}\lambda_{C}\lambda_{B}\lambda_{D}}-\str{\lambda_{A}\lambda_{B}\lambda_{C}\lambda_{D}}+\str{\lambda_{A}\lambda_{D}\lambda_{C}\lambda_{B}}\right)+\order{1}
        \right]\ ,
        \label{eq:colororderedfermions}
    \end{align}
    where we used the fact that, thanks to the completeness relation, we can write
    \begin{align}
        f_{ABI}f_{CDJ}g^{IJ}=2\str{\comm{\lambda_C}{\lambda_D}_{\degf}\comm{\lambda_A}{\lambda_B}_{\degf}}\ .
    \end{align}
    The amplitude now grows with $s$, signaling the breakdown of perturbative unitarity at high energies. Soft masses for supergroup vector multiplets seem to require spontaneous symmetry breaking. 
    
	\subsection{Adding spinors}
    \label{subsec:addingfermions1}
	Finally, let us consider the effects of Yukawa couplings between spinor multiplets and our scalar multiplet in the fundamental representation.  As a first example, consider a model in which the Yukawa couplings involve a spinor, $\Theta^{I}$, in the adjoint of $SU(N|M)$ and a spinor, $\xi_i$, in the fundamental.
	Besides the kinetic terms, the Lagrangian contains a Yukawa interaction term
	\begin{align}
		\mathcal{L}_{Yuk}=-y\Phi_i\bar{\xi}^j(\lambda_I)_{i}^j\Theta^I+\text{ h.c.}\ ,
	\end{align}
	implying the Feynman rule
	\begin{table}[H]
		\centering
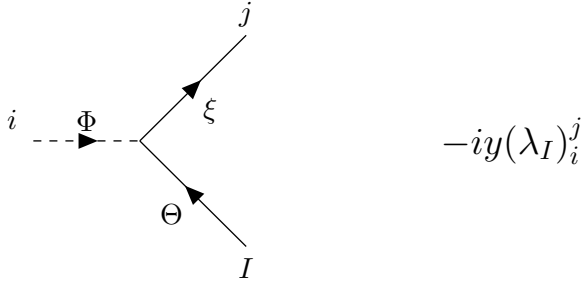
	\begin{tabular}{c}
		\begin{minipage}[c][4cm]{0.8\textwidth}
				\begin{tikzpicture}[scale=1.4]
						\begin{feynhand}
								\vertex (a) at (1,1); \vertex (a1) at (0,1); 
								\vertex (a2) at (2,0); 
								\vertex (a4) at (2,2); 
								\propag[chasca] (a1) to [edge label = $\Phi$] (a); \propag[fer] (a2) to [edge label = $\Theta$] (a);
								\propag[antfer] (a4) to [edge label = $\xi$] (a);
								\node at (4.5,1) {\Large{$-i y(\lambda_I)_{i}^j$}};
								\node at (-0.2,1.2) {$i$}; \node at (2,2.2) {$j$};
								\node at (2,-0.2) {$I$};
							\end{feynhand} 
					\end{tikzpicture}
			\end{minipage} 
	\end{tabular}
	\end{table}
	
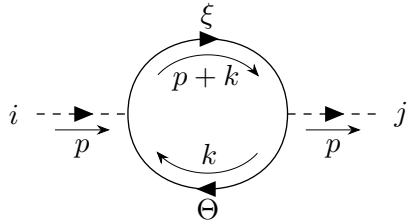
\begin{figure}[H]
        \centering
        \renewcommand{\arraystretch}{1}
        \begin{tabular}{ccc}
            \begin{minipage}[c][3cm]{0.4\textwidth}
                \begin{tikzpicture}[scale=1.5]
                    \begin{feynhand}
                        \vertex (a) at (0.8,1); \vertex (a1) at (0,1); 
                        \vertex (b) at (2.2,1); 
                        \vertex (b1) at (3,1); 
                        \propag[chasca, mom'={$p$}] (a1) to (a); 
                        \propag[fer, mom' ={$k$}] (b) to [edge label = $\Theta$, in=270, out=270,looseness= 1.62] (a);
                        \propag[antfer, revmom={$p+k$}] (b) to [edge label' = $\xi$, in=90, out=90, looseness= 1.62] (a);
                        \propag[chasca, mom'={$p$}] (b) to (b1); 
                        \node at (-0.2,1) {$i$}; \node at (3.2,1) {$j$};
                    \end{feynhand} 
                \end{tikzpicture}
            \end{minipage}
        \end{tabular}
        \caption{One-loop contribution to the $\Phi$ mass from the Yukawa interaction.}
        \label{eq:oneloopfromfermions}
    \end{figure}

	At one-loop there is only one diagram contributing to the mass renormalization of $\Phi$, shown in Fig.~\ref{eq:oneloopfromfermions}.
    Its contribution is
    \begin{align}
        i\Sigma(p)=-(-iy)^2(i\delta_{k}^l)(ig^{IJ})(\lambda_I)_{i}^k(\lambda_J)_{l}^j\int\frac{\dd[d]{k}}{(2\pi)^k}\frac{\Tr\left[(\slashed{p}+\slashed{k}+m_\xi)(\slashed{k}+m_\Theta)\right]}{\left[(p+k)^2-m_{\xi}^2+i\epsilon \right]\left[k^2-m_{\Theta}^2+i\epsilon\right]}\ .
    \end{align}
    Again, the prefactor of this diagram vanishes for $M=N\pm1$.
    
     \subsubsection{Soft mass for some $\Theta^I$ components}
    Now suppose we wish to give a soft mass to some of the components of (say) the spinor $\Theta^I$ introduced in Section~\ref{subsec:addingfermions1}. We do this by splitting the metric $g_{IJ}=g_{IJ}^{(1)}+g_{IJ}^{(2)}$ and giving an additional soft mass to the components corresponding to $g_{IJ}^{(2)}$, i.e. 
    \begin{align}
    	\mathcal{L}\to\mathcal{L}-m_{\Theta,\text{soft}}\bar{\Theta}^Ig_{IJ}^{(2)}\Theta^J\ .
    \end{align}
    Then there are two diagrams responsible for the correction to the $\Phi$ mass; both of them are of the same form as Fig.~\ref{eq:oneloopfromfermions}, but with the modes relative to $g_{IJ}^{(1)}$ and $g_{IJ}^{(2)}$ running in the loop, respectively. Defining
    \begin{align}
    	\mathcal{I}_{\text{Yuk}}(p,m_\Theta,m_\xi)\equiv\int\frac{\dd[d]{k}}{(2\pi)^k}\frac{\Tr\left[(\slashed{p}+\slashed{k}+m_\xi)(\slashed{k}+m_\Theta)\right]}{\left[(p+k)^2-m_{\xi}^2+i\epsilon \right]\left[k^2-m_{\Theta}^2+i\epsilon\right]}\ ,
    \end{align}
    we have for the two contributions
    \begin{align}
    	i\Sigma^{(1)}(p)+i\Sigma^{(2)}(p)=-y^2\delta^k_l(\lambda_I)_{i}^k(\lambda_J)_l^j&\left[g^{(1)IJ}\mathcal{I}_{\text{Yuk}}(p,m_\Theta,m_\xi) \right. \nonumber \\
        &\left. \quad +g^{(2)IJ}\mathcal{I}_{\text{Yuk}}(p,m_\Theta+m_{\Theta,\text{soft}},m_\xi)\right]\ .
    \end{align}
    For $M=N+1$ we can use that
    \begin{align}
    	0=(\lambda_I)_{i}^k(\lambda_J)_k^jg^{IJ}\Longrightarrow (\lambda_I)_{i}^k(\lambda_J)_k^jg^{(2)IJ}=-(\lambda_I)_{i}^k(\lambda_J)_k^jg^{(1)IJ}\ ,
    \end{align}
    meaning
    \begin{align}
    	i(\Sigma^{(1)}(p))^j_i+i(\Sigma^{(2)}(p))^j_i=-y^2(\lambda_I)_{i}^k(\lambda_J)_k^jg^{(1)IJ} &\left(\mathcal{I}_{\text{Yuk}}(p,m_\Theta,m_\xi) \right. \nonumber \\
        &\left. \quad -\mathcal{I}_{\text{Yuk}}(p,m_\Theta+m_{\Theta,\text{soft}},m_\xi)\right)\ .
    \end{align}
    Then the correction to the physical mass of the component $\Phi_i$, in the limit of $m_{\Theta,\text{soft}}\gg m_{\Theta},m_\xi,m_{\Phi}$ is
    \begin{align}
    	\delta m_{\Phi_i}^2=-(\lambda_I)_{i}^k(\lambda_J)_k^jg^{(1)IJ}\frac{y^2}{32 \pi^2}m_{\Theta,\text{soft}}^2\left(3-2\log(\frac{\tilde{\mu}^2}{m_{\Theta,\text{soft}}^2})\right)\ .
    \end{align}
    For example, giving a soft mass to the wrong-statistics components of $\Theta$ means taking $g^{(2)IJ}=g^{(F)IJ}$, where $g^{(F)IJ}$ is the fermionic part of the metric we already used in Section~\ref{sec:softmassesgaugebosons}. Then we can use Eq.~\eqref{eq:fermionicorbosoniccomplrel} to get
    \begin{align}
    	\delta m_{\Phi_i}^2=-\frac{y^2}{32 \pi^2}m_{\Theta,\text{soft}}^2\left(3-2\log(\frac{\tilde{\mu}^2}{m_{\Theta,\text{soft}}^2})\right)\times\begin{cases}
    		\frac{N+1}{2} \qquad\text{if $\degf(i)=0$}\\
    		-\frac{N}{2} \qquad\text{if $\degf(i)=1$}\ .
    	\end{cases}
    \end{align}
    
    \subsection{Adding spinors II}
    The result of the previous section relied crucially on $\Theta$ belonging to the adjoint representation. It would also be interesting to find an example where a similar cancellation for the correction to the scalar mass arises for a spinor belonging to the fundamental representation of $SU(N|M)$. 
    While this is not the case for a spinor transforming in the fundamental coupled to a spinor transforming as a singlet, a slight addition to our construction will help in reaching our goal. 
    Consider a theory of two spinors belonging to the fundamental of $SU(N|M)$, $\xi_i$ and $\tilde{\xi}_i$. In addition, we add two spinors, $\chi$ and $\tilde{\chi}$, that are singlets of $SU(N|M)$. However, we assign the correct, fermionic statistic to $\xi_i$ and $\chi$, but wrong, bosonic statistic to $\tilde{\xi}_i$ and $\tilde{\chi}$. We can then write the Lagrangian as
    \begin{align}
        \mathcal{L}=\mathcal{L}_{\Phi}+i \bar{\xi}^i\slashed{\partial}\xi_i+i \bar{\tilde{\xi}}^i\slashed{\partial}\tilde{\xi}_i+i\bar{\chi}\slashed{\partial}\chi+i\bar{\tilde{\chi}}\slashed{\partial}\tilde{\chi}-y\Phi_i(\bar{\xi}^i\chi+\bar{\tilde{\xi}}^i
        \tilde{\chi})\ .
    \end{align}
    This Lagrangian is symmetric under 
    \begin{align}
        \xi_i&\to \tilde{\xi}_i &\text{ and}&&\chi&\to\tilde{\chi}\ .
        \label{eq:discretetransformation}
    \end{align}
    Of course, we could have packed the two pairs of spinors into (super-)vectors
    \begin{align}
        \Xi_i&=\begin{pmatrix}
            \xi_i\\
            \tilde{\xi}_i
        \end{pmatrix}&
        X&=\begin{pmatrix}
            \chi\\
            \tilde{\chi}
        \end{pmatrix}\ ,
    \end{align}
    and written
    \begin{align}
        \mathcal{L}=\mathcal{L}_{\Phi}+\bar{\Xi}^i\slashed{\partial}\Xi_i+i\bar{X}\slashed{\partial}X-y\Phi_i\bar{\Xi}^iX\ .
    \end{align}
    Seen in this form, the Lagrangian is actually invariant under a continuous $SU(1|1)$. However, the construction of $SU(N|N)$ entails some complications, linked to the fact that the identity matrix is supertraceless when $N=M$ and the matrix $g_{IJ}$ is singular \cite{Bars1984, Arnone:2001iy}. As such, we content ourselves with the discrete transformation properties in Eq.~\eqref{eq:discretetransformation}. In this case, the one-loop correction to the mass coming from the $\order{y^2}$ diagrams cancels because of the difference in sign between the loop containing $\xi$ and $\chi_i$, and that containing $\tilde{\xi}$ and $\tilde{\chi}_i$.

    At this stage, we have seen that the mass of a scalar multiplet in the fundamental of $SU(N|M)$ is not renormalized by its own quartic at one-loop provided $M=N+1$; is not renormalized by $SU(N|M)$ gauge interactions at one-loop provided $M=N \pm 1$; and is not renormalized by $SU(N|M)$-symmetric Yukawa interactions when $M=N\pm 1$ for select representations of the spinor multiplets. Turning on soft supergroup symmetry-breaking masses induces one-loop corrections proportional to the soft terms with only logarithmic cutoff sensitivity, much as in the soft breaking of spacetime supersymmetry. In contrast to spacetime supersymmetry, however, turning on soft masses in the supergroup vector multiplet necessarily leads to tree-level unitarity violation (above and beyond the unitarity issues posed by the negative-norm states themselves). This motivates the exploration of spontaneous symmetry breaking.
    
    \section{Breaking $SU(N|M)$}
    \label{sec:breakingSUNM}
    The surprising one-loop properties of theories with global or local $SU(N|N+1)$ symmetry explored in Section~\ref{sec:model} warrant further study despite the unitarity challenges posed by the wrong-statistics and wrong-sign ghosts. As a first step, turning on soft masses for these problematic fields raises the possibility that they might be partially decoupled or rendered unstable, opening the door to a unitary interpretation {\it a la} Lee \& Wick \cite{Lee:1969fy}. As we have already seen, soft terms in the vector multiplet seem to require UV completion in the form of spontaneous symmetry breaking. More broadly, it would be satisfying to interpret all soft terms as low-energy remnants of spontaneous breaking of the supergroup symmetry. 
    
    Here we present a way to break the $SU(N|M)$ symmetry down to its bosonic subgroup $SU(N)\times SU(M) \times U(1)$ with a Higgs-like mechanism. If the $SU(N|M)$ symmetry is gauged, this provides a mass for the $B^i_\mu$ fields, i.e. the wrong-statistics components of the vector multiplet. While one might be tempted to obtain this pattern of symmetry breaking from the vev of a scalar field belonging to the adjoint representation of $SU(N|M)$, it turns out that the allowed potential for this multiplet does not lead to the desired vacuum. To obtain the desired pattern of symmetry breaking, we can instead add to the theory a scalar field belonging to the product of a fundamental and an antifundamental representation, without the constraint of (super)tracelessness. This example will turn out to have the desired properties, and a local minimum with the right symmetry breaking pattern can be found.

    \subsection{Adjoint of $SU(N|M)$}
    As anticipated, we first introduce a scalar field $\Sigma^i_j$ belonging to the adjoint representation of $SU(N|M)$. To avoid cubic terms in the potential, we enforce on it a $\mathbb{Z}_2$ symmetry $\Sigma^i_j\to-\Sigma^i_j$. Viewed as a matrix, $\Sigma^i_j$ is hermitian and supertraceless, $\str{\Sigma}=0$. Its Lagrangian reads
		\begin{align}
			\mathcal{L}_{\Sigma}=\str{\comm{\nabla_\mu}{\Sigma}^2}+\mu^2\str{\Sigma^2}-\quarticSigmaA\str{\Sigma^2}^2-\frac{1}{4}\quarticSigmaB \str{\Sigma^4}\ ,
		\end{align}
		where we added all renormalizable terms allowed by symmetry.
    By dimensional analysis
		\begin{align}
			[\mu]&=1 & [\quarticSigmaA]=[\quarticSigmaB ]=4-d\ ,
		\end{align}
		with $d$ being the number of space-time dimensions.
    The kinetic term can be rewritten in a more familiar notation using\footnote{Here and in the following we go back and forth between the picture where objects like $\Sigma$ belonging to the adjoint representation are seen as matrices with one index in the fundamental and one in the anti-fundamental, and the one where they are seen as vectors with one index belonging to the adjoint representation. The mapping between the two pictures is done via the generators $(\lambda_I)^i_j$ so that $\Sigma^i_j=\Sigma^I(\lambda_I)^i_j$.\label{note:adjointvsfund}}
        \begin{align}
			\comm{\nabla_\mu}{\Sigma}^{i}_j&\equiv\dlo{\mu}\Sigma^i_j+ig\comm{\mathcal{A}_\mu}{\Sigma}^i_j=\left(\dlo{\mu}\Sigma^K-g \mathcal{A}_{\mu}^I\tensor{f}{_I_J^K}\Sigma^J\right)(\lambda_K)^i_j\ ,
		\end{align}
		and
		\begin{align}
			\str{\comm{\nabla_\mu}{\Sigma}^2}&=\str{\left(\dlo{\mu}\Sigma^K-g\mathcal{A}_{\mu}^I\tensor{f}{_I_J^K}\Sigma^J\right)\lambda_K\left(\dup{\mu}\Sigma^L-g\mathcal{A}^{\mu M}\tensor{f}{_M_N^L}\Sigma^N\right)\lambda_L}\nonumber\\
			&=\frac{1}{2}\left(\dlo{\mu}\Sigma^K-g\mathcal{A}_{\mu}^I\tensor{f}{_I_J^K}\Sigma^J\right)g_{KL}\left(\dup{\mu}\Sigma^L-g\mathcal{A}^{\mu^M}\tensor{f}{_M_N^L}\Sigma^N\right)\ .
		\end{align}
        In particular, the pure kinetic term is
		\begin{align}
			\mathcal{L}_{\Sigma,\text{kin}}=\frac{1}{2}\dlo{\mu}\Sigma^I\dup{\mu}\Sigma^Jg_{IJ}
		\end{align}
		where $g_{IJ}$ is the metric in Eq.~\eqref{eq:metric}. Again, $\Sigma^I$ contains both wrong-statistics components, corresponding to $g_{IJ}=\pm i$, and wrong-sign ones, corresponding to $g_{IJ}=-1$. 
    To check the behaviour of the potential along different directions in field space, we need to rephase the bosonic components of $\Sigma$ so that they all have the right sign for the kinetic term. (In other words, we are interested in finding extrema of the potential that have ghosts but do not have tachyons or tachyonic ghosts.) We can do that by using a diagonal matrix $\tensor{A}{^I_J}$
		\begin{align}
			\Sigma^I \to \tensor{A}{^I_J}\Sigma^J\equiv \tilde{\Sigma}^I,
            \label{eq:rephasing}
		\end{align}
            and defining
            \begin{align}
                g_{IJ}\to \tensor{A}{_I^K}\tensor{A}{_J^L}g_{KL}\equiv\tilde{g}_{IJ}\ ,
            \end{align}
		so that the kinetic term
		\begin{align}
			\frac{1}{2}\dlo{\mu}\tilde{\Sigma}^I\dup{\mu}\tilde{\Sigma}^J\tilde{g}_{IJ},
		\end{align}
		has the right signs for the bosonic part. 
  More specifically, we pick
		\begin{align}
			\tensor{A}{^I_J}=\text{diag }(\underbrace{1,1,\dots, 1}_\textrm{$N^2$ times},\underbrace{i,i,\dots, i}_\textrm{$M^2-1$ times},\underbrace{1,1,\dots, 1}_\textrm{$2N M$ times}),
		\end{align}
		where the first $N^2$ terms correspond to the $N^2-1$ generators of the upper $SU(N)$ bosonic block, plus one $U(1)$ generator (which, with our normalization, has the correct sign for $M>N$), the following $M^2-1$ terms to the generators of the bosonic $SU(M)$, and finally the last part is picked to leave the fermionic generators untouched. 
		With this transformation, the mass term for $\Sigma$ becomes
		\begin{align}
			\mathcal{L}_{\Sigma}&\supset \mu^2\str{\Sigma^2}=\frac{1}{2}\mu^2\Sigma^Ig_{IJ}\Sigma^J \underset{\Sigma \rightarrow \tilde{\Sigma}}{\rightarrow}\nonumber\\
			&=\frac{1}{2}\mu^2\tilde{\Sigma}^I\tilde{g}_{IJ}\tilde{\Sigma}^J\ ,
		\end{align}
		showing that the mass term for the bosonic fields keeps its tachyonic sign once we perform the rephasing.
		This suggests that $\expval{\Sigma}=0$ should represent a local maximum for the potential, and a minimum must be looked for somewhere else.
		\subsection{Stationary point of $V[\Sigma]$}
        \label{sec:stationarypoint1}
        As we detail in Appendix~\ref{app:wrongpotential}, the potential $V[\Sigma]$ does not have minima that induce spontaneous symmetry breaking in the pattern $SU(N|M)\to H\supset SU(N)$. Then, to reach our goal, we need to be slightly more daring. We relax one of our assumptions and consider a field $\Sigma^i_j$ transforming as a direct product of a fundamental and antifundamental representations, but without the constraint of it being supertraceless. Again, we impose on $\Sigma^i_j$ a $\mathbb{Z}_2$ symmetry so that we can avoid odd terms in the potential. 
		The decomposition of $\Sigma$ in terms of generators can still be done provided we extend the list of generators to include the identity
		\begin{align}
			\lambda_I &\to \lambda_{\tilde{I}}=\left\{\lambda_I,\, \lambda_T\equiv \frac{1}{\sqrt{2(N-M)}}\mathbb{I}\right\}\ .
		\end{align}
        This amounts to extending the algebra of $SU(N|M)$ to $U(N|M)$, as we are adding back the supertraceful generator $\lambda_T$. 
        We choose the potential to still be
		\begin{align}
			V[\Sigma]&=-\frac{1}{2}\mu^2\Sigma^{\tilde{I}}g_{\tilde{I}\tilde{J}}\Sigma^{\tilde{J}}+\frac{1}{4}\quarticSigmaA \left(\Sigma^{\tilde{I}}g_{\tilde{I}\tilde{J}}\Sigma^{\tilde{J}}\right)^2+\frac{1}{4}\quarticSigmaB \Sigma^{\tilde{I}}\Sigma^{\tilde{J}}\Sigma^{\tilde{K}}\Sigma^{\tilde{L}}T_{\tilde{I}\tilde{J}\tilde{K}\tilde{L}}\ .
            \label{eq:scalarpotential2}
		\end{align}
        where we have implicitly defined 
        \begin{align}
            g_{\tilde{I}\tilde{J}}&\equiv2\str{\lambda_{\tilde{I}}\lambda_{\tilde{J}}}\\
            T_{\tilde{I}\tilde{J}\tilde{K}\tilde{L}}&\equiv\str{\lambda_{\tilde{I}}\lambda_{\tilde{J}}\lambda_{\tilde{K}}\lambda_{\tilde{L}}}\ .
        \end{align}
        In particular, $g_{\tilde{I}\tilde{J}}$ is the same as $g_{IJ}$ but for an additional $1$ in the diagonal corresponding to $\lambda_T$.
		
  Notice that Eq.~\eqref{eq:scalarpotential2} is not the most general form of the potential anymore, as we have set to zero all terms $\propto \str{\Sigma}$. This is certainly allowed at tree level, but since there is no symmetry protecting this choice we expect it to be lifted at one-loop and beyond. We will first confirm that the tree-level vacuum is viable before proceeding to check stability at one-loop.
  
        \subsection{Runaway directions}
        As a first check, we need to assure ourselves that the potential in Eq.~\eqref{eq:scalarpotential2} is bounded from below.
        Since the potential is gauge invariant, we can always evaluate it on a diagonal $\Sigma^i_j$. We can then parametrize the independent directions spanning $V$ with a $\Sigma^{\tilde{I}}$ of the form
		\begin{align}
			\Sigma^i_j=\Sigma^{(D)\tilde{I}} \left(\lambda^{(D)}_{\tilde{I}}\right)^i_j
		\end{align}
		where $ \lambda^{(D)}_{\tilde{I}}$ are the diagonal generators. To consider the physical directions, we rephase with an $i$ the components with the wrong sign, so that $g_{\tilde{I}\tilde{J}}^{(D)}\to \delta_{\tilde{I}\tilde{J}}$. 
		After this rephasing, we choose spherical coordinates on the space spanned by the new $\Sigma^{(D)\tilde{I}} $. If we call $\rho$ the radial coordinate, we get
		\begin{align}
			V[\Sigma]\to -\frac{1}{2}\mu^2 \rho^2+\frac{\quarticSigmaA}{4}\rho^4+\frac{\quarticSigmaB }{4}\rho^4 T(\theta_i)
		\end{align}
		where we extracted a $\rho^4$ from the $\quarticSigmaB $ terms on dimensional grounds and called $T(\theta_i)$ the remaining, $\rho$-independent function, where $\theta_i$ are the angular coordinates of our spherical parametrization. $T(\theta_i)$ is just a polynomial in $\cos(\theta_i)$ and $\sin(\theta_i)$. Since $\cos(\theta_i),\, \sin(\theta_i)\in [-1,1]$, $T(\theta_i)$ is bounded from above and below, meaning there exists one (or more) $\theta_{i,\text{max}}$ such that $\max(T(\theta_i))=T(\theta_{i,\text{max}})$ is at its maximum, and conversely one (or more) $\theta_{i,\text{min}}$ such that $\min(T(\theta_i))=T(\theta_{i,\text{min}})$. $T(\theta_{i,\text{max}})$ and $T(\theta_{i,\text{min}})$ are then just numbers fixed by the group structure. It is then clear that there always exist a large portion of parameter space such that $V[\rho \gg 1]>0$. 
    	\subsubsection{Minimization of the potential}\label{subsubsec:minimizingPotential}
		While the generic structure of the potential is quite nontrivial to study, we may content ourselves into looking for minima in specific directions. In particular, let us pick the ansatz
		\begin{align}
			\expval{\Sigma^I}=\rho_1\delta^I_U+\rho_2\delta^I_T\ .
		\end{align}
        
		Inside this subspace, the gradient of the potential is
		\begin{align}
			\partial_{\tilde{A}}V[\expval{\Sigma^I}]=&-\mu^2(\rho_1g_{\tilde{A}U}+\rho_2g_{\tilde{A}T})+\quarticSigmaA(\rho_1g_{\tilde{A}U}+\rho_2g_{\tilde{A}T})(\rho_1^2+\rho_2^2)\nonumber\\
			&+\quarticSigmaB \left(\hat{T}_{\tilde{A}UUU}\rho_1^3+3 \hat{T}_{\tilde{A}UUT}\rho_1^2\rho_2+3\hat{T}_{\tilde{A}TTU}\rho_2^2\rho_1+\hat{T}_{\tilde{A}TTT}\rho_2^3\right)\ .
		\end{align}
        where we defined for brevity $\hat{T}_{ I J K L}\equiv\str{\lambda_{\{I}\lambda_{J}\lambda_{K}\lambda_{L\}_{\degf}}}$.
		Let us analyze the different possibilities
		\begin{itemize}
			\item $\tilde{A}$ fermionic: first of all $g_{\tilde{A}T}=g_{\tilde{A}U}=0$. Moreover, for any matrix $M$, $\str{M}$ can only be nonzero if M is bosonic. Indeed it is easy to convince oneself that e.g. $\lambda_{\tilde{A}}\lambda_T\lambda_T\lambda_U$ only has non-zero components in the off-diagonal blocks, for $\tilde{A}$ fermionic. Thus, all pieces in the $\quarticSigmaB $ term vanish: $\partial_{\tilde{A}}V[\expval{\Sigma^I}]=0$ for $\tilde{A}$ fermionic. 
			\item $\tilde{A}$ bosonic but $\tilde{A}\neq T, U$: again $g_{\tilde{A}T}=g_{\tilde{A}U}=0$. Moreover, since $\lambda_U$ and $\lambda_T$ both act as a multiple of the identity on the bosonic generators which are not themselves, we get  $\hat{T}_{\tilde{A}UUU}\propto\hat{T}_{\tilde{A}UUT}\propto \hat{T}_{\tilde{A}TTU}\propto\hat{T}_{\tilde{A}TTT}\propto\str{\lambda_{\tilde{A}}}=0$. 
		\end{itemize}
		To check the remaining two cases it is first useful to compute
		\begin{align}
			\hat{T}_{UUUU}&=T_{UUUU}=-\frac{M^2+NM+N^2}{4(N-M)NM}\nonumber\\
			\hat{T}_{TUUU}&=T_{TUUU}=\frac{i}{4\sqrt{NM}}\frac{N+M}{N-M}\nonumber\\
			\hat{T}_{TTUU}&=T_{TTUU}=\frac{1}{4(N-M)}\nonumber\\
			\hat{T}_{TTTU}&=T_{TTTU}=0\nonumber\\
			\hat{T}_{TTTT}&=T_{TTTT}=\frac{1}{4(N-M)}.
		\end{align}
		Then we get the two conditions
		\begin{align}
			\partial_{U}V[\expval{\Sigma^I}]=&-\mu^2\rho_1+\quarticSigmaA \rho_1(\rho_1^2+\rho_2^2)\nonumber\\
			&+\quarticSigmaB (T_{UUUU}\rho_1^3+3 T_{UUUT}\rho_1^2\rho_2+3T_{UTTU}\rho_2^2\rho_1+\hat{T}_{UTTT}\rho_2^3)\nonumber\\
			=&-\mu^2\rho_1+\quarticSigmaA \rho_1(\rho_1^2+\rho_2^2)\nonumber\\
			&+\quarticSigmaB \left(-\frac{M^2+NM+N^2}{4(N-M)NM}\rho_1^3+3 \frac{i}{4\sqrt{NM}}\frac{N+M}{N-M}\rho_1^2\rho_2+3\frac{1}{4(N-M)}\rho_2^2\rho_1\right)=0\label{eq:vanishingfirstdev1}\\
			\partial_{T}V[\expval{\Sigma^I}]=&-\mu^2\rho_2+\quarticSigmaA\rho_2(\rho_1^2+\rho_2^2)\nonumber\\
			&+\quarticSigmaB \left(T_{TUUU}\rho_1^3+3 T_{UUTT}\rho_1^2\rho_2+3T_{TTTU}\rho_2^2\rho_1+\hat{T}_{TTTT}\rho_2^3\right)\nonumber\\
			=&-\mu^2\rho_2+\quarticSigmaA \rho_2(\rho_1^2+\rho_2^2)\nonumber\\
			&+\quarticSigmaB \left(\frac{i}{4\sqrt{NM}}\frac{N+M}{N-M}\rho_1^3+3\frac{1}{4(N-M)} \rho_1^2\rho_2+\frac{1}{4(N-M)}\rho_2^3\right)=0
			\label{eq:vanishingfirstdev2}
		\end{align}
        There are four solutions to the constraint of Eqs.~\eqref{eq:vanishingfirstdev1} and \eqref{eq:vanishingfirstdev2}. For one of them, $\expval{\Sigma}$ is non-zero only on the upper $\sim SU(N)$ diagonal, for a second one only on the lower $\sim SU(M)$ diagonal. In the third one, it is proportional to the identity $\sim \lambda_T$, while for the fourth it is proportional to $\sigma_3$. The latter solution corresponds to 
		\begin{align}
			\rho_1=\frac{4 i \mu \sqrt{NM}}{\sqrt{N-M} \sqrt{\quarticSigmaB +4 \quarticSigmaA (N-M)}}\nonumber\\
			\rho_2=\frac{2 \mu (N+M)}{\sqrt{N-M} \sqrt{\quarticSigmaB +4\quarticSigmaA (N-M)}}\ ,
		\end{align}
		meaning
		\begin{align}
			\Sigma^i_j=\frac{\sqrt{2} \mu }{\sqrt{\quarticSigmaB +4 \quarticSigmaA (N-M)}}(\sigma_3)^i_j\equiv \rho (\sigma_3)^i_j\ .
		\end{align}
		Since a vacuum $\propto \sigma_3$ is the only one that guarantees a symmetry breaking pattern of the type $SU(N|M)\to H \supset SU(N)\times SU(M)$, we focus on it from now on.
        \subsubsection{Mass matrix}
		The Hessian matrix is 
		\begin{align}
			\partial_{\tilde{B}}\partial_{\tilde{A}}V=-\mu^2g_{\tilde{A}\tilde{B}}+\quarticSigmaA \left(g_{\tilde{A}\tilde{B}}\left(\Sigma^{\tilde{K}}g_{\tilde{K}\tilde{L}}\Sigma^{\tilde{L}}\right)+2g_{\tilde{B}\tilde{K}}\Sigma^{\tilde{K}}g_{\tilde{A}\tilde{J}}\Sigma^{\tilde{J}}\right)+3\quarticSigmaB  \hat{T}_{\tilde{A}\tilde{B}\tilde{K}\tilde{L}}\Sigma^{\tilde{K}}\Sigma^{\tilde{L}}\ .
		\end{align}
            Plugging the result for the vacuum we get
		\begin{align}
			\partial_{\tilde{B}}\partial_{\tilde{A}}V=&-\mu^2g_{\tilde{A}\tilde{B}}+\quarticSigmaA (g_{\tilde{A}\tilde{B}}2\rho^2(N-M)+8\rho^2\str{\lambda_{\tilde{A}}\sigma_3}\str{\lambda_{\tilde{B}}\sigma_3})\nonumber\\
			&+\quarticSigmaB \rho^2\left(g_{\tilde{A}\tilde{B}}+ 
            \str{\lambda_{\{\tilde{A}}\sigma_3\lambda_{\tilde{B}\}_{\degf}}\sigma_3}\right)\ .
		\end{align}
		Then
		\begin{align}
			\label{eq:masses}
   \partial_{\tilde{B}}\partial_{\tilde{A}}V=\mu^2\begin{cases}
				g_{\tilde{A}\tilde{B}}\frac{(-2\quarticSigmaB )}{4(M-N)\quarticSigmaA-\quarticSigmaB }\qquad\text{$\tilde{A}$ and $\tilde{B}$ bosonic, $\tilde{A},\tilde{B}\neq U, T$}\\
				0\qquad\text{$\tilde{A}$ and $\tilde{B}$ fermionic}\\
				\frac{2(\quarticSigmaB  (N-M)-16\quarticSigmaA N M)}{(N-M)(\quarticSigmaB +4\quarticSigmaA(N-M))}\qquad \tilde{A}=\tilde{B}=U\\
				\frac{2  \left(4\quarticSigmaA (M+N)^2+\quarticSigmaB  (N-M)\right)}{(N-M) (\quarticSigmaB +4\quarticSigmaA (N-M))}\qquad \tilde{A}=\tilde{B}=T\\
				\frac{-16 i \quarticSigmaA   \sqrt{N M} (M+N)}{(N-M) (\quarticSigmaB +4 \quarticSigmaA (N-M))}\qquad \tilde{A}=U\ ,\tilde{B}=T
			\end{cases}
		\end{align}
		
  Note that there are massless fermionic scalars; these are precisely the Goldstone modes expected from the spontaneous supergroup breaking pattern $SU(N|M) \rightarrow SU(N) \times SU(M) \times U(1).$ 
  As for the rest, the submatrix given by the restriction to the two indices $T$ and $U$ has eigenvalues $
			\left\{2,\ \frac{-2\quarticSigmaB }{4(M-N)\quarticSigmaA -\quarticSigmaB }\right\}$.
		Thus, after the rephasing that gives the correct sign to all kinetic terms, if we impose that 
		\begin{align}
			\frac{-2\quarticSigmaB }{4(M-N)\quarticSigmaA -\quarticSigmaB }>0\ ,
		\end{align}
		we get that the mass matrix around the vacuum for the states with the correct-sign kinetic term is positive.

    \subsection{One-loop potential}
    \label{eq:onelooppotential}
		While we have found a viable vacuum at tree level, it is natural to wonder whether this remains true at one-loop. A direct way to check if our conclusions are robust is by computing the one-loop effective potential {\it à la} Coleman-Weinberg \cite{Coleman:1973jx}, using the path integral. With respect to the standard procedure (see for example \cite{schwartz2014quantum}, Section 34.2) here we are dealing with a scalar field $\Sigma^{\tilde{I}}$ defined on a flat, non-positive definite $(N+M)^2-1$-dimensional manifold with metric $g_{\tilde{I}\tilde{J}}$. As such, we need to deal with some additional subtleties, which we treat at length in Appendix \ref{app:cwpotential}. 

 For simplicity, we will restrict ourselves to the Coleman-Weinberg potential arising from the interactions of scalar multiplets transfoming under a global $SU(N|M)$ symmetry. Given a Lagrangian for a scalar field $\Sigma^{\tilde{I}}$ of the form
		\begin{align}
			\mathcal{L}=-\frac{1}{2} g_{\tilde{I}\tilde{J}}\Sigma^{\tilde{I}} \Box\Sigma^{\tilde{J}}-V[\Sigma]
		\end{align}
		where $V[\Sigma]$ is a generic potential, the one-loop Coleman-Weinberg potential is
  \begin{align} \label{eq:cwpot}
			V_{\text{eff}}=V+\frac{1}{64\pi^2}\text{str}\left[\left(\tensor{\tilde{V}}{^{\tI}_{\tJ}}\right)^2\ln(\frac{\tensor{\tilde{V}}{^{\tI}_{\tJ}}}{\tilde{m}^2})\right]\ .
		\end{align}

Here $\tilde m$ is an arbitrary renormalization scale and 
  \begin{align}
			\tensor{\tilde{V}}{^\tL_\tJ}\equiv g^{{\tL\tK}}\partial_{\tJ}\partial_{\tK} V[\Sigma](-1)^{ \left(\degf(\tI)+\degf(\tJ) \right)\degf(\tJ)} \ .
		\end{align} 

For details of the derivation, see Appendix \ref{app:cwpotential}.

Among other things, the one-loop effective potential allows us to check the fate of tree-level flat directions. In particular, as we saw in Eq.~\ref{eq:masses}, at tree level there are massless fermionic scalars. The masslessness of these states beyond tree level follows from Goldstone's theorem, but it is gratifying to verify this explicitly at one-loop. To this end, let us proceed as follows: We want to exploit the invariance of the potential under $SU(N|M)$ to reduce perturbations along fermionic directions to perturbations along bosonic, diagonal directions. We expect that, if there is a mass, it can only appear as an off-diagonal, imaginary piece in the mass matrix in correspondence of fermionic directions such that $g_{IJ}=\pm i$. Thus, we perturb the vacuum in two fermionic directions $F_{1,2}$ such that $g_{F_1F_2}=i$
		\begin{align}
			\Sigma=\rho \sigma_3+\epsilon_1 \lambda_{F_1}+\epsilon_2\lambda_{F_2}\ ,
		\end{align}
		where $\epsilon_{1,2}$ are Grassmann numbers. 
		This matrix looks like
		\begin{align}
			\Sigma^i_j=\left(\begin{array}{c|c}
				\begin{matrix}
					\rho& & 0\\
					&\ddots & \\
					0  & & \rho
				\end{matrix}
				&0\\ \hline \\
				0 &\begin{matrix}
					-\rho& & 0\\
					&\ddots & \\
					0  & & -\rho
				\end{matrix}
			\end{array}\right)+
			\frac{1}{2}\left(\begin{array}{ccc|ccc}
				& & & 0 & \dots & 0\\
				&0& & \dots &  \epsilon_1+ i \epsilon_2 &  \dots\\
				& & & 0 & \dots & 0\\
				\hline
				0 & \dots & 0& & & \\
				\dots &  \epsilon_1-i \epsilon_2  &  \dots&&0 & \\
				0 & \dots & 0& & & \\
			\end{array}\right)\ .
		\end{align}
		Now we can exploit that $V_{\text{eff}}$ is $SU(N|M)$-invariant to evaluate it on the diagonalized $\Sigma^i_j$. 
		To compute the eigenvalues we need the generalization of the determinant to supermatrices, namely the Berezinian. For a supermatrix $X$ whose form is
		\begin{align}
			X=\left(\begin{array}{cc}
				A & B\\
				C& D
			\end{array}\right)\ ,
		\end{align}
		with $A,D$ bosonic and $C,B$ fermionic, the Berezinian reads
		\begin{align}
			\text{Ber}(X)=\det{A-BD^{-1} C}\det{D}^{-1}.
		\end{align}
		The eigenvalues of $\Sigma$ are just $\pm \rho$ except for those corresponding to the block
		\begin{align}
			(M_\Sigma)^i_j\equiv \left(\begin{array}{cc}
				\rho & \frac{\epsilon_1+i \epsilon_2}{2}\\
				\frac{\epsilon_1-i \epsilon_2}{2}& -\rho
			\end{array}\right)\ .
            \label{eq:2x2blockmatrix}
		\end{align}
		If $X$ is a $2\times2$ matrix, 
		\begin{align}
			X=\left(\begin{array}{cc}
				a & b\\
				c& d
			\end{array}\right)\ ,
		\end{align}
		with $a,d$ bosonic and $c,b$ fermionic, we can find its eigenvalues $\quarticSigma_{1,2}$ by requiring that the diagonalized matrix has the same Berezinian and supertrace:
		\begin{align}
			\begin{cases}
				\quarticSigmaA  (\quarticSigmaB )^{-1}=( a - b d^{-1} c )d^{-1}\\
				\quarticSigmaA -\quarticSigmaB =a-d\ .
			\end{cases}
		\end{align}
		Solving this system, and using that $(b c)^n=0$ for $n>1$, we get 
		\begin{align}
			\quarticSigmaA &=a+\frac{b c}{a-d}=\rho- i\frac{\epsilon_1\epsilon_2}{4\rho}  & \quarticSigmaB &=d+\frac{b c}{a-d}=-\rho- i\frac{\epsilon_1\epsilon_2}{4\rho}\ ,
		\end{align}
		where we already plugged the explicit values from Eq.~\eqref{eq:2x2blockmatrix}.
		This means that, after diagonalization
		\begin{align}
			M_\Sigma \to 
			\left(\begin{array}{cc}
				\rho- i\frac{\epsilon_1\epsilon_2}{4\rho} & 0\\
				0& -\rho- i\frac{\epsilon_1\epsilon_2}{4\rho}
			\end{array}\right)\ .
		\end{align}
		\begin{align}
			\Sigma=\rho \sigma_3-i\frac{\epsilon_1\epsilon_2}{4\rho} a^{I}\lambda^{(D)}_I
		\end{align}
		where $a^{I}\lambda^{(D)}_I$ is some linear combination of diagonal generators.
		Defining $c^{I}\equiv -i\frac{\epsilon_1\epsilon_2}{4\rho} a^{I}$, the mass term is 
		\begin{align}
			\frac{\partial^2}{\partial \epsilon_2\partial \epsilon_1}V[\rho \sigma_3+c^I\lambda^{(D)}_I]\eval_{\epsilon_{1,2}=0}=\frac{\partial^2 c_I}{\partial \epsilon_2\partial \epsilon_1}\frac{\partial V}{\partial c^I}\eval_{\epsilon_{1,2}=0}+\frac{\partial c^I}{\partial\epsilon_2}\frac{\partial c^J}{\partial\epsilon_1}\frac{\partial^2 V}{\partial c^I\partial c^J}\eval_{\epsilon_{1,2}=0}\ .
		\end{align}
		The second term vanishes since $\frac{\partial c^J}{\partial \epsilon_{1,2}}\propto \epsilon_{2,1}$, and we get
		\begin{align}
			\frac{\partial^2}{\partial \epsilon_2\partial \epsilon_1}V[\rho \sigma_3+c^I\lambda^{(D)}_I]\eval_{\epsilon_{1,2}=0}=-i\frac{a^I}{4\rho}\frac{\partial V}{\partial c^I}\eval_{\epsilon_{1,2}=0}\ .
		\end{align}
		However, $\frac{\partial V}{\partial c^I}\eval_{\epsilon_{1,2}=0}$ is just the gradient of the potential (only along the directions corresponding to diagonal generators) evaluated on the minimum, and must thus vanish. The fermionic scalars corresponding to broken supergroup generators remain massless at one-loop, consistent with our expectations from Goldstone's theorem.

		\subsection{Mass spectrum in the broken phase}
        \label{sec:1loopafterSSB}
		As we have seen, it is possible to spontaneously break the $SU(N|M)$ symmetry down to the bosonic subgroup $SU(N)\times SU(M)\times U(1)$. When $SU(N|M)$ is a global symmetry, there are massless fermionic scalars corresponding to the broken fermionic generators. When $SU(N|M)$ is a local symmetry, we expect these fermionic scalars to be eaten to become the longitudinal modes of the massive fermionic vectors.
  
  To obtain the mass spectrum after spontaneous symmetry breaking, we expand $\Sigma\to \rho \sigma_3+\Sigma$. The kinetic term for $\Sigma$ becomes
		\begin{align}
			\mathcal{L}_{k,\Sigma}=\str{ \gradedcomm{\nabla_\mu}{\Sigma}^2}=&\str{\partial_\mu\Sigma \partial^\mu\Sigma}-g^2\str{\gradedcomm{\mathcal{A}_\mu}{\Sigma}^2}-g^2\rho^2\str{\gradedcomm{\mathcal{A}_\mu}{\sigma_3}^2}\nonumber\\
			&+ig\left[\str{\partial^\mu\Sigma \gradedcomm{\mathcal{A}_\mu}{\Sigma}}+\str{\gradedcomm{\mathcal{A}_\mu}{\Sigma}\partial^\mu\Sigma}\right]\nonumber\\
			&+ig\rho\left[\str{\partial^\mu\Sigma \gradedcomm{\mathcal{A}_\mu}{\sigma_3}}+\str{\gradedcomm{\mathcal{A}_\mu}{\sigma_3}\partial^\mu\Sigma}\right]\nonumber\\
			&-g^2\rho \left[\str{\gradedcomm{\mathcal{A}_\mu}{\Sigma}\gradedcomm{\mathcal{A}^\mu}{\sigma_3}}+\str{\gradedcomm{\mathcal{A}_\mu}{\sigma_3}\gradedcomm{\mathcal{A}^\mu}{\Sigma}}\right]\ .
		\end{align}
		As usual (see e.g. \cite{Weinberg_1996}, Chapter 21), we remove the mixing between $\Sigma$ and $\mathcal{A}_\mu$ by modifying the gauge-fixing function. In particular, we pick
        \begin{align}
			\mathcal{L}_{GF}=&-\frac{1}{\alpha}\str{\left(\partial_\mu \mathcal{A}^\mu-ig \alpha \rho \gradedcomm{\sigma_3}{\Sigma}\right)^2}\nonumber\\
			=&-\frac{1}{\alpha}\str{[\partial_\mu \mathcal{A}^\mu]^2}-ig\rho \str{\left[\mathcal{A}^\mu\gradedcomm{\sigma_3}{\partial_\mu\Sigma}+\gradedcomm{\sigma_3}{\partial_\mu \Sigma}\mathcal{A}^\mu\right]}\nonumber\\
			&+g^2\alpha \rho^2\str{\gradedcomm{\sigma_3}{\Sigma}^2}\ .
		\end{align}
		Summing them, we get
		\begin{align}
			\mathcal{L}_{k,\Sigma}+\mathcal{L}_{GF}=&\str{\partial_\mu\Sigma \partial^\mu\Sigma}-g^2\str{\gradedcomm{\mathcal{A}_\mu}{\Sigma}^2}-g^2\rho^2\str{\gradedcomm{\mathcal{A}_\mu}{\sigma_3}^2}\nonumber\\
			&+ig\left[\str{\partial^\mu\Sigma \gradedcomm{\mathcal{A}_\mu}{\Sigma}}+\str{\gradedcomm{\mathcal{A}_\mu}{\Sigma}\partial^\mu\Sigma}\right]\nonumber\\
			&-g^2\rho\left[\str{\gradedcomm{\mathcal{A}_\mu}{\Sigma}\gradedcomm{\mathcal{A}^\mu}{\sigma_3}}+\str{\gradedcomm{\mathcal{A}_\mu}{\sigma_3}\gradedcomm{\mathcal{A}^\mu}{\Sigma}}\right]\nonumber\\
			&-\frac{1}{\alpha}\str{[\partial_\mu \mathcal{A}^\mu]^2}+g^2\alpha \rho^2\str{\gradedcomm{\sigma_3}{\Sigma}^2}\ ,
		\end{align} \label{eq:kinwithGFafterSSB}
		where now the kinetic mixing between $\Sigma$ and $\mathcal{A}_\mu$ has disappeared and we have a mass term for the fermionic components of $\mathcal{A}_\mu$ as well as for those of $\Sigma$.

  Finally, we'd like to work out the one-loop corrections to the mass of a scalar in the fundamental of a spontaneously broken local $SU(N|M)$ symmetry. To compute the mass correction coming from the gauge coupling of $\Phi$ to $\mathcal{A}_\mu$, we only need to specify how the Lagrangian above modifies the propagator of $\mathcal{A}_\mu$. As a consequence, we only keep the quadratic terms of Eq.~\eqref{eq:kinwithGFafterSSB} and sum them to the quadratic terms from the $\mathcal{A}_\mu$ kinetic term, Eq.~\eqref{eq:gaugeselfint}, to get the full $\order{\mathcal{A}^2}$ Lagrangian
		\begin{align}
			\mathcal{L}_{G}^0=&\left(-\frac{1}{2}\partial_\mu\mathcal{A}^I_\nu\partial^\mu\mathcal{A}^{J\nu}+\frac{1}{2}\partial_\mu\mathcal{A}^I_\nu\partial^\nu\mathcal{A}^{J\mu}\right)g_{IJ}-\frac{1}{2\alpha}\partial_\mu\mathcal{A}^{I\mu}\partial_\nu\mathcal{A}^{J\nu}g_{IJ}+\nonumber\\
			&-g^2\rho^2\mathcal{A}^{I\mu}\mathcal{A}^{J}_\mu\str{\comm{\lambda_I}{\sigma_3}\comm{\lambda_J}{\sigma_3}}\ .
		\end{align}
		The last term can be rearranged to give
		\begin{align}
			-g^2\rho^2\mathcal{A}^{I\mu}\mathcal{A}^{J}_\mu\str{\comm{\lambda_I}{\sigma_3}\comm{\lambda_J}{\sigma_3}}=2g^2\rho^2\mathcal{A}^{I\mu}\mathcal{A}^J_\mu g_{IJ}^{(F)}\ ,
		\end{align}\
		where $g_{IJ}^{(F)}$ and $g_{IJ}^{(B)}$ have been defined in Sec~\ref{sec:softmassesgaugebosons}. The Lagrangian can be split between fermionic and the bosonic components of $\mathcal{A}_\mu$:
		\begin{align}
			\mathcal{L}_{G, B}^0&=-\frac{1}{2}\partial_\mu\mathcal{A}^I_\nu\partial^\mu\mathcal{A}^{J\nu}g^{(B)}_{IJ}+\frac{1}{2}\left(1-\frac{1}{\alpha}\right)\partial_\mu\mathcal{A}^I_\nu\partial^\nu\mathcal{A}^{J\mu}g^{(B)}_{IJ}
            \label{eq:quadraticlagrangianamuboson} 
            \\
			\mathcal{L}_{G, F}^0&=-\frac{1}{2}\partial_\mu\mathcal{A}^I_\nu\partial^\mu\mathcal{A}^{J\nu}g^{(F)}_{IJ}+\frac{1}{2}\left(1-\frac{1}{\alpha}\right)\partial_\mu\mathcal{A}^I_\nu\partial^\nu\mathcal{A}^{J\mu}g^{(F)}_{IJ}+\frac{m^2_{\mathcal{A}}}{2}\mathcal{A}^{I\mu}\mathcal{A}^{J}_\mu g_{IJ}^{(F)}\ ,
            \label{eq:quadraticlagrangianamufermion}
		\end{align}
		where we defined $m^2_{\mathcal{A}}\equiv 4\rho^2 g^2$. This is just the Lagrangian we studied in Section~\ref{sec:softmassesgaugebosons}, so we get that the mass of the scalar field $m^2_{\Phi}$ gets a quadratic correction as in Eq.~\eqref{eq:masscorrectionfrombosons}.

       \section{Conclusions} \label{sec:conc}

In this paper we have explored diverse aspects of theories with global or local $SU(N|M)$ symmetries, with a particular interest in theories with $M \neq N$ that admit the fundamental representation. Surprisingly, despite the mismatch between the number of even- and odd-graded generators, the one-loop corrections to the mass of a scalar multiplet transforming in the fundamental of $SU(N|M)$ from its own quartic coupling, gauge couplings to $SU(N|M)$ vectors, and yukawa couplings to select representations of $SU(N|M)$ spinor multiplets vanish when $M = N+1$. Soft breaking of the $SU(N|M)$ symmetry induces at most logarithmic dependence on the cutoff, although soft masses for fields in an $SU(N|M)$ vector multiplet require UV completion via spontaneous symmetry breaking. Indeed, $SU(N|M)$ may be broken to its bosonic $SU(N) \times SU(M) \times U(1)$ subgroup via a scalar multiplet transforming as the direct product of the fundamental and antifundamental representation. The vacuum is free of both tachyons and tachyonic ghosts provided certain constraints between tree-level parameters in the scalar potential hold, and remains stable at one-loop. For a spontaneously broken global $SU(N|M)$ symmetry, Goldstone's theorem is satisfied by massless fermionic scalars. When $SU(N|M)$ is gauged, these scalars are eaten to become the longitudinal modes of massive fermionic vectors, providing a satisfactory UV completion of soft masses in the vector multiplet.

There are a variety of open questions. While the vanishing one-loop corrections to a fundamental scalar multiplet's mass are remarkable, it is less clear what happens beyond one-loop. The vanishing supertraces that ensure the all-loop finiteness of pure $SU(N|N)$ gauge theories \cite{Arnone:2000bv, Arnone:2000qd, Arnone:2001iy} do not necessarily extend to loops involving quartic and Yukawa couplings, or to fields transforming in representations other than the adjoint. Needless to say, it would be interesting to understand what couplings and representations enjoy finiteness beyond one-loop. Spontaneous symmetry breaking also warrants further study. Here we have focused exclusively on the spontaneous breaking of $SU(N|M)$ to its bosonic $SU(N) \times SU(M) \times U(1)$ subgroup by scalars transforming in the direct product of the fundamental and anti-fundamental representation. It would be interesting to study other patterns of symmetry breaking involving the bosonic subgroup as well. 

More broadly, it remains to be seen whether supergroup internal symmetries are in any way relevant to the real world. The remarkable radiative properties of these theories would make them compelling candidates for physics beyond the Standard Model were it not for the obvious challenges to unitarity posed by the proliferation of wrong-sign and wrong-statistics fields. Even so, there is a sense in which supergroup internal symmetries can provide a satisfying symmetry-based organizing principle for Lee-Wick models. It may be the case that the arguments for perturbative unitarity in Lee-Wick models (or other theories with apparent negative-norm states) can be extended to supergroup internal symmetries. 

Given the appeal of finiteness (whether at one-loop or all loops), the possibility of a unitary interpretation certainly warrants further exploration. Should such a unitary interpretation exist, then the phenomenological aspects of supergroup internal symmetries would become quite compelling. A phenomenological model for electroweak symmetry breaking involving supergroup symmetries would not merely be a convoluted variation on the familiar story of spacetime supersymmetry. Supergroup internal symmetries allow much greater flexibility, as they may involve only a subset of the fields in the theory. But let us not get too far ahead of ourselves. While this work highlights a number of fun properties of supergroup theories, further phenomenological applications require a plausible unitary interpretation that is presently lacking.

\section*{Acknowledgements}

We would like to thank Hsin-Chia Cheng, Savas Dimopoulos, Florian Nortier, Roni Harnik, Jay Hubisz, Graham Kribs, Markus Luty, Surjeet Rajendran, Lisa Randall, John Terning, and Giovanni Villadoro for useful conversations and healthy skepticism. This work was supported in part by the U.S. Department of Energy under the grant DE-SC0011702 and performed in part at the Kavli Institute for Theoretical Physics, supported by the National Science Foundation under Grant No.~NSF PHY-1748958. JNH was supported by the National Science Foundation under Grant No. NSF PHY-1748958 and by the Gordon and Betty Moore Foundation through Grant No. GBMF7392. EG is supported by the Collaborative Research Center SFB1258 and the Excellence Cluster ORIGINS, which is funded by the Deutsche Forschungsgemeinschaft (DFG, German Research Foundation) under Germany’s Excellence Strategy – EXC-2094-390783311.

		\appendix
		\section{Potential for $\str{\Sigma}=0$ } \label{app:wrongpotential}
    		In Section~\ref{sec:breakingSUNM} we considered a scalar sector whose vacuum structure allowed for a symmetry breaking pattern of the form $SU(N|M)\to SU(N)\times SU(M) \times U(1)$. To this end, we introduced a scalar field transforming as the reducible representation built by taking the tensor product of a fundamental and anti-fundamental irrep. This representation can of course be decomposed into a supertraceless component, i.e. the adjoint irrep, and a singlet, represented by the supertrace itself. However, we mentioned how the structure of the potential for a field transforming into an adjoint representation only did not allow for a vacuum that produces the desired SSB pattern. Here we wish to justify that statement. For a scalar field $\Sigma^I$ transforming in the adjoint representation the most general renormalizable potential looks like
		\begin{align}
			V[\Sigma]&=-\mu^2\str{\Sigma^2}+\quarticSigmaA \str{\Sigma^2}^2+\frac{1}{4}\quarticSigmaB \str{\Sigma^4}\nonumber\\
			&=-\frac{1}{2}\mu^2\Sigma^Ig_{IJ}\Sigma^J+\frac{1}{4}\quarticSigmaA \left(\Sigma^Ig_{IJ}\Sigma^J\right)^2+\frac{1}{4}\quarticSigmaB \Sigma^{I}\Sigma^{J}\Sigma^{K}\Sigma^{L}T_{IJKL}\ ,
			\label{eq:potential}
		\end{align}
		with $T_{IJKL}\equiv \str{\lambda_I\lambda_J\lambda_K\lambda_L}$. Moreover $\Sigma^{I}\Sigma^{J}\Sigma^{K}\Sigma^{L}=\Sigma^{\{I}\Sigma^{J}\Sigma^{K}\Sigma^{L\}_\degf}$, where the $\degf$-symmetrization notation has been introduced in Section~\ref{sec:rewievsunm}. Since $T_{IJKL}$ is contracted with $\Sigma^{I}\Sigma^{J}\Sigma^{K}\Sigma^{L}$ in Eq.~\eqref{eq:potential}, its only surviving component is
		\begin{align}
			T_{IJKL}\rightarrow \hat{T}_{IJKL}\equiv T_{\{IJKL\}_\degf}\ ,
		\end{align}
		so that we can rewrite
		\begin{align}
			V[\Sigma]&=-\frac{1}{2}\mu^2\Sigma^Ig_{IJ}\Sigma^J+\frac{1}{4}\quarticSigmaA \left(\Sigma^Ig_{IJ}\Sigma^J\right)^2+\frac{1}{4}\quarticSigmaB \Sigma^{I}\Sigma^{J}\Sigma^{K}\Sigma^{L}\hat{T}_{IJKL}\ .
			\label{eq:potential2}
		\end{align}
		Notice that by construction, $g_{IJ}=g_{\{IJ\}_\degf}$, so no symmetrization is needed for it. 
		
		While a full study of the potential is possible, along the lines of e.g. \cite{Li:1973mq}, here we are only interested in checking whether we can find a minimum that breaks $SU(N|M)$ down to its bosonic subgroup. 
		As such, we make the ansatz 
		\begin{align}
			\expval{\Sigma}=\mathcal{N}\times  \text{diag }(\underbrace{N+1, \dots, N+1}_\textrm{$N$ times},\underbrace{N,\dots, N}_\textrm{$N+1$ times})\equiv v\times \lambda_{U}\ ,
			\label{eq:vev}
		\end{align}
        as $\lambda_U$ is the only generator that would grant the desired pattern, and we content ourselves with the possibility that, if a minimum of this kind is indeed found, it could just be a local one. Here, $\lambda_U$ is the generator relative to the bosonic $U(1)$ subgroup, see Eq.~\eqref{eq:sugenerators}. To determine the value of $v$, we look at the gradient of $V$
		\begin{align}
			\partial_AV\equiv	\pdv{V}{\Sigma^A}&=-\frac{1}{2}\mu^2\left(g_{AJ}\Sigma^J+(-1)^{\degf(A)\degf(I)}\Sigma^I g_{IA}\right)\nonumber \\
            &+\frac{1}{2}\quarticSigmaA \left(g_{AJ}\Sigma^J+(-1)^{\degf(A)\degf(I)}\Sigma^I g_{IA}\right)\left(\Sigma^Kg_{KL}\Sigma^L\right)\nonumber\\
			&+\frac{1}{4}\quarticSigmaB \left[T_{AIJK}+(-1)^{\degf(A)\degf(I)}T_{IAJK}+(-1)^{\degf(A)(\degf(I)+\degf(J))}T_{IJAK}\right.\nonumber\\
			&\left.+(-1)^{\degf(A)(\degf(I)+\degf(J)+\degf(K))}T_{IJKA}\right]\Sigma^I\Sigma^J\Sigma^K\nonumber\\
			&=-\mu^2\left(g_{AJ}\Sigma^J\right)+\quarticSigmaA g_{AJ}\Sigma^J\left(\Sigma^Kg_{KL}\Sigma^L\right)+\quarticSigmaB \hat{T}_{AIJK}\Sigma^I\Sigma^J\Sigma^K,
			\label{eq:gradient1}
		\end{align}
		where the factors of $(-1)$ are the consequence of the bosonic/fermionic nature of the $\Sigma$ field, i.e. of 
		\begin{align}
			\partial_A\left(\Sigma^B\dots\right) =\delta_A^B\dots +(-1)^{\degf(A)\degf(B)}\Sigma^B\left(\partial_A\dots\right)\ ,
		\end{align} 
		and we showed explicitly how everything can be rewritten in terms of the $\degf$-symmetrized quantity $\hat{T}_{IJKL}$.
		Calling $U$ the index corresponding to the $U(1)$ generator, we can write the ansatz from Eq.~\eqref{eq:vev} as
		\begin{align}
			\expval{\Sigma^I}=v\times \delta^I_U\ ,
		\end{align}
		so that\footnote{Recall $\degf(U)=0$}
		\begin{align}
			\partial_AV|_{min}=-\mu^2g_{AU}v+\quarticSigmaA g_{AU}v^3+\quarticSigmaB \hat{T}_{AUUU}v^3\ .
		\end{align}
		Now we can evaluate this expression for $A=U$
		\begin{align}
			\partial_UV|_{min}=-\mu^2 v+\quarticSigmaA v^3+\quarticSigmaB \hat{T}_{UUUU}v^3 ,
		\end{align}
		where we used that, with our normalization, $g_{UU}=1$.
		Now
		\begin{align}
			\hat{T}_{UUUU}=T_{UUUU}&=\str{\lambda_U^4}=\frac{N^2M^2}{4(M-N)^2}\left[\frac{1}{N^4}N-\frac{1}{M^4}M\right]\nonumber\\
			&=\frac{M^2+MN+N^2}{4MN(M-N)}>0  &\forall N,\, M\in \mathbb{N}^+\ ,
		\end{align}
		so that 
		\begin{align}
			\partial_UV=v&\left\{-\mu^2+\left[\quarticSigmaA +\quarticSigmaB \frac{M^2+MN+N^2}{4MN(M-N)}\right]v^2\right\}=0 \nonumber\\
             &\Longrightarrow \begin{cases}
				v=0\\
				v^2=\mu^2\left(\quarticSigmaA  +\quarticSigmaB \frac{M^2+MN+N^2}{4MN(M-N)}\right)^{-1}\end{cases}\ .
		\end{align}
		For $A\neq U$, instead, the first two terms vanish since $g_{AU}=0$, and there only remains
		\begin{align}
			\partial_AV=v^3\quarticSigmaB \hat{T}_{AUUU}\ .
		\end{align}
		All terms in $\hat{T}_{AUUU}$ are proportional to $T_{AUUU}=\str{\lambda_A\lambda_U^3}$.
		However, $\lambda_U^3$ can be written as a linear combination of $\lambda_U$ and the identity $\mathbb{I}$:
		\begin{align}
			\lambda_U^3=a\lambda_U+b\mathbb{I}\ .
		\end{align}
		But then
		\begin{align}
			T_{AUUU}=a\str{\lambda_A\lambda_U}+b\str{\lambda_A}=0\ ,
		\end{align}
		where we used that $\str{\lambda_A\lambda_U}=\frac{1}{2}g_{AU}=0$ for $A\neq U$ and that $\str{\lambda_A}=0$ since the generators are supertraceless.
		So this is a stationary point of the potential.
  
		\subsection{Around the vacuum}
		To check the nature of the above stationary point, we compute the eigenvalues of the Hessian matrix around it. To this end, we need to evaluate the second derivative of $V[\Sigma]$ with respect to $\Sigma$, $\partial_A\partial_BV[\Sigma]$, and evaluate it on the vacuum $\expval{\Sigma}\propto \delta^I_U$. It is then easy to see that there is going to be a piece $\propto \hat{T}_{ABUU}$. This will turn out to be the piece requiring the most work, so we start by computing it.
		The fact that the generators are in block form (see their definition, Eq.~\eqref{eq:sugenerators}) allows for some simplifications. In particular, we will need the following results\footnote{Note we here use regular, non-graded anticommutators.}:
		\begin{align}
			\acomm{\lambda_U}{T^a_N}&=\frac{1}{2}\sqrt{\frac{2MN}{M-N}}\left[\begin{pmatrix}
				1/N &0\\
				0 &1/M
			\end{pmatrix}\begin{pmatrix}
				t^a_n &0\\
				0 &0
			\end{pmatrix}+\begin{pmatrix}
				t^a_n &0\\
				0 &0
			\end{pmatrix}\begin{pmatrix}
				1/N &0\\
				0 &1/M
			\end{pmatrix}\right]\nonumber\\
            &=\frac{1}{N}\sqrt{\frac{2MN}{M-N}}T^a_N\nonumber\\
			\acomm{\lambda_U}{T^b_M}&=\frac{1}{M}\sqrt{\frac{2MN}{M-N}}T^b_M\nonumber\\
			\acomm{\lambda_U}{S_i}&=\frac{1}{4}\sqrt{\frac{2MN}{M-N}}\left[\begin{pmatrix}
				1/N &0\\
				0 &1/M
			\end{pmatrix}\begin{pmatrix}
				0 &s_i\\
				s_i^\dagger &0
			\end{pmatrix}+\begin{pmatrix}
				0 &s_i\\
				s_i^\dagger &0
			\end{pmatrix}\begin{pmatrix}
				1/N &0\\
				0 &1/M
			\end{pmatrix}\right]\nonumber\\
			&=\left(\frac{1}{M}+\frac{1}{N}\right)\frac{1}{2}\sqrt{\frac{2MN}{M-N}}S_i\nonumber\\
			\acomm{\lambda_U}{\tilde{S}_i}&=\left(\frac{1}{M}+\frac{1}{N}\right)\frac{1}{2}\sqrt{\frac{2MN}{M-N}}\tilde{S}_i\ .
			\label{eq:anticommutators}
		\end{align}
		The last anticommutator we need is
		\begin{align}
			\acomm{\lambda_U}{\lambda_U}=\frac{1}{2}\frac{2MN}{M-N}\begin{pmatrix}
				1/N^2 &0\\
				0 &1/M^2
			\end{pmatrix}\ .
			\label{eq:lambdauanticomm}
		\end{align}
		This matrix can be expanded as a linear combination of the $2\times2$ identity and $\lambda_U$ itself. Specifically,
		\begin{align}
			\acomm{\lambda_U}{\lambda_U}=\left(\frac{1}{M}+\frac{1}{N}\right)\sqrt{\frac{2MN}{M-N}}\lambda_U-\frac{1}{M-N}\mathbf{1}\ .
		\end{align}
		From Eq.~\eqref{eq:anticommutators} we see that the anticommutator of any generator but $\lambda_U$ with $\lambda_U$ is proportional to the generator itself, times some fixed numerical factor depending only on the class the generators belongs to. To ease the notation, we can then define
		\begin{align}
			\kappa(A)\equiv \begin{cases}
				\frac{1}{N}\sqrt{\frac{2MN}{M-N}}, &A=\text{bosonic, correct sign}\\
				\frac{1}{M}\sqrt{\frac{2MN}{M-N}}, &A=\text{bosonic, wrong sign}\\
				\frac{1}{2}\left(\frac{1}{N}+\frac{1}{M}\right)\sqrt{\frac{2MN}{M-N}}, &A=\text{fermionic}\\
			\end{cases}\ ,
		\end{align}
		so that we can write in general
		\begin{align}
			\acomm{\lambda_U}{\lambda_A}&=\kappa(A)\lambda_A&A&\neq U\ .
		\end{align}
		Our next step is then to compute $\hat{T}_{ABUU}$. Explicitly,
		\begin{align}
			\hat{T}_{ABUU}=&\frac{1}{12}\Big[T_{ABUU}+T_{AUBU}+T_{AUUB}+T_{UABU}+T_{UAUB}+T_{UUAB}\nonumber\\
			&+(-1)^{\degf(A)\degf(B)}\left(T_{BAUU}+T_{BUAU}+T_{BUUA}+T_{UBAU}+T_{UBUA}+T_{UUBA}\right)\Big]\ .
		\end{align}
		Now we can exploit the anticommutation relations we found above to rework some of these pieces. Specifically,
		\begin{align}
			T_{AUBU}+T_{ABUU}&=\kappa(B)T_{ABU}\\
			T_{AUUB}+T_{UAUB}&=\kappa(A)T_{AUB}\\
			T_{UABU}&=-T_{AUBU}+\kappa(A)T_{ABU}=T_{ABUU}-\kappa(B)T_{ABU}+\kappa(A)T_{ABU}\\
			T_{UUAB}&=-T_{UAUB}+\kappa(A)T_{UAB}\nonumber\\
			&=T_{ABUU}-\kappa(B)T_{ABU}+\kappa(B)T_{AUB}-\kappa(A)T_{AUB}+\kappa(A)T_{UAB}\ ,
		\end{align}
		meaning
		\begin{align}
			12\hat{T}_{ABUU}=&2T_{ABUU}+T_{ABU}(\kappa(A)-\kappa(B))+T_{AUB}\kappa(B)+T_{UAB}\kappa(A)\nonumber\\
            &+(-1)^{\degf(A)\degf(B)}(A\leftrightarrow B)\nonumber\\
			=&2T_{ABUU}+T_{ABU}(\kappa(A)-\kappa(B))+T_{AUB}\kappa(B)-T_{AUB}\kappa(A)+T_{AB}\kappa(A)^2\nonumber\\
			&+(-1)^{\degf(A)\degf(B)}(A\leftrightarrow B)\nonumber\\
			=&2T_{ABUU}+T_{ABU}(\kappa(A)-\kappa(B))-T_{ABU}(\kappa(B)-\kappa(A))\nonumber\\
			&+T_{AB}(\kappa(B)-\kappa(A))\kappa(B)+T_{AB}\kappa(A)^2+(-1)^{\degf(A)\degf(B)}(A\leftrightarrow B)\nonumber\\
			=&2T_{ABUU}+2T_{ABU}(\kappa(A)-\kappa(B))+T_{AB}(\kappa(A)^2-\kappa(A)\kappa(B)+\kappa(B)^2)\nonumber\\
            &+(-1)^{\degf(A)\degf(B)}(A\leftrightarrow B)\ ,
		\end{align}
		where we repeatedly used that the anticommutation relations found above let us write e.g. 
		\begin{align}
			T_{AUBU}=-T_{ABUU}+\kappa(B)T_{ABU} \ ,
		\end{align}
		and so on.
		
		Moreover, using Eq.~\eqref{eq:lambdauanticomm} and $\lambda_U\lambda_U=\frac{1}{2}\acomm{\lambda_U}{\lambda_U}$ we can write
		\begin{align}
			T_{ABUU}=\frac{1}{2}\left(\frac{1}{M}+\frac{1}{N}\right)\sqrt{\frac{2MN}{M-N}}T_{ABU}-\frac{1}{2}\frac{1}{M-N}T_{AB}\ ,
		\end{align}
		implying
		\begin{align}
    	12\hat{T}_{ABUU}=&T_{ABU}\left(2\kappa(A)-2\kappa(B)+\left(\frac{1}{M}+\frac{1}{N}\right)\sqrt{\frac{2MN}{M-N}}\right)\nonumber\\
			&+T_{AB}\left(\kappa(A)^2-\kappa(A)\kappa(B)+\kappa(B)^2-\frac{1}{M-N}\right)+(-1)^{\degf(A)\degf(B)}(A\leftrightarrow B) \ ,
			\label{eq:hattabuu}
		\end{align}
		so we only need to study $T_{ABU}$ and $T_{AB}$ (and their counterparts with $A$ and $B$ exchanged).
		First of all, both the matrix $\lambda_A\lambda_B\lambda_U$ and $\lambda_A\lambda_B$ can have nonzero supertrace only if they have diagonal components, i.e. if they are "bosonic". This means that either $A$ and $B$ are both fermionic, or they are both bosonic, i.e.
		\begin{align}
			T_{ABU}&=T_{AB}=0&&\text{if $A$ bosonic and $B$ fermionic or viceversa,}
		\end{align}
		and we get the first result
		\begin{align}
			\hat{T}_{ABUU}=0&&\text{if $A$ bosonic and $B$ fermionic or viceversa.}
		\end{align}
		We next study the different cases where $A$ and $B$ are either both fermionic or both bosonic separately.
		\paragraph{$A$ and $B$ both fermionic}
		If both $A$ and $B$ correspond to fermionic generators of the kind $S_i$, or both $\tilde{S}_i$, then, by direct inspection, $T_{ABU}$ and $T_{AB}$ are nonzero only if $A=B$. This is simply a consequence of $S_iS_j$ and $\tilde{S}_i\tilde{S}_j$ being off-diagonal for $i\neq j$. Then, on this subset, $T_{ABU}=a \delta_{AB}$ and $T_{AB}=b \delta_{AB}$ for some constants $a$ and $b$. However, we need at the same time both tensors to be antisymmetric in $A\leftrightarrow B$, since both indices are fermionic. Thus $a=b=0$, and we get the additional result
		\begin{align}
			\hat{T}_{ABUU}=0&&\text{if $A$ and $B$ are either both of type $S_i$ or type $\tilde{S}_i$.}
		\end{align}
		We meet the first nontrivial case when $A$ is of type $\tilde{S}_i$ and $B$ of type $S_i$ (the opposite case being the same up to a minus sign). Then, both $T_{ABU}$ and $T_{AB}$ are non-zero only for $A$ and $B$ such that $g_{AB}=2T_{AB}=i$.
		For these values, we get 
		\begin{align}
			\tilde{S}_{i}S_j=\left(\begin{array}{ccc|ccc}
				\ddots &  & & & & \\
				&  \frac{i}{4}  & & &0 & \\
				&  & \ddots & & & \\
				\hline
				& & & \ddots &  &\\
				&0& & &  -\frac{i}{4}  &\\
				& & & &  & \ddots\\
			\end{array}\right)\ ,
		\end{align}
		so that
		\begin{align}
			T_{ABU}&=\str{\tilde{S}_iS_j\lambda_U}=\frac{1}{2}\sqrt{\frac{2NM}{M-N}}\frac{i}{4}\left(\frac{1}{N}+\frac{1}{M}\right)\\
			T_{AB}&=\frac{1}{2}g_{AB}=\frac{i}{2}\ .
		\end{align}
		Plugging into Eq.~\eqref{eq:hattabuu} and using the value of $\kappa(A)$ and $\kappa(B)$ for $A$ and $B$ both fermionic we get
		\begin{align}
			\hat{T}_{ABUU}&=\frac{i\left(M^2+MN+N^2\right)}{12 M N (M-N)} &&\text{if $A$ is type $\tilde{S}_i$ and $B$ type $S_i$ s.t. $g_{AB}\neq 0$.}
		\end{align}
		\paragraph{$A$ and $B$ both bosonic}
		If $A$ is bosonic with correct sign and $B$ bosonic with wrong sign, then  the matrix $\lambda_A\lambda_B$ vanishes identically. Thus we can have either $A$ and $B$ both bosonic with correct or both with wrong sign. Then, in each subspace, $\lambda_U$ acts as a multiple of the identity, and we get
		\begin{align}
			T_{ABU}=\begin{cases}
				\frac{1}{2}\sqrt{\frac{2NM}{M-N}}\frac{1}{N}T_{AB} \quad \text{$A$ and $B$ correct sign}\\
				\frac{1}{2}\sqrt{\frac{2NM}{M-N}}\frac{1}{M}T_{AB} \quad \text{$A$ and $B$ wrong sign}\\
			\end{cases}.
		\end{align}
		Then, using again $T_{AB}=\frac{1}{2}g_{AB}$ and the values of $\kappa(A)$ for bosonic indices, we get
		\begin{align}
			\hat{T}_{ABUU}&=\frac{M}{4N(M-N)}\delta_{AB}&& \text{$A$ and $B$ correct sign}\\
			\hat{T}_{ABUU}&=-\frac{N}{4M(M-N)}\delta_{AB}&& \text{$A$ and $B$ wrong sign.}
		\end{align}
		Finally, we assumed until now that $A,B\neq U$. The remaining possibility is $\hat{T}_{UUUU}$, that we already found in Section~\ref{subsubsec:minimizingPotential}. In summary, we have
		\begin{align}
			\hat{T}_{ABUU}=
			\begin{cases}
				\frac{i\left(M^2+MN+N^2\right)}{12 M N (M-N)}, &\text{$A$ and $B$ fermionic with $g_{AB}=i$}\\
				\frac{-i\left(M^2+MN+N^2\right)}{12 M N (M-N)}, &\text{$A$ and $B$ fermionic with $g_{AB}=-i$}\\
				\frac{M}{4N(M-N)}\delta_{AB}, &\text{$A$ and $B$ bosonic with $g_{AB}=1$, $A,\neq U$}\\
				-\frac{N}{4M(M-N)}\delta_{AB}, & \text{$A$ and $B$ bosonic with $g_{AB}=-1$, $A,\neq U$}\\
				\frac{M^2+MN+N^2}{4 MN (M-N)}, & A=B=U\\
				0 , & \text{otherwise}
			\end{cases}.
			\label{eq:hattabuufinal}
		\end{align}
		
		Now we will see how our hard work pays off. Let us recall the expression for the first derivative of the potential
		\begin{align}
			\partial_AV=-\mu^2(g_{AJ}\Sigma^J)+\quarticSigmaA g_{AJ}\Sigma^J(\Sigma^Kg_{KL}\Sigma^L)+\quarticSigmaB \hat{T}_{AIJK}\Sigma^I\Sigma^J\Sigma^K\ .
		\end{align}
		Then
		\begin{align}
			\partial_{B}\partial_AV=-\mu^2g_{AB}+\quarticSigmaA (g_{AB}(\Sigma^Kg_{KL}\Sigma^L)+2g_{BK}\Sigma^Kg_{AJ}\Sigma^J)+3\quarticSigmaB  \hat{T}_{ABKL}\Sigma^K\Sigma^L\ .
		\end{align}
		Evaluated at the minimum $\expval{\Sigma^I}=v \delta_U^I$, it becomes
		\begin{align}
			\partial_{B}\partial_AV|_{min}=-\mu^2g_{AB}+\quarticSigmaA v^2(g_{AB}+2g_{BU}g_{AU})+3\quarticSigmaB v^2 \hat{T}_{ABUU}\ .
		\end{align}
		As we have seen, this matrix has only 5 different possible entries, since the non-zero entries of $\hat{T}_{ABUU}$ only appear in correspondence to the non-zero entries of $g_{AB}$. Plugging the values found in Eq.~\eqref{eq:hattabuufinal} and the values of $g_{AB}$ we get
		\begin{align}
			\partial_{B}\partial_AV|_{min}=\mu^2\begin{cases}
				0, &\text{$A$ and $B$ fermionic}\\
				\frac{ \quarticSigmaB (M-N) (2 M+N)}{\quarticSigmaB \left(M^2+M N+N^2\right)+4 \quarticSigmaA  M N (M-N)}, &\text{$A=B\neq U$ bosonic with $g_{AB}=1$}\\
				\frac{ \quarticSigmaB (M-N) (M+2 N)}{\quarticSigmaB \left(M^2+M N+N^2\right)+4 \quarticSigmaA  M N (M-N)}, & \text{$A=B\neq U$ bosonic with $g_{AB}=-1$}\\
				2 , & A=B=U\\
				0 , & \text{otherwise}
			\end{cases}.
			\label{eq:quadraticpotential}
		\end{align}
		Notice that the potential is flat in the fermionic directions, again consistent with Goldstone's theorem. The symmetry breaking pattern implied by this vacuum is such that the fermionic directions are broken, and the massless modes develop exactly in those directions. 
		
		While this is what we wanted, an issue now arises when we focus on the bosonic part. The eigenvalues for $A=B\neq U$ bosonic with $g_{AB}=1$ and those for $A=B\neq U$ bosonic with $g_{AB}=-1$ have the same sign. As such, after the rephasing in Eq.~\eqref{eq:rephasing}, one of the two will be negative. This shows that what we found is actually just a saddle point, and the minimum must lie somewhere else. This justifies going beyond the adjoint representation and relying on a non-supertraceless field as we did in Section~\ref{sec:breakingSUNM}. 
		
  \section{Coleman-Weinberg Potential}
  \label{app:cwpotential}

In this appendix, we derive the one-loop Coleman-Weinberg potential in Eq.~\eqref{eq:cwpot}. Our starting point is a Lagrangian for a scalar field $\Sigma^{\tilde{I}}$ of the form
		\begin{align}
			\mathcal{L}=-\frac{1}{2} g_{\tilde{I}\tilde{J}}\Sigma^{\tilde{I}} \Box\Sigma^{\tilde{J}}-V[\Sigma]
		\end{align}
		where $V[\Sigma]$ is some potential which for now we keep generic. 
		As usual, we shift $\Sigma\to \Sigma_b+\Sigma$ and drop tadpole terms to get, at one-loop
		\begin{multline}
			e^{i\Gamma(\Sigma_b)}=e^{i\int \dd[4]x\left(-\frac{1}{2} g_{\tilde{I}\tilde{J}}\Sigma_b^{\tilde{I}} \Box\Sigma_b^{\tilde{J}}-V[\Sigma_b]\right)}\nonumber\\
			\times\int_{\text{restr.}}\mathcal{D}\Sigma \exp{i\int\dd[4]x\left(-\frac{1}{2} g_{\tilde{I}\tilde{J}}\Sigma^{\tilde{I}} \Box\Sigma^{\tilde{J}}-\frac{1}{2}\Sigma^{\tilde{I}}\Sigma^{\tilde{J}}\partial_{\tilde{J}}\partial_{\tilde{I}}V[\Sigma_b]\right)}
		\end{multline}
		where $\Gamma(\Sigma_b)$ is the effective action we are after. Notice that the ordering of the derivative in the Taylor expansion is the correct one, namely $\partial_{\tilde{J}}\partial_{\tilde{I}}V$ rather than $\partial_{\tilde{I}}\partial_{\tilde{J}}V$. To convince ourselves that this is the case, let us set for example $V[\Sigma]=\Sigma^{\tilde{I}}g_{\tilde{I}\tilde{J}}\Sigma^{\tilde{J}}$. Then
		\begin{align}
			\frac{1}{2}\Sigma^{\tI}\Sigma^{\tJ}\partial_{\tJ}\partial_{\tI} V[\Sigma]=\Sigma^{\tI}\Sigma^{\tJ}\frac{1}{2}\left(2g_{{\tI \tJ}}\right)=V[\Sigma]\ ,
		\end{align}
		while the opposite convention would give an additional $(-1)^{\degf(\tI)\degf({\tJ})}$. To perform the gaussian integral, however, we first need to commute $\Sigma^{\tJ}$ all the way to the right of the expansion, in order to have the desired $\sim \exp{-\Sigma^{\tI}M_{\tI\tJ}\Sigma^{\tJ}/2}$. This means we pick an additional factor of $(-1)^{(\degf(\tI)+\degf(\tJ))\degf(\tJ)}$. We can now do the integral to obtain 
		\begin{align}
			e^{i\Gamma(\Sigma_b)}=\text{const.}\times e^{i\int \dd[4]x\left(-\frac{1}{2} g_{\tI\tJ}\Sigma_b^\tI \Box \Sigma_b^\tJ-V[\Sigma_b]\right)}\frac{1}{\sqrt{\text{Ber}\left(\Box g_{\tI\tJ}+\partial_\tJ\partial_\tI V[\Sigma_b](-1)^{(\degf(\tI)+\degf(\tJ))\degf(\tJ)}\right)}}\ .
		\end{align}
        where $\text{Ber}$ indicates the Berezinian or superdeterminant, which is the natural generalization of the determinant to supermatrices and is defined as
        \begin{align}
            \text{Ber}(X)= \exp(\str{\log{X}})\ ,
        \end{align}
        with the property $\text{Ber}(X Y)=\text{Ber}(X )\text{Ber}(Y )$\footnote{Notice that, for a supermatrix with zero fermionic components, $B=\begin{pmatrix}
            A&0\\
            0&D
        \end{pmatrix}$, the superdeterminant reduces to \begin{align}
            \text{Ber}(X)=\det{A}\det{D}^{-1}\ ,
        \end{align} reproducing the familiar result that for integration over fermionic variables the determinant appears in the numerator.}.
		Let us define 
		\begin{align}
			\tensor{\tilde{V}}{^\tL_\tJ}\equiv g^{{\tL\tK}}\partial_{\tJ}\partial_{\tK} V[\Sigma_b](-1)^{(\degf(\tK)+\degf(\tJ))\degf(\tJ)},
		\end{align}
		to simplify the notation.
        Remembering that $g^{\tI\tJ}$ is the inverse of $g_{\tI\tJ}$, we can massage the square root into
  		\begin{align}
			\frac{1}{\sqrt{\text{Ber}\left(\Box g_{\tI\tJ}+\partial_\tJ\partial_\tI V[\Sigma_b](-1)^{(\degf(I)+\degf(J))\degf(\tJ)}\right)}}=\frac{1}{\sqrt{\text{Ber}(g_{\tI {\tL}})\text{Ber}\left(\Box \delta^{\tL}_\tJ+\tensor{V}{^{\tL}_{\tJ}}\right)}}\ ,
		\end{align}
		meaning
		\begin{align}
			e^{i\Gamma(\Sigma_b)}=\text{const.}' \times e^{i\int \dd[4]x\left(-\frac{1}{2} g_{\tI\tJ}\Sigma^\tI_{b} \Box \Sigma^\tJ_b -V[\Sigma_b]\right)}\frac{1}{\sqrt{\text{Ber}(\Box \delta^{\tL}_\tJ+\tensor{V}{^{\tL}_{\tJ}})}}\ ,
		\end{align}
		where we reabsorbed the $\Sigma_b$-independent $\frac{1}{\sqrt{\text{Ber}(g_{\tI{\tL}})}}$ into the overall constant.
		Defining 
		\begin{align}
			\Gamma[\Sigma_b]=\int \dd[4]x\left(-\frac{1}{2} g_{\tI\tJ}\Sigma^\tI_{b} \Box\Sigma^\tJ_{b}-V[\Sigma_{b}]\right)+\Delta\Gamma[\Sigma_b],
		\end{align} 
		we can rewrite the correction to the tree-level potential $\Delta\Gamma[\Sigma_b]$ as
		\begin{align}
			i\Delta\Gamma[\Sigma_b]=-\frac{1}{2}\Str \ln\left(\Box \delta^{\tL}_\tJ+\tensor{V}{^{\tL}_{\tJ}}\right)+ {\ln \left(\text{const.}'\right)},
		\end{align}
		where the trace $\Str$ is over position eigenstates $\ket{x}$ and over group indices. Now we can pull out the $\Sigma_b$-independent integral over $\ln[\Box]$ and go to momentum space. Here we rotate to Euclidean metric and define $\Delta V_{\text{eff}}[\Sigma_b]=-\frac{\Delta\Gamma(\Sigma_b)}{VT}$ to get rid of a factor of space-time volume $VT$
		\begin{align}
			\Delta V_{\text{eff}}[\Sigma_b]=\frac{1}{16 \pi ^2}\strsans\int_m^\Lambda\dd k k^3\ln (\delta^{\tL}_\tJ+\frac{\tensor{V}{^{\tL}_{\tJ}}}{k^2})\ 
		\end{align}
		where we regulated the integral with both a UV-regulator $\Lambda$ and a IR one $m$, and the trace $\strsans$ now is only to be taken over the internal indices. 
        The $\log$ of a matrix is defined via its Taylor expansion, so
        \begin{align}
			\Delta V_{\text{eff}}[\Sigma_b]=&\frac{1}{16\pi^2}\strsans\left\{\tensor{\tilde{V}}{^\tL_\tJ}\int_m^\Lambda \dd k k-\frac{(\tensor{\tilde{V}}{^\tL_\tJ})^2}{2}\int^\Lambda_m\dd k\frac{1}{k}+(-1)\sum_{j=3}^{\infty}(-\tensor{\tilde{V}}{^\tL_\tJ})^j\left.\frac{k^{4-2j}}{j(4-2j)}\right|^\Lambda_m\right\}\nonumber\\
			\overset{m\to 0,\, \Lambda \to \infty}{\sim}&\frac{1}{32\pi^2}\strsans(\tensor{\tilde{V}}{^\tL_\tJ})\Lambda^2+\frac{1}{64\pi^2}\strsans\left[\left(\tensor{\tilde{V}}{^\tL_\tJ}\right)^2\ln(\frac{e^{-\frac{1}{2}}\tensor{\tilde{V}}{^\tL_\tJ}}{\Lambda^2})\right]\ .
			\label{eq:deltaveffbosons}
		\end{align} 
        Note that the $\text{str}$ here differs from that defined in Eq.~\eqref{eq:original_str} since it is taken over indices in the adjoint representation and should be interpreted as $\text{str} \left( \tensor{\tilde{V}}{^\tL_\tJ} \right) \equiv \sum_{\tL} (-1)^{f(\tL)} \tensor{\tilde{V}}{^\tL_\tL}$, where the sum runs over the generators of $U(N | M)$.
	In summary,
        \begin{align}
			V_{\text{eff}}[\Sigma_b]\equiv V[\Sigma_b]+\Delta V_{\text{eff},1}+\Delta V_{\text{eff},2}\ ,
		\end{align}
		where 
		\begin{align}
			\Delta V_{\text{eff},1}&\equiv \frac{e^{-\frac{1}{2}}}{32\pi^2}\text{str}\left(\tensor{\tilde{V}[\Sigma_b]}{^\tL_\tJ}\right)\tilde{\Lambda}^2,\,&
			\Delta V_{\text{eff},2}&\equiv \frac{1}{64\pi^2}\text{str}\left[\left(\tensor{\tilde{V}[\Sigma_b]}{^\tL_\tJ}\right)^2\ln(\frac{\tensor{\tilde{V}[\Sigma_b]}{^\tL_\tJ}}{\tilde{\Lambda}^2})\right]\ .
		\end{align}
        and we defined $\tilde{\Lambda}^2\equiv e^{\frac{1}{2}}\Lambda^2$.
		We will drop the tilde from now on and just replace $\tilde{\Lambda}\to \Lambda$.

		For a potential of the form
		\begin{align}
			V[\Sigma]=-\frac{1}{2}\mu^2\Sigma^\tI g_{\tI\tJ}\Sigma^\tJ+\frac{\quarticSigma}{4}A_{\tI\tJ\tK\tL}\Sigma^\tI\Sigma^\tJ\Sigma^\tK\Sigma^\tL\ ,
		\end{align}
        we get
		\begin{align}
			\tensor{\tilde{V}}{^\tI_\tJ}=-\mu^2\delta^\tI_\tJ+3\quarticSigma g^{\tI\tK}A_{\tK\tA\tB\tJ} \Sigma_b^{\tA}\Sigma_b^{\tB}\ ,
		\end{align}
		and 
		\begin{align}
			\text{str}{\tensor{\tilde{V}}{^\tI_\tJ}}=&-\mu^2 (N-M)^2+3(-1)^{\degf(\tI)}\quarticSigma g^{\tI\tK}\tensor{A}{_{\tK\tA\tB\tI}}\Sigma_b^\tA\Sigma_b^\tB \\
			\text{str}{\left(\tensor{(\tilde{V}^2)}{^\tI_\tJ}\right)}=&\mu^4(N-M)^2-6\mu^2\quarticSigma (-1)^{\degf(\tI)}g^{\tI\tK}\tensor{A}{_{\tK\tA\tB\tI}}\Sigma_b^\tA\Sigma_b^\tB\nonumber\\
        &+9\quarticSigma^2(-1)^{\degf(\tI)}g^{\tI\tK}\tensor{A}{_{\tK\tL\tA\tB}}g^{\tL\tM}\tensor{A}{_{\tM\tI\tC\tD}}\Sigma_b^\tC\Sigma_b^\tD \Sigma_b^\tA\Sigma_b^\tB\ .
		\end{align}
        We are now ready to plug in the explicit form of the potential in Eq.~\eqref{eq:scalarpotential2}.
		Defining 
        \begin{align}
            \hat{G}_{\tI\tJ\tK\tL}&\equiv\frac{1}{3} \left[g_{\tI\tJ}g_{\tK\tL}+(-1)^{\degf(\tJ)\degf(\tK)}g_{\tI\tK}g_{\tJ\tL}+(-1)^{\degf(\tJ)\degf(\tL)+\degf(\tK)\degf(\tL)}g_{\tI\tL}g_{\tJ\tK}\right]\nonumber\\
            \hat{\Sigma}^{\tI\tJ\tK\tL}&\equiv\Sigma^{\tI}\Sigma^{\tJ}\Sigma^{\tK}\Sigma^{\tL}\nonumber\\
            \hat{T}_{\tI\tJ\tK\tL}&\equiv\str{\lambda_{\{\tI}\lambda_{\tJ}\lambda_{\tK}\lambda_{\tL\}_{\degf}}},
        \end{align}
		we have
		\begin{align}
			A_{\tI\tJ\tK\tL}=a_1\hat{G}_{\tI\tJ\tK\tL}+a_2\hat{T}_{\tI\tJ\tK\tL}, \qquad a_{1,2}=\frac{\quarticSigma_{1,2}}{\quarticSigma}\ .
		\end{align}
        Then, we need to compute\footnote{Note that in the following the $\text{str}$ on the LHS (e.g. $\text{str}{\tensor{\tilde{V}}{^\tI_\tJ}}$) is distinct from that on the RHS (e.g. \str{\lambda_\tL}). As previously stated, the former should be interpreted as $\text{str} \tensor{\tilde{V}}{^\tL_\tJ}  \equiv \sum_{\tL} (-1)^{f(\tL)} \tensor{\tilde{V}}{^\tL_\tL}$ whereas the latter is defined in Eq.~\ref{eq:original_str}.} 
        \begin{align}
			\text{str}{\tensor{\tilde{V}}{^\tI_\tJ}}
			=& -\mu^2 (N-M)^2+G_{2,1} \Sigma_b^\tK g_{\tK\tL}\Sigma_b^\tL+G_{2,2} \Sigma_b^\tK\str{\lambda_\tK}\str{\lambda_\tL}\Sigma_b^\tL\\
			\text{str}{\left(\tensor{(\tilde{V}^2)}{^\tI_\tJ}\right)}
			=&\mu^4(N-M)^2-2\mu^2\left[G_{2,1} \Sigma_b^\tK g_{\tK\tL}\Sigma_b^\tL+G_{2,2} \Sigma_b^\tK\str{\lambda_\tK}\str{\lambda_\tL}\Sigma_b^\tL\right]\nonumber\\
			&+G_{4,1}\hat{G}_{\tI\tJ\tK\tL}\hat{\Sigma_b}^{\tI\tJ\tK\tL}+G_{4,2}\hat{T}_{\tI\tJ\tK\tL}\hat{\Sigma_b}^{\tI\tJ\tK\tL}\nonumber\\
			&+G_{4,3}\str{\lambda_\tI}\str{\lambda_\tJ}g_{\tK\tL}\hat{\Sigma_b}^{\tI\tJ\tK\tL}+G_{4,4}\str{\lambda_\tI}\hat{T}_{\tJ\tK\tL}\hat{\Sigma_b}^{\tI\tJ\tK\tL}\ ,
		\end{align}
		where we defined the group-dependent constants
		\begin{align}
			G_{2,1}&=((N-M)^2+2)\quarticSigmaA +\frac{\quarticSigmaB }{2(N-M)}&
			G_{2,2}&=\frac{1}{2}\quarticSigmaB \\
			G_{4,1}&=\quarticSigmaA ^2\left[(N-M)^2+8\right]+\frac{\quarticSigmaA \quarticSigmaB }{N-M}+\frac{3\quarticSigmaB ^2}{16}\\
			G_{4,2}&=12 \quarticSigmaA \quarticSigmaB  +\frac{\quarticSigmaB ^2}{2}(N-M)&
			G_{4,3}&=\quarticSigmaA \quarticSigmaB &
			G_{4,4}&=\quarticSigmaB ^2\ ,
            \label{eq:renormgroupdependent}
		\end{align}
        and used the results of Appendix~\ref{app:UNMrelations}.
        Eq.~\eqref{eq:renormgroupdependent} also shows that, as expected, the terms we put to zero by hand at tree level, i.e. those proportional to $\tr{\Sigma}$, are generated at one-loop.
		Then, when we write the one-loop effective potential, we have to include counterterms for them, too. Explicitly,
        \begin{align}
			V_{\text{eff}}
			=&V+\frac{e^{-\frac{1}{2}}}{32\pi^2}\Lambda^2 \left(-\mu^2 (N-M)^2+G_{2,1} \Sigma_b^\tK g_{\tK\tL}\Sigma_b^\tL+G_{2,2} \Sigma_b^\tK\str{\lambda_\tK}\str{\lambda_\tL}\Sigma_b^\tL \right)\nonumber\\
			&+\frac{1}{64\pi^2}\ln(\frac{\bar{m}^2}{\Lambda^2})\Big\{
			\mu^4(N-M)^2-2\mu^2\left[G_{2,1} \Sigma_b^\tK g_{\tK\tL}\Sigma_b^\tL+G_{2,2} \Sigma_b^\tK\str{\lambda_\tK}\str{\lambda_\tL}\Sigma_b^\tL\right]\nonumber\\
			&\qquad \qquad \qquad \qquad +G_{4,1}\hat{G}_{\tI\tJ\tK\tL}\hat{\Sigma_b}^{\tI\tJ\tK\tL}+G_{4,2}\hat{T}_{\tI\tJ\tK\tL}\hat{\Sigma_b}^{\tI\tJ\tK\tL}\nonumber\\
			&\qquad \qquad \qquad \qquad +G_{4,3}\str{\lambda_\tI}\str{\lambda_\tJ}g_{\tK\tL}\hat{\Sigma_b}^{\tI\tJ\tK\tL}+G_{4,4}\str{\lambda_\tI}\hat{T}_{\tJ\tK\tL}\hat{\Sigma_b}^{\tI\tJ\tK\tL}
			\Big\}\nonumber\\
                &+\frac{1}{64\pi^2}\text{str}\left[\left(\tensor{\tilde{V}}{^\tI_\tJ}\right)^2\ln(\frac{\tensor{\tilde{V}}{^\tI_\tJ}}{\bar{m}^2})\right]\nonumber\\
			&+A_{0}+A_{2,1}\Sigma_b^\tI g_{\tI\tJ}\Sigma_b^\tJ+A_{2,2}\str{\lambda_\tI}\str{\lambda_\tJ}\Sigma_b^\tI\Sigma_b^\tJ\nonumber\\
			&+A_{4,1}\hat{G}_{\tI\tJ\tK\tL}\hat{\Sigma_b}^{\tI\tJ\tK\tL}+A_{4,2}\hat{T}_{\tI\tJ\tK\tL}\hat{\Sigma_b}^{\tI\tJ\tK\tL}\nonumber\\
			&+A_{4,3}\str{\lambda_\tI}\str{\lambda_\tJ}g_{\tK\tL}\hat{\Sigma_b}^{\tI\tJ\tK\tL}+A_{4,4}\str{\lambda_\tI}\hat{T}_{\tJ\tK\tL}\hat{\Sigma_b}^{\tI\tJ\tK\tL}\ ,
		\end{align}
        where we have introduced the arbitrary scale $\bar{m}$.
		Choosing
		\begin{align}
			A_0&=\frac{e^{-\frac{1}{2}}}{32\pi^2}\Lambda^2\mu^2(N-M)^2-\frac{1}{64\pi^2}\ln\left(\frac{\bar{m}^2}{\Lambda^2}\right)\mu^4(N-M)^2\\
			A_{2,1}&=-\frac{e^{-\frac{1}{2}}}{32\pi^2}\Lambda^2G_{2,1}+2\mu^2\frac{1}{64\pi^2}\ln\left(\frac{\bar{m}^2}{\Lambda^2}\right)G_{2,1}\\
			A_{2,2}&=-\frac{e^{-\frac{1}{2}}}{32\pi^2}\Lambda^2G_{2,2}+2\mu^2\frac{1}{64\pi^2}\ln\left(\frac{\bar{m}^2}{\Lambda^2}\right)G_{2,2}\\
			A_{4,i}&=-\frac{1}{64\pi^2}\ln\left(\frac{\bar{m}^2}{\Lambda^2}\right)G_{4,i},\qquad i=1,\,2\,3\,4\ ,
		\end{align}
		we can remove all $\Lambda$-dependent terms and get to the final form
		\begin{align}
			V_{\text{eff}}=V+\frac{1}{64\pi^2}\text{str}\left[\left(\tensor{\tilde{V}}{^I_J}\right)^2\ln(\frac{\tensor{\tilde{V}}{^I_J}}{\bar{m}^2})\right]\ .
		\end{align}

  \section{Useful relations in $SU(N|M)$}
		\label{app:sunmcontractions}
        In this Section, we obtain and summarize a series of results for $SU(N|M)$ and $U(N|M)$, i.e. the extension used in Section~\ref{sec:breakingSUNM}. Some of these relations are used in the text. 
        \subsection{$SU(N|M)$ identities}
        To make this section more self contained, we first recap some properties of $SU(N|M)$.
		The algebra of the group is defined by the commutation relation
		\begin{align}
			\comm{\lambda_I}{\lambda_J}_{\degf}=i\tensor{f}{_I_J^K}\lambda_K\ ,
		\end{align}
		where $\tensor{f}{_I_J^K}$ are the structure constant. We take the generators to be normalized to 
		\begin{align}
			\str{\lambda_{I}\lambda_J}=\frac{1}{2}g_{IJ}\ ,
		\end{align}
		where $g_{IJ}$ is as in Eq.~\eqref{eq:metric}. With this normalization, the completeness relation reads
		\begin{align}
			(\lambda_I)_{i}^jg^{IJ}(\lambda_J)_{k}^l=\frac{1}{2}\left(\delta_{i}^l\delta_{k}^j(-1)^{\degf(j)\degf{(k)}}-\frac{1}{N-M}\delta_{i}^j\delta_{k}^l\right)\ .
		\end{align}
		Since the generators of $SU(N|M)$ form, together with the identity, a complete basis of hermitian matrices, we can always decompose the product of two of them as
		\begin{align}
			\lambda_I\lambda_J=\frac{1}{2}\left[\frac{1}{N-M}g_{IJ}+\left(\tensor{d}{_I_J^K}+i\tensor{f}{_I_J^K}\right)\lambda_K\right]\ .
			\label{eq:prodofgenerators}
		\end{align}
		Eq.~\eqref{eq:prodofgenerators} can be taken as a definition of the tensor $\tensor{d}{_I_J^K}$. It is useful to define
		\begin{align}
			\acomm{X}{Y}_{\degf} \equiv XY+(-1)^{\degf(X)\degf(Y)}YX\ ,
		\end{align}
		i.e. the generalization of the anticommutator to our case. Then
		\begin{align}
			\acomm{\lambda_I}{\lambda_J}_{\degf}=\frac{1}{N-M}g_{IJ}+\tensor{d}{_I_J^K}\lambda_K \\
            \comm{\lambda_I}{\lambda_J}_{\degf}=\tensor{f}{_I_J^K}\lambda_K
		\end{align}
		and
		\begin{align}
			d_{IJL}&\equiv\tensor{d}{_I_J^K}g_{KL}=2\str{\acomm{\lambda_I}{\lambda_J}_{\degf}\lambda_L}\\
			\tensor{d}{_I_J^K}&=d_{IJL}g^{LK}\\
            f_{IJL}&\equiv\tensor{f}{_I_J^K}g_{KL}=2\str{\comm{\lambda_I}{\lambda_J}_{\degf}\lambda_L}\\
            \tensor{f}{_I_J^K}&=f_{IJL}g^{LK}\ .
		\end{align}
		An important property we will use later is that, since the product of two fermionic or bosonic generators can only be bosonic, while the product of one fermionic and one bosonic generator is fermionic, then
		\begin{align}
			\tensor{d}{_I_J^K}\neq 0 \text{ only when } \degf(K)= \degf(I)+\degf(J) \mod{2}\ ,
			\label{eq:dtensorproperty}
		\end{align}
        and similarly for $\tensor{f}{_I_J^K}$.
        Using this, we can check that $f_{IJK}$ and $d_{IJK}$ are fully $\degf$-antisymmetric and $\degf$-symmetric respectively, using the generalized cyclicity of traces involving generators in Eq.~\eqref{eq:gradedciclicity}.
		Using this decomposition, we can compute
		\begin{align}
			\str{\lambda_I\lambda_J\lambda_K\lambda_L}=&\frac{1}{4}\text{str}\left\{\left[\frac{1}{N-M}g_{IJ}+(\tensor{d}{_I_J^P}+i\tensor{f}{_I_J^P})\lambda_P\right]\right.\times \nonumber \\
   &\times
   \left.
   \left[\frac{1}{N-M}g_{KL}+(\tensor{d}{_K_L^Q}+i\tensor{f}{_K_L^Q})\lambda_Q\right]\right\}=\nonumber\\
			=&\frac{1}{4}\frac{1}{N-M}g_{IJ}g_{KL}+\frac{1}{8}(\tensor{d}{_I_J^P}+i\tensor{f}{_I_J^P})(\tensor{d}{_K_L^Q}+i\tensor{f}{_K_L^Q})g_{PQ}\ .
		\end{align}
		We can use this expression to compute $\hat{T}_{IJKL}$, i.e. the fully f-symmetrized version of $T_{IJKL}=\str{\lambda_I\lambda_J\lambda_K\lambda_L}$. 
		Since the structure constants are f-antisymmetric under the exchange of their first two indices, they will drop out when computing $\hat{T}_{IJKL}$. Thus we need to f-symmetrize only the terms containing $g_{IJ}$ and $\tensor{d}{_I_J^K}$. Of the 24 terms built out of $g_{IJ}g_{KL}$ by permuting the 4 indices, only three are independent, as the other ones can be brought to these three by using the f-symmetry properties of $g_{IJ}$. Putting the appropriate combinatorics factor and the minus signs we get
		\begin{align}
			g_{IJ}g_{KL}\xrightarrow{\text{f-symm}} \frac{1}{3} \left[g_{IJ}g_{KL}+(-1)^{\degf(J)\degf(K)}g_{IK}g_{JL}+(-1)^{\degf(J)\degf(L)+\degf(K)\degf(L)}g_{IL}g_{JK}\right]=\hat{G}_{IJKL}\ ,
		\end{align}
		which one can verify has the right f-symmetry properties. 
		The second term is a bit longer to check. Indeed, the first two indices of $\tensor{d}{_I_J^P}$ are still f-symmetric as those of $g_{IJ}$. However, now we need to take into account that, while $g_{IJ}g_{KL}$ is clearly the same as $g_{KL}g_{IJ}$, $\tensor{d}{_I_J^P}\tensor{d}{_K_L^Q}g_{PQ}\neq \tensor{d}{_K_L^P}\tensor{d}{_I_J^Q}g_{PQ}$. Thus there will be 6 independent terms. Accounting for the minus signs we pay for moving indices past each other we get
		\begin{align}
			\tensor{d}{_I_J^P}\tensor{d}{_K_L^Q}g_{PQ}\xrightarrow{\text{f-symm}}& \frac{1}{6}\left[\tensor{d}{_I_J^P}\tensor{d}{_K_L^Q}\left(g_{PQ}+(-1)^{(\degf(I)+\degf(J))(\degf(K)+\degf(L))}g_{QP}\right)\right.\nonumber\\
			&(-1)^{\degf(J)\degf(K)}\tensor{d}{_I_K^P}\tensor{d}{_J_L^Q}\left(g_{PQ}+(-1)^{(\degf(I)+\degf(K))(\degf(J)+\degf(L))}g_{QP}\right)\nonumber\\
			&\left.(-1)^{\degf(J)\degf(L)+\degf(K)\degf(L)}\tensor{d}{_I_L^P}\tensor{d}{_J_K^Q}\left(g_{PQ}+(-1)^{(\degf(I)+\degf(L))(\degf(J)+\degf(K))}g_{QP}\right)\right]\ .
		\end{align}
		However, we can now use the property in Eq.~\eqref{eq:dtensorproperty} to simplify this a bit. Indeed, we can rewrite e.g. 
		\begin{align}
			&g_{PQ}+(-1)^{(\degf(I)+\degf(J))(\degf(K)+\degf(L))}g_{QP}=g_{PQ}\left[1+(-1)^{(\degf(I)+\degf(J))(\degf(K)+\degf(L))}(-1)^{\degf(P)\degf(Q)}\right]=\nonumber\\
			&=g_{PQ}\left[1+(-1)^{(\degf(I)+\degf(J))(\degf(K)+\degf(L))}(-1)^{(\degf(I)+\degf(J))(\degf(K)+\degf(L))}\right]=2g_{PQ}\ ,
		\end{align}
		since the only non-zero pieces come from $\degf(P)= \degf(I)+\degf(J) \mod{2}$ and $\degf(Q)= \degf(K)+\degf(K) \mod{2}$. Then
		\begin{align}
			\tensor{d}{_I_J^P}\tensor{d}{_K_L^Q}g_{PQ}\xrightarrow{\text{f-symm}}&\frac{1}{3}\left[\tensor{d}{_I_J^P}\tensor{d}{_K_L^Q}g_{PQ}+(-1)^{\degf(J)\degf(K)}\tensor{d}{_I_K^P}\tensor{d}{_J_L^Q}g_{PQ}+\right.\nonumber
   \\ &
   \left. +(-1)^{\degf(J)\degf(L)+\degf(K)\degf(L)}\tensor{d}{_I_L^P}\tensor{d}{_J_K^Q}g_{PQ}\right]\ ,
		\end{align}
		meaning
		\begin{align}
			\hat{T}_{IJKL}=&\frac{1}{12(N-M)}\left[g_{IJ}g_{KL}+(-1)^{\degf(J)\degf(K)}g_{IK}g_{JL}+(-1)^{\degf(J)\degf(L)+\degf(K)\degf(L)}g_{IL}g_{JK}\right]+\nonumber\\
			&+	\frac{1}{24}\left[\tensor{d}{_I_J^P}\tensor{d}{_K_L^Q}g_{PQ}+(-1)^{\degf(J)\degf(K)}\tensor{d}{_I_K^P}\tensor{d}{_J_L^Q}g_{PQ}+\right.\nonumber\\
   &+\left.(-1)^{\degf(J)\degf(L)+\degf(K)\degf(L)}\tensor{d}{_I_L^P}\tensor{d}{_J_K^Q}g_{PQ}\right]=\nonumber\\
			=&\frac{1}{12(N-M)}\left[g_{IJ}g_{KL}+(-1)^{\degf(J)\degf(K)}g_{IK}g_{JL}+(-1)^{\degf(J)\degf(L)+\degf(K)\degf(L)}g_{IL}g_{JK}\right]+\nonumber\\
			&+	\frac{1}{24}\left[d_{IJP}d_{KLQ}g^{QP}+(-1)^{\degf(J)\degf(K)}d_{IKP}d_{JLQ}g^{QP}+\right.\nonumber\\
   &+\left.(-1)^{\degf(J)\degf(L)+\degf(K)\degf(L)}d_{ILP}d_{JKQ}g^{QP}\right]\ .
		\end{align}
		Some additional important identites are
		\begin{align}
			(-1)^{\degf(J)\degf(K)}d_{IKP}d_{JLQ}g^{QP}g^{IJ}=d_{IKP}d_{LJQ}g^{QP}g^{IJ}=\frac{(N-M)^2-4}{N-M}g_{KL}
			\label{eq:dtenssq}\\
			d_{CLF}d_{IDG}g^{GF}=2\left(\str{\acomm{\lambda_C}{\lambda_L}_{\degf}\acomm{\lambda_I}{\lambda_D}_{\degf}}-\frac{1}{N-M}g_{CL}g_{ID}\right)\ ,
		\end{align}
		implying
		\begin{align}
			g^{IJ}\tensor{\hat{T}}{_{JKLI}}(-1)^{\degf(I)}=\frac{1}{12(N-M)}g_{KL}(2(N-M)^2-3)\ .
		\end{align}
		Notice that Eq.~\eqref{eq:dtenssq} reproduces the $SU(N)$ result
		\begin{align}
			d_{abc}d_{abd}=\frac{N^2-4}{N}\delta_{cd},
		\end{align}
		in the $M\to 0$ limit, as it should. 
        In order to obtain the one-loop potential, the last identity that we need is the one involving terms of order $\order{\hat{T}_{IJKL}^2}$. More explicitly, defining $\tensor{\hat{T}}{^I_{JAB}}\equiv g^{IL}\hat{T}_{LJAB}$, we have
		\begin{align}
			(-1)^{\degf(I)}\tensor{\hat{T}}{^I_{JAB}}\tensor{\hat{T}}{^J_{ICD}}=&\frac{1}{(4!)^2}\left[\frac{4(2(N-M)^2+9)}{(N-M)^2}g_{CD}g_{AB}+\right.\nonumber\\
			&\left.+2\left(g_{CA}g_{DB}(-1)^{\degf(A)\degf(D)}+g_{CB}g_{DA}\right)+\right.\nonumber\\
			&\left. +8\frac{(N-M)^2-9}{(N-M)}\left(T_{CDAB}+(-1)^{\degf(A)\degf(B)}T_{CDBA}+\right.\right. \nonumber\\
			&\left. \left.+(-1)^{\degf(C)\degf(D)}T_{DCAB}+(-1)^{\degf(A)\degf(B)+\degf(C)\degf(D)}T_{DCBA}\right)\right].
		\end{align}
		After f-symmetrization this becomes
		\begin{align}
			&\frac{1}{3}(-1)^{\degf(I)}\left(\tensor{\hat{T}}{^I_{JAB}}\tensor{\hat{T}}{^J_{ICD}}+\tensor{\hat{T}}{^I_{JDB}}\tensor{\hat{T}}{^J_{ICA}}(-1)^{\degf(A)\degf(D)}+\tensor{\hat{T}}{^I_{JDA}}\tensor{\hat{T}}{^J_{ICB}}(-1)^{\degf(B)(\degf(D)+\degf(A))}\right)\nonumber\\
			&=\frac{1}{3 (4!)}\left[4\frac{(N-M)^2+3}{(N-M)^2}\hat{G}_{CDAB}+\frac{96((N-M)^2-9)}{N-M}\hat{T}_{CDAB}\right].
		\end{align}
        For comparison, the analogous result for $SU(N)$ is
		\begin{align}
			\hat{T}_{ijab}\hat{T}_{jicd}=&\frac{1}{(4!)^2}\left[\frac{4(2N^2+9)}{N^2}\delta_{ab}\delta_{cd}+2(\delta_{ac}\delta_{bd}+\delta_{ad}\delta_{cb})+\right.\nonumber\\
        &\left.+\frac{8(N^2-9)}{N} (T_{abcd}+T_{bacd}+T_{abdc}+T_{badc})\right],
		\end{align}
		meaning
		\begin{align}
			&\frac{1}{3}(\hat{T}_{ijab}\hat{T}_{jicd}+\hat{T}_{ijac}\hat{T}_{jibd}+\hat{T}_{ijad}\hat{T}_{jibc})=\nonumber\\
			&=\frac{1}{3(4!)^2}\left[\frac{4 (9 + 3 N^2)}{N^2}(\delta_{ab}\delta_{cd}+\delta_{ac}\delta_{bd}+\delta_{ad}\delta_bc)+\frac{96 \left(N^2-9\right)}{N} \hat{T}_{abcd}\right].
		\end{align}
		\subsection{$U(N|M)$ identities}
        \label{app:UNMrelations}
        Similar identities as in the main text can be obtained for $U(N|M)$, i.e. the extension of the $SU(N|M)$ group and algebra we needed in Section.~\ref{sec:breakingSUNM}. More specifically, we performed the replacing
		\begin{align}
			SU(N|M)&\to U(N|M) & \lambda_I& \to \lambda_{\tilde{I}}\ ,	
		\end{align}
		where $\lambda_{\tilde{I}}=\left\{\lambda_I,\, \lambda_T\equiv \frac{1}{\sqrt{2(N-M)}}\mathbb{I}\right\}$. The normalization factor for $\lambda_T$ is chosen so that 
		\begin{align}
			g_{TT}=2\str{\lambda_T\lambda_T}=1\ ,
		\end{align}
		while $g_{N\tilde{I}}=0$, $\tilde{I}\neq N$. This means that the symmetric two form gets modified as $g_{IJ}\to{g_{\tilde{I}\tilde{J}}}$, where the only difference is the addition of a 1 in the diagonal in correspondence of $\lambda_T$. However, the normalization is also exactly the one needed to remove the second piece in the completeness relation, meaning that
		\begin{align}
			(\lambda_{\tilde{I}})_{i}^jg^{\tilde{I}\tilde{J}}(\lambda_{\tilde{J}})_{k}^l=\frac{1}{2}\delta_{i}^l\delta_{k}^j(-1)^{\degf(j)\degf{(k)}}\ .
		\end{align}
		The identities involving $\hat{T}_{\tilde{I}\tilde{J}\tilde{K}\tilde{L}}$ are modified and become
		\begin{align}
			(-1)^{\degf(I)}g^{\tilde{I}\tilde{J}}\hat{T}_{\tilde{J}\tilde{K}\tilde{L}\tilde{I}}&=\frac{1}{6(N-M)}g_{\tilde{K}\tilde{L}}+\frac{1}{6}\str{\lambda_{\tilde{K}}}\str{\lambda_{\tilde{L}}}\ ,
		\end{align}
        and
		\begin{align}
			(-1)^{\degf(\tilde{I})}\tensor{\hat{T}}{^{\tilde{I}}_{\tilde{J}\tilde{A}\tilde{B}}}&\tensor{\hat{T}}{^{\tilde{J}}_{\tilde{I}\tilde{C}\tilde{D}}}=\frac{1}{(4!)^2}\left[2(4g_{\tilde{A}\tilde{B}}g_{\tilde{C}\tilde{D}}+(-1)^{\degf(\tilde{A})\degf(\tilde{D})}g_{\tilde{C}\tilde{A}}g_{\tilde{D}\tilde{B}}+g_{\tilde{C}\tilde{D}}g_{\tilde{D}\tilde{A}})\right.\nonumber\\
           &\left.+8(N-M)\left(\str{\lambda_{\tilde{\{C}}\lambda_{\tilde{D\}_{\degf}}} \lambda_{\tilde{\{A}}\lambda_{\tilde{B\}_{\degf}}}}\right)\right.\nonumber\\
			&+4\str{\lambda_{\tilde{A}}}\left(\str{\lambda_{\{\tilde{C}}\lambda_{\tilde{D}\}_{\degf}}\lambda_{\tilde{B}}}+(-1)^{\degf(\tilde{B})(\degf(\tilde{C})+\degf(\tilde{D}))}\str{\lambda_{\tilde{B}}\lambda_{\{\tilde{C}}\lambda_{\tilde{D}\}_{\degf}}}\right)\nonumber\\
			&+4\str{\lambda_{\tilde{B}}}\left(\str{\lambda_{\{\tilde{C}}\lambda_{\tilde{D}\}_{\degf}}\lambda_{\tilde{A}}}+(-1)^{\degf(\tilde{A})(\degf(\tilde{C})+\degf(\tilde{D}))}\str{\lambda_{\tilde{A}}\lambda_{\{\tilde{C}}\lambda_{\tilde{D}\}_{\degf}}}\right)\nonumber\\
			  &+4\str{\lambda_{\tilde{C}}}\left(\str{\lambda_{\tilde{D}}\lambda_{\{\tilde{A}}\lambda_{\tilde{B}\}_{\degf}}}+(-1)^{\degf(\tilde{D})(\degf(\tilde{A})+\degf(\tilde{B}))}\str{\lambda_{\{\tilde{A}}\lambda_{\tilde{B}\}_{\degf}}\lambda_{\tilde{D}}}\right)\nonumber\\
			  &\left.+4\str{\lambda_{\tilde{D}}}\left(\str{\lambda_{\tilde{C}}\lambda_{\{\tilde{A}}\lambda_{\tilde{B}\}_{\degf}}}+(-1)^{\degf(\tilde{C})(\degf(\tilde{A})+\degf(\tilde{B}))}\str{\lambda_{\{\tilde{A}}\lambda_{\tilde{B}\}_{\degf}}\lambda_{\tilde{C}}}\right)\right]\\
             (-1)^{\degf(\tilde{I})}\tensor{\hat{T}}{^{\tilde{I}}_{\tilde{J}\tilde{A}\tilde{B}}}
             &\tensor{\hat{G}}{^{\tilde{J}}_{\tilde{I}\tilde{C}\tilde{D}}}=\frac{1}{3}\left(\frac{1}{6(N-M)}g_{\tilde{C}\tilde{D}}g_{\tilde{A}\tilde{B}}+\frac{1}{6}\str{\lambda_{\tilde{A}}}\str{\lambda_{\tilde{B}}}g_{\tilde{C}\tilde{D}}+2\hat{T}_{\tilde{C}\tilde{D}\tilde{A}\tilde{B}}\right)\\
		(-1)^{\degf(\tilde{I})}\tensor{\hat{G}}{^{\tilde{I}}_{\tilde{J}\tilde{A}\tilde{B}}}&\tensor{\hat{T}}{^{\tilde{J}}_{\tilde{I}\tilde{C}\tilde{D}}}=\frac{1}{3}\left(\frac{1}{6(N-M)}g_{\tilde{C}\tilde{D}}g_{\tilde{A}\tilde{B}}+\frac{1}{6}\str{\lambda_{\tilde{C}}}\str{\lambda_{\tilde{D}}}g_{\tilde{A}\tilde{B}}+2\hat{T}_{\tilde{C}\tilde{D}\tilde{A}\tilde{B}}\right)\\
           (-1)^{\degf(\tilde{I})}\tensor{\hat{G}}{^{\tilde{I}}_{\tilde{J}\tilde{A}\tilde{B}}}&\tensor{\hat{G}}{^{\tilde{J}}_{\tilde{I}\tilde{C}\tilde{D}}}=\frac{1}{9}\left(\left[(N-M)^2+4\right]g_{\tilde{A}\tilde{B}}g_{\tilde{C}\tilde{D}}+2\left((-1)^{\degf(\tilde{A})\degf(\tilde{D})}g_{\tilde{C}\tilde{A}}g_{\tilde{D}\tilde{B}}+g_{\tilde{C}\tilde{B}}g_{\tilde{D}\tilde{A}}\right)\right)
		\end{align}
		meaning, after f-symmetrization they each respectively become,
		\begin{align}
			\frac{1}{3}(-1)&^{\degf(\tilde{I})}\left(\tensor{\hat{T}}{^{\tilde{I}}_{\tilde{J}\tilde{A}\tilde{B}}}\tensor{\hat{T}}{^J_{\tilde{I}\tilde{C}\tilde{D}}}+\tensor{\hat{T}}{^{\tilde{I}}_{\tilde{J}\tilde{D}\tilde{B}}}\tensor{\hat{T}}{^J_{\tilde{I}\tilde{C}\tilde{A}}}(-1)^{\degf(\tilde{A})\degf(\tilde{D})}+\tensor{\hat{T}}{^{\tilde{I}}_{\tilde{J}\tilde{D}\tilde{A}}}\tensor{\hat{T}}{^J_{\tilde{I}\tilde{C}\tilde{B}}}(-1)^{\degf(\tilde{B})(\degf(\tilde{D})+\degf(\tilde{A}))}\right)\nonumber\\
			&=\frac{1}{3 (4!)^2}\left[12(g_{\tilde{C}\tilde{D}}g_{\tilde{A}\tilde{B}}+(-1)^{\degf(\tilde{A})\degf(\tilde{D})}g_{\tilde{C}\tilde{A}}g_{\tilde{D}\tilde{B}}+g_{\tilde{C}\tilde{B}}g_{\tilde{D}\tilde{A}})+96(N-M)\hat{T}_{\tilde{C}\tilde{D}\tilde{A}\tilde{B}}\right.\nonumber\\
			&\qquad +\left.48 \left(\str{\lambda_{\tilde{A}}}\hat{T}_{\tilde{C}\tilde{D}\tilde{B}}+\str{\lambda_{\tilde{B}}}\hat{T}_{\tilde{C}\tilde{D}\tilde{A}}+\str{\lambda_{\tilde{D}}}\hat{T}_{\tilde{C}\tilde{A}\tilde{B}}+\str{\lambda_{\tilde{C}}}\hat{T}_{\tilde{D}\tilde{A}\tilde{B}}\right)\right]\nonumber\\
			&=\frac{1}{3 (4!)^2}\left[36\hat{G}_{\tilde{C}\tilde{D}\tilde{A}\tilde{B}}+96(N-M)\hat{T}_{\tilde{C}\tilde{D}\tilde{A}\tilde{B}}\right.\nonumber\\
			&\qquad +\left.48 \left(\str{\lambda_{\tilde{A}}}\hat{T}_{\tilde{C}\tilde{D}\tilde{B}}+\str{\lambda_{\tilde{B}}}\hat{T}_{\tilde{C}\tilde{D}\tilde{A}}+\str{\lambda_{\tilde{D}}}\hat{T}_{\tilde{C}\tilde{A}\tilde{B}}+\str{\lambda_{\tilde{C}}}\hat{T}_{\tilde{D}\tilde{A}\tilde{B}}\right)\right]
		\end{align}
	    \begin{align}
	        \frac{1}{3}(-1)&^{\degf(\tilde{I})}\left(\tensor{\hat{T}}{^{\tilde{I}}_{\tilde{J}\tilde{A}\tilde{B}}}\tensor{\hat{G}}{^J_{\tilde{I}\tilde{C}\tilde{D}}}+\tensor{\hat{T}}{^{\tilde{I}}_{\tilde{J}\tilde{D}\tilde{B}}}\tensor{\hat{G}}{^J_{\tilde{I}\tilde{C}\tilde{A}}}(-1)^{\degf(\tilde{A})\degf(\tilde{D})}+\tensor{\hat{T}}{^{\tilde{I}}_{\tilde{J}\tilde{D}\tilde{A}}}\tensor{\hat{G}}{^J_{\tilde{I}\tilde{C}\tilde{B}}}(-1)^{\degf(\tilde{B})(\degf(\tilde{D})+\degf(\tilde{A}))}\right)\nonumber\\
            =&\frac{1}{9}\Bigg(\frac{1}{6}\left[\str{\lambda_{\tilde{A}}}\str{\lambda_{\tilde{B}}}g_{\tilde{C}\tilde{D}}+\str{\lambda_{\tilde{D}}}\str{\lambda_{\tilde{B}}}g_{\tilde{C}\tilde{A}}+\str{\lambda_{\tilde{D}}}\str{\lambda_{\tilde{A}}}g_{\tilde{C}\tilde{B}}\right]\nonumber\\
            &\qquad+\frac{1}{2(N-M)}\hat{G}_{\tilde{C}\tilde{D}\tilde{A}\tilde{B}}+6\hat{T}_{\tilde{C}\tilde{D}\tilde{A}\tilde{B}}\Bigg)
	    \end{align}
         \begin{align}
    	        \frac{1}{3}(-1)&^{\degf(\tilde{I})}\left(\tensor{\hat{G}}{^{\tilde{I}}_{\tilde{J}\tilde{A}\tilde{B}}}\tensor{\hat{T}}{^J_{\tilde{I}\tilde{C}\tilde{D}}}+\tensor{\hat{G}}{^{\tilde{I}}_{\tilde{J}\tilde{D}\tilde{B}}}\tensor{\hat{T}}{^J_{\tilde{I}\tilde{C}\tilde{A}}}(-1)^{\degf(\tilde{A})\degf(\tilde{D})}+\tensor{\hat{G}}{^{\tilde{I}}_{\tilde{J}\tilde{D}\tilde{A}}}\tensor{\hat{T}}{^J_{\tilde{I}\tilde{C}\tilde{B}}}(-1)^{\degf(\tilde{B})(\degf(\tilde{D})+\degf(\tilde{A}))}\right)\nonumber\\
                =&\frac{1}{9}\Bigg(\frac{1}{6}\left[\str{\lambda_{\tilde{C}}}\str{\lambda_{\tilde{D}}}g_{\tilde{A}\tilde{B}}+\str{\lambda_{\tilde{C}}}\str{\lambda_{\tilde{A}}}g_{\tilde{D}\tilde{B}}+\str{\lambda_{\tilde{C}}}\str{\lambda_{\tilde{B}}}g_{\tilde{D}\tilde{A}}\right]\nonumber\\
                &\qquad +\frac{1}{2(N-M)}\hat{G}_{\tilde{C}\tilde{D}\tilde{A}\tilde{B}}+6\hat{T}_{\tilde{C}\tilde{D}\tilde{A}\tilde{B}}\Bigg)
    	    \end{align}
         \begin{align}
    	        \frac{1}{3}(-1)&^{\degf(\tilde{I})}\left(\tensor{\hat{G}}{^{\tilde{I}}_{\tilde{J}\tilde{A}\tilde{B}}}\tensor{\hat{G}}{^J_{\tilde{I}\tilde{C}\tilde{D}}}+\tensor{\hat{G}}{^{\tilde{I}}_{\tilde{J}\tilde{D}\tilde{B}}}\tensor{\hat{G}}{^J_{\tilde{I}\tilde{C}\tilde{A}}}(-1)^{\degf(\tilde{A})\degf(\tilde{D})}+\tensor{\hat{G}}{^{\tilde{I}}_{\tilde{J}\tilde{D}\tilde{A}}}\tensor{\hat{G}}{^J_{\tilde{I}\tilde{C}\tilde{B}}}(-1)^{\degf(\tilde{B})(\degf(\tilde{D})+\degf(\tilde{A}))}\right)\nonumber\\
                =&\frac{1}{9}\left[(N-M)^2+8\right]\hat{G}_{\tilde{C}\tilde{D}\tilde{A}\tilde{B}}\ .
    	    \end{align}
        These formulas are the ones we use in Section~\ref{eq:onelooppotential} to get the explicit expression of the one-loop potential.
		\clearpage
		\bibliographystyle{JHEP}
		\bibliography{biblio.bib}
	\end{document}